\documentclass[11pt,document,nofootinbib,superscriptaddress,onecolumn,preprintnumbers,balancelastpage]{article}

\pdfoutput=1
\usepackage{jheppub}

\usepackage[section]{placeins}
\usepackage{graphicx,color}
\usepackage{amsmath,amssymb,amsfonts}
\usepackage{textcomp}
\usepackage{stackrel}
\usepackage{enumitem}
\usepackage{tikz}
\usepackage[english]{babel}
\usepackage{verbatim}
\usepackage{hyperref}
\usepackage{comment}
\usepackage[normalem]{ulem}
\usepackage{xcolor}
\usepackage{subcaption}
\usepackage{multirow, float}
\usepackage[vcentermath,enableskew]{youngtab}
\usepackage{ytableau}
\usepackage{dcolumn}
\usepackage{physics}
\usepackage[bottom]{footmisc}
\usepackage[toc,page]{appendix}
\usepackage{bm}
\usepackage{dsfont}
\usepackage{float}
\usepackage{graphicx}
\usepackage{tabularx}
\usepackage{caption}
\graphicspath{
  {./}
  {figures/}
}

\setlist[enumerate,1]{label={(\arabic*)}}

\usepackage{eso-pic}

\newcommand{\tf}{\texorpdfstring}
\newcommand{\gev}{~\text{GeV}}

\newcommand{\onbb}{0\nu\beta\beta}

\def\nn{\nonumber}

\definecolor{byzantine}{rgb}{0.74, 0.2, 0.64}

\newcount\Comments  
\newcommand{\kibitz}[2]{\ifnum\Comments=1\textcolor{#1}{#2}\fi}

\preprint{}

\makeatletter
\gdef\@fpheader{}
\makeatother

\title{
Matching from quark to hadronic operators: external source vs spurion methods
}

\author[a,b]{Gang~Li,}
\author[c,d,e]{Chuan-Qiang Song,}
\author[c,d,e]{Jiang-Hao Yu}

\affiliation[a]{\footnotesize School of Physics and Astronomy, Sun Yat-sen University, Zhuhai 519082, P. R. China}
\affiliation[b]{
Guangdong Provincial Key Laboratory of Quantum Metrology and Sensing, Sun Yat-Sen University, Zhuhai 519082, China}
\affiliation[c]{\footnotesize School of Fundamental Physics and Mathematical Sciences, Hangzhou Institute for Advanced Study, UCAS, Hangzhou 310024, China}
\affiliation[d]{\footnotesize School of Physical Sciences, University of Chinese Academy of Sciences,   Beijing 100049, P.R. China}
\affiliation[e]{\footnotesize Institute of Theoretical Physics, Chinese Academy of Sciences,   Beijing 100190, P. R. China}

\emailAdd{ligang65@mail.sysu.edu.cn}
\emailAdd{songchuanqiang21@mails.ucas.ac.cn}
\emailAdd{jhyu@itp.ac.cn}

\abstract{

Various weak processes at the hadronic scale have been utilized to search for new physics at high energy scale, which can be described by the QCD chiral Lagrangian matched from the low-energy effective theory (LEFT). Utilizing the chiral symmetry $SU(2)_L\times SU(2)_R \to SU(2)_V$ at the quark and hadronic levels, we make a comprehensive comparison of various matching methods, including the external source method, conventional spurion method, and our systematic spurion method. Although different methods show agreements for dimension-6 LEFT operator matching, we find that for higher-dimensional operators, the external source method is quite limited or inapplicable, the conventional spurion methods needs to introduce more and more spurions, while our spurion method does not need to introduce any new spurions than the ones in the dimension-6 matching. Using minimal set of spurions, we thus establish an one-to-one correspondence between the LEFT and chiral operators, with several examples, such as derivative operators at dimensions 7 and 8 with a single bilinear, four-quark operators at the dimension-9 level with two quark bilinears, which can be applied to the study of neutrino/electron scatterings and neutrinoless double beta decay.

}

\begin{document}
\emergencystretch 3em 
\maketitle
\newpage


\section{Introduction}

Effective field theories (EFTs) provide a systematic and model-independent way to study physics beyond the Standard Model (BSM)~\cite{Georgi:1993mps}.
While new particles may appear at TeV scale, their effects can be explored indirectly.
Depending on the energy scale of the processes studied, various degrees of freedom and gauge symmetries within the SM are employed to construct the Standard Model Effective Field Theory (SMEFT)~\cite{Grzadkowski:2010es}, Low-Energy Effective Theory (LEFT)~\cite{Jenkins:2017jig}, and Chiral Perturbation Theory ($\chi$PT)~\cite{Weinberg:1968de,Weinberg:1978kz}. They are valid at different energy ranges varying from the electroweak scale to the hadronic scale, and matchings among them are necessary in order to derive constraints from experimental measurements.

The matching of the SMEFT operators onto the LEFT operators has been studied in Refs.~\cite{Jenkins:2017jig,Carmona:2021xtq,Fuentes-Martin:2022jrf,Scholer:2023bnn,Graf:2025cfk,Liao:2025qwp}. 
As shown in the seminal works~\cite{Gasser:1983yg,Gasser:1984gg,Gasser:1987rb}, by treating the lepton currents as the external sources, one can match the SM electromagnetic and weak interactions to $\chi$PT, which was originally formulated through the chiral symmetry of the quantum chromodynamics (QCD) Lagrangian.
Furthermore, this method enables the matching of all semileptonic LEFT operators at the dimension-6 level to the chiral Lagrangian, which has been applied to the studies of neutrino
non-standard  interactions~\cite{Lindner:2016wff,AristizabalSierra:2018eqm,Farzan:2018gtr,Altmannshofer:2018xyo,Bischer:2019ttk,Hoferichter:2020osn,Du:2020dwr,Li:2024iij}, $\mu-e$ conversion~\cite{Crivellin:2017rmk,Cirigliano:2017azj,Bartolotta:2017mff,Dekens:2018pbu,Rule:2021oxe,Cirigliano:2022ekw,Haxton:2022piv,Haxton:2024lyc}, and $\mu \to e\gamma$~\cite{Dekens:2018pbu} etc.

While the external source method provides a convenient framework for matching, it is limited by the interactions under consideration. 
In scenarios where the external source method becomes inapplicable,
the spurion method is used for the matching between the LEFT and $\chi$PT,
which has been applied to the LEFT operators for four-quark~\cite{Grinstein:1985ut,Savage:1998yh,Prezeau:2003xn,Graesser:2016bpz,Cirigliano:2017ymo,Cirigliano:2018yza,Liao:2019gex,Akdag:2022sbn},
six-quark~\cite{Buchoff:2015qwa, Bijnens:2017xrz,He:2021mrt}, tensor~\cite{Mertens:2011ts}, or derivative interactions~\cite{Akdag:2022sbn} in the studies of
neutrinoless double beta $(0\nu\beta\beta)$ decay, neutron-antineutron oscillation, dinucleon decays, $s\to d\gamma$ and nucleon electric dipole moments. 
In this method, the spurion fields\,\footnote{The external sources can be considered specific spurion fields~\cite{Dekens:2018pbu}.} transform non-trivially under chiral symmetry, such that the LEFT operators are invariant under chiral transformations.

Conventionally, the mesonic chiral Lagrangian is constructed by mapping each chiral quark field to the Goldstone matrix field according to their transformation properties under chiral symmetry~\cite{Liao:2019gex}.  Complications arise in eliminating redundant operators particularly if the tensor and derivative interactions are considered, or when constructing chiral Lagrangians beyond the leading order (LO). 
Besides, it is essential to ensure that the charge conjugation $(C)$ and parity $(P)$ symmetries of the chiral operators are the same as those of the LEFT operators.

The LEFT operators involving a single quark bilinear can be directly matched to the chiral Lagrangian in the above way. However, for LEFT operators with two or more quark bilinears, the quark operators are usually decomposed into a sum of irreducible representations of the chiral group, before mapping the quark fields to 
the hadronic counterparts~\cite{Liao:2019gex}. For instance, in case of the dimension-9 operators contributing to $K^\pm\rightarrow \pi^\mp l^\pm l^\pm$, and focusing on the quark interaction $(\bar u_L\gamma^\mu d_L)(\bar u_L\gamma_\mu s_L) $, one can introduce a spurion $T$ into four-quark operators $\mathcal{O}_{LL}=T_{a\,c}^{b\,d}(\bar{q}_L^a\gamma^\mu q_{Lb})(\bar q_L^c\gamma_\mu q_{Ld})$, where $a,b,c,d$ are the $SU(3)$ flavor indices. 
The spurion $T$ transforms under the $SU(3)_L\times SU(3)_R$ group representation $T\in (\mathbf{8}_L\otimes\mathbf{8}_L)\otimes \mathbf{1}_R =  (\mathbf{27}_L\oplus\mathbf{10}_L\oplus\bar{\mathbf{10}}_L\oplus \mathbf{8}_L\oplus\mathbf{8}_L\oplus\mathbf{1}_L)\otimes \mathbf{1}_R$. For the process under consideration,
only the irreducible representation $\mathbf{27}_L \otimes \mathbf{1}_R$ remains~\cite{Liao:2019gex}. 
Analogous discussions of $K \to \pi \pi$ can be found in Refs.~\cite{Savage:1998yh,Cirigliano:2017ymo}.

However, as the number of quark bilinears increases, the irreducible decomposition of direct product representations becomes more complicated. For instance, this complexity is evident in the construction of six-quark operators governing neutron-antineutron oscillation~\cite{Buchoff:2015qwa, Bijnens:2017xrz}. 
Additionally, the process of eliminating redundant operators to derive the minimal form of the chiral Lagrangian using integration by parts (IBP), equations of motion (EOMs) and other identities in the conventional spurion method is notably cumbersome.

Recently, a systematic spurion method based on $SU(3)_L \times SU(3)_R \to SU(3)_V$ was proposed to establish matching between the LEFT and $\chi$PT~\cite{Song:2025snz}, 
which enables the construction of complete chiral Lagrangian using the Young tensor technique~\cite{Li:2020gnx,Li:2020xlh,Li:2022tec} with all redundant terms being eliminated. See also Ref.~\cite{Low:2022iim} for the construction of the QCD mesonic chiral Lagrangian of $\mathcal{O}(p^6)$ and $\mathcal{O}(p^8)$ without external sources.  
Unlike the external source method, the spurion approach treats the lepton and photon fields as degrees of freedom in both the LEFT the $\chi$PT. This enables the matching to be easily extended to higher-dimensional LEFT operators even for complicated Lorentz structures and $\chi$PT at higher orders by using the Young tensor technique~\cite{Sun:2025zuk}. Besides, in comparison with the conventional spurion method, the product of spurions is not decomposed into a direct sum of irreducible representations\,\footnote{
It is noted that Ref.~\cite{Graesser:2016bpz} considered the matching between four-quark operators responsible for generating $\onbb$ decay and the $\chi$PT in the conventional spurion method without employing irreducible decomposition.}
Consequently, regardless of the number of quark bilinear considered, no additional types of spurions are required in the LEFT operators.

In this work, we establish an one-to-one correspondence between the LEFT operators and the chiral operators in the systematic spurion method. The LEFT operators are initially constructed in the chiral basis. Their invariance under the $SU(2)_L \times SU(2)_R$ chiral symmetry can be ensured by introducing spurion fields dependent on the quark bilinear structure. Subsequently, these LEFT operators are reorganized in the $CP$ eigenstate in order to be matched to the chiral operators in the $\chi$PT using the dressed spurions, which are covariant under the subgroup $SU(2)_V$, 
using the Young tensor technique~\cite{Li:2020gnx,Li:2020xlh,Li:2022tec}. Ref.~\cite{Song:2025snz} relies on the $SU(3)_V$ symmetry and introduces a single spurion $\mathbf{T}$, and is thus restricted to operators with an even number of quark fields. In comparison, our new method is not subject to this limitation, and can be directly applied to the study of baryon number violation~\cite{Wilczek:1979hc,Ellis:1979hy,Weinberg:1979sa,Weinberg:1980bf,Abbott:1980zj,Kaymakcalan:1983uc}.

We explicitly demonstrate that the chiral operators derived from the external source method, the conventional spurion method, and our systematic spurion method are consistent when considering scalar/pseudo-scalar, vector/axial-vector, and tensor interactions. Besides, in the context of the conventional spurion method, we show the equivalence of
constructing the chiral Lagrangian in different bases. Furthermore, we emphasize that for the derivative interactions, which cannot be matched using the external source method, the systematic spurion method provides a more effective and convenient framework compared to the conventional spurion approach. Notably, we find that using a minimal set of fundamental spurions are sufficient in the systematic spurion method, even for the derivative and four-quark LEFT operators at high dimensions. 

It should be emphasized that although the external source method is convenient and the low-energy constants (LECs) for vector and axial-vector currents are related via hidden local symmetry, redundancies in the construction of chiral operators for scalar, pseudo-scalar, and tensor currents must be further reduced.
In contrast, our systematic spurion method does not rely on the hidden local symmetry, and all chiral operators are independent by construction using the Young tensor technique. Besides, the relations among LECs in this method are determined from $U(1)_{\rm em}$ gauge invariance (which relates LECs of the operators involving vector quark current to that of the kinetic term), together with the properties of QCD and additional constraints such as parity, which applies not only to vector and axial-vector interactions but also to other types of interactions.

The remainder of the paper is organized as follows. In Sec.~\ref{sec:external_source}, we review the $\chi\rm PT$ and the external source method. In Sec.~\ref{sec:spurion-conventional}, we employ the conventional spurion method to obtain the matching between LEFT and $\chi \rm PT$. Then we match the LEFT operators at dimension-6 level to chiral Lagrangian using the systematic spurion method and give the conversion relations of the three methods in Sec.~\ref{sec:Young tensor}. In Sec.~\ref{sec:dim7-9}, we consider the matching beyond dimension-6 LEFT level, including derivative and four-quark operators. We conclude in Sec.~\ref{sec:conclusion}. In the appendices, we provide some useful identities and the Cayley-Hamilton relation, expanded discussions and illustrative examples of tensor interactions, basics and example of the Young tensor technique, details of matching for derivative operators to chiral Lagrangian in the conventional spurion method, and effective operators with photon fields.

\section{Basics of \tf{$\chi$PT}{ChPT} and external source method}
\label{sec:external_source}

In this section, we will briefly review the basics of $\chi$PT, and the external source method. 
The Lagrangian is
\begin{align}
    \mathcal{L} = \mathcal{L}_{\rm QCD}^0 + \mathcal{L}_m + \mathcal{L}_{\rm ext}\;.
\end{align}
Here, $\mathcal{L}_{\rm QCD}^0$ denotes the masslesss QCD Lagrangian, and $\mathcal{L}_{\rm ext}$ describes the interactions with external sources, which are given by~\cite{Scherer:2002tk,Scherer:2012xha,Cata:2007ns}
\begin{align}
\label{eq:ext-interactions}
    \mathcal{L}_{\rm ext} = \bar{q} \gamma^\mu\left(v_\mu+\gamma^5 a_\mu\right) q-\bar{q}\left(s-i \gamma^5 p\right) q + \bar{q} \sigma_{\mu \nu} \bar{t}^{\mu \nu} q\;,
\end{align}
with the quark isospin doublet $q= (u,d)^T$. The quark mass term is written as
\begin{align}
    \mathcal{L}_m = -\bar{q} m_q q\;,\quad 
    m_q=\left(\begin{array}{cc}
    m_u & 0 \\
    0 & m_d
    \end{array}\right)\;,
\end{align}
where $m_u$ and $m_d$ denote the masses of $u$ and $d$ quarks, respectively.

In the chiral basis, we have
\begin{align}
    \mathcal{L}_{\rm ext} &= \bar{q}_L \gamma^\mu \ell_\mu q_L+\bar{q}_R \gamma^\mu r_\mu q_R-\left[\bar{q}_L (s - ip) q_R + {\rm h.c.} \right]  \nn\\
    &\quad +\bar{q}_L \sigma_{\mu \nu} t^{\mu \nu \dagger} q_R+\bar{q}_R \sigma_{\mu \nu} t^{\mu \nu} q_L\;.
\end{align}
The chiral fields $q_{L/R} = P_{L/R}~ q$ with the projectors $P_{L/R} \equiv (1\mp \gamma^5)/2 $, and the external sources
\begin{align}
   &\ell_\mu=v_\mu-a_\mu \;, \quad r_\mu=v_\mu+a_\mu\;.
\end{align}
The chiral quark fields transform under the chiral symmetry $SU(2)_L \times SU(2)_R$ as
\begin{align}
\label{eq:q-transform}
    q_L \to L q_L\;,\quad q_R \to R q_R\;,
\end{align}
where $L \in SU(2)_L$ and $R \in SU(2)_R$.

The matrix field $U(x)$ is defined as
\begin{align}
\label{eq:Umatrix}
    U(x) = \exp\left(i \frac{\vec{\phi}\cdot\vec{\tau}}{F_0}\right),
\end{align}
where $\vec{\tau}$ are Pauli matrices with $F_0 \simeq 92.4$ MeV the pion decay constant and $\vec{\phi}$ represent the three Nambu-Goldstone bosons from the symmetry breaking. Under chiral transformation, $U \to R U L^\dagger$. The covariant derivative for fields transforming as $A \to R A L^\dagger$ is given by~\cite{Fearing:1994ga,Ebertshauser:2001nj}
\begin{align}
\label{D}
    D_\mu A \equiv \partial_\mu A - i r_\mu A + i A \ell_\mu\;.
\end{align}
The other building blocks that involve the external source are
\begin{align}
    \chi &= 2B(s + m_q +ip)\;,\notag\\
    F_L^{\mu\nu} &= \partial^\mu \ell^\nu - \partial^\nu \ell^\mu - i[\ell^\mu,\ell^\nu]\;,\label{msl}\\
    F_R^{\mu\nu} &= \partial^\mu r^\nu - \partial^\nu r^\mu - i[r^\mu,r^\nu]\;.\notag
\end{align}
Here, the LEC $B \simeq 2.8\gev$ describes the quark condensation.
The chiral Lagrangian in the $U$-parameterization~\cite{Sun:2025zuk} can be constructed~\cite{Gasser:1983yg,Gasser:1984gg,Fearing:1994ga,Ebertshauser:2001nj} using the building blocks $(U\;,U^\dagger\;,\chi\;,\chi^\dagger\;, F_{L}^{\mu\nu}\;,F_{R}^{\mu\nu})$ and the covariant derivative $D_\mu$ in Eq.~\eqref{D}. 

The chiral Lagrangian of $\mathcal{O}(p^2)$ takes the following form
\begin{align}
    \mathcal{L}_2 &= \frac{F_0^2}{4}\Tr\left[D_\mu U(D^\mu U)^\dagger\right] + \frac{F_0^2}{4}\Tr\left(\chi U^\dagger + U\chi^\dagger\right)\;,\label{eq:source_vector}
\end{align}
with the Hermitian conjugate
\begin{align}
    (D^\mu U)^\dagger &= \partial^\mu U^\dagger + i U^\dagger r^\mu - i \ell^\mu U^\dagger\;.
\end{align}
The first term in Eq.~\eqref{eq:source_vector} includes vector and axial-vector contributions
\begin{align}
    \Tr\left[D_\mu U(D^\mu U)^\dagger\right] &\supset -2i\Tr\left(r^\mu U\partial_\mu U^\dagger - \ell^\mu \partial_\mu U^\dagger U\right)\;,\label{eq:source_lr}\\
    &= -2i\Tr\left(v^\mu[U,\partial_\mu U^\dagger] + a^\mu\{U,\partial_\mu U^\dagger\}\right)\;.\label{eq:source_va}
\end{align}
The second term can be expanded as
\begin{align}
    \Tr\left(\chi U^\dagger + U\chi^\dagger\right) &= 2B\Tr\left\{\left[(s+ m_q) U^\dagger + U(s+m_q)\right] + i\left[pU^\dagger - Up\right]\right\}\;,
\end{align}
which describe the external scalar/pseudo-scalar sources. 

At $\mathcal{O}(p^4)$, the chiral Lagrangian is expressed as~\cite{Gasser:1983yg} 
\begin{equation}
\label{eq:source_p4}
\begin{aligned}
\mathcal{L}_4= & \frac{l_1}{4}\left\{\operatorname{Tr}\left[D_\mu U\left(D^\mu U\right)^{\dagger}\right]\right\}^2+\frac{l_2}{4} \operatorname{Tr}\left[D_\mu U\left(D_\nu U\right)^{\dagger}\right] \operatorname{Tr}\left[D^\mu U\left(D^\nu U\right)^{\dagger}\right] \\
& +\frac{l_3}{16} \left[\operatorname{Tr}\left(\chi U^\dagger+U\chi^\dagger\right)\right]^2+\frac{l_4}{4} \operatorname{Tr}\left[D_\mu U(D^\mu \chi)^\dagger+D_\mu\chi(D^\mu U)^\dagger\right]\\
& +l_5 \left[\operatorname{Tr}(F_{R\mu\nu}UF_{L}^{\mu\nu} U^\dagger)-\frac{1}{2}\operatorname{Tr}(F_{L\mu\nu}F_L^{\mu\nu}+F_{R\mu\nu}F_R^{\mu\nu})\right]\\
&+i\frac{l_6}{2}\operatorname{Tr}\left[F_{R\mu\nu}D^\mu U(D^\nu U)^\dagger+F_{L\mu\nu}(D^\mu U)^\dagger D^\nu U\right] \\
& -\frac{l_7}{16}\left[\operatorname{Tr}\left(\chi U^{\dagger}-U \chi^{\dagger}\right)\right]^2+\frac{h_1+h_3}{16} \operatorname{Tr}\left(U \chi^{\dagger} U \chi^{\dagger}+\chi U^{\dagger} \chi U^{\dagger}\right) \\
& -\frac{h_1-h_3}{16} \left\{\left[\operatorname{Tr}\left(\chi U^\dagger+U\chi^\dagger\right)\right]^2+\left[\operatorname{Tr}\left(\chi U^\dagger-U\chi^\dagger\right)\right]^2-2\operatorname{Tr}\left(\chi U^\dagger\chi U^\dagger+U\chi^\dagger U\chi^\dagger\right)\right\}\\
&-2h_2 \operatorname{Tr}\left(F_{L\mu\nu}F_L^{\mu\nu}+F_{R\mu\nu}F_R^{\mu\nu}\right)\;,
\end{aligned}
\end{equation}
where $l_i$ $(i=1,\cdots,7)$ and $h_i$ $(i=1,2,3)$ denote LECs of the chiral operators.

On the other hand, square root of the matrix field $U$ is given by
\begin{align}
\label{eq:u-matrix}
    u(x) = \exp\left(i \frac{\vec{\phi}\cdot\vec{\tau}}{2F_0}\right) = 1 + i \frac{\vec{\phi}\cdot\vec{\tau}}{2F_0} + \cdots\;.
\end{align}
Under the chiral transformation,
\begin{align}
\label{eq:u}
    u \to R u K^\dagger = K u L^\dagger, \quad u^\dagger \to L u^\dagger K^\dagger = K u^\dagger R^\dagger\;, 
\end{align}
where $K$ is an $SU(2)$-valued function that defines the non-linear realization of the chiral symmetry  breaking $SU(2)_L \times SU(2)_R \to SU(2)_V$.

To systematically construct the chiral Lagrangian in the $u$-parameterization~\cite{Sun:2025zuk}, the following building blocks are introduced
\begin{align}
    \chi_\pm &= u^\dagger \chi u^\dagger \pm u \chi^\dagger u\;,\label{mass}\\
    u_\mu &= i\left[u^\dagger (\partial_\mu - i r_\mu)u - u(\partial_\mu - i \ell_\mu)u^\dagger\right] = i u^\dagger D_\mu U u^\dagger\;,\label{eq:vielbein}\\
    f_{\pm}^{\mu\nu} &= u F_L^{\mu\nu} u^\dagger \pm u^\dagger F_R^{\mu\nu} u\;,\label{eq:fmunu}\\
    \label{eq:tensor}
    t_\pm^{\mu\nu} &= u^\dagger t^{\mu\nu} u^\dagger \pm u t^{\mu\nu\dagger} u\;,
\end{align}
where the constituent fields $\chi$, $F_L^{\mu\nu}$ and $F_R^{\mu\nu}$ are defined in Eq.\eqref{msl}. The properties of building blocks and covariant derivative under the $C$ and $P$ transformations can be found in Refs.~\cite{Fettes:2000gb,Song:2024fae}.

The building blocks $X = \chi_\pm$, $u^\mu$, $f^{\mu\nu}_\pm$, and $t_\pm^{\mu\nu}$ transform under
$SU(2)_L \times SU(2)_R$
as 
\begin{align}
    X \to K X K^\dagger\;,\quad K \in SU(2)_V\;.
\end{align}
It should be noted that in the $u$-parameterization, it is $SU(2)_V$ to be respected while $SU(2)_L \times SU(2)_R$ invariance is kept automatically by the definitions of $X$~\cite{Bijnens:2023hyv,Sun:2025zuk}.  
The covariant derivative acting on $X$ is defined as~\cite{Fettes:2000gb,Mateu:2007tr,Bijnens:1999sh}:
\begin{align}
\label{eq:covariant derivative}
    \nabla_\mu X &= \partial_\mu X + [\Gamma_\mu,X]\;,\quad
    \Gamma_\mu \equiv \frac{1}{2}\left[u^\dagger(\partial_\mu - i r_\mu)u + u(\partial_\mu - i \ell_\mu)u^\dagger\right]\;.
\end{align}
The combination of $\nabla_\mu u_\nu $ and $\nabla_\nu u_\mu$ yields
\begin{align}
\label{eq:hmunu-fmunu}
    h_{\mu\nu} &= \nabla_\mu u_\nu + \nabla_\nu u_\mu\;,\nn\\
    f_{-\mu\nu} &= \nabla_\mu u_\nu - \nabla_\nu u_\mu\;.
\end{align}

In parallel to the above construction, the chiral Lagrangian can be equivalently constructed~\cite{Bijnens:1999sh,Bijnens:2001bb,Cata:2007ns,Bijnens:2018lez,Bijnens:2023hyv,Li:2024ghg} through the building blocks $(u^\mu\;,\chi_\pm\;,f_\pm^{\mu\nu}\;,t_\pm^{\mu\nu})$ and the covariant derivative $\nabla_\mu$ in Eq.~\eqref{eq:covariant derivative}. Both bases of building blocks have been used in the external source method, referred to as the LR basis and $u$ basis, respectively, in Ref.~\cite{Bijnens:2023hyv}.

In the $u$ basis, the chiral Lagrangian of $\mathcal{O}(p^2)$ is given by~\cite{Bijnens:1999sh}
\begin{equation}
\label{eq:ubasis2}
    \mathcal{L}_2^\prime=\frac{F_0^2}{4}\langle u^\mu u_\mu+\chi_+\rangle\;.
\end{equation}
While at $p^4$ order, the chiral Lagrangian can be expressed as~\cite{Li:2024ghg}\,\footnote{In Ref.~\cite{Li:2024ghg}, the external sources are taken into account, while the complete and independent chiral operators are constructed using the Young tensor technique~\cite{Li:2020gnx,Li:2020xlh,Li:2022tec}. }
\begin{align}
\label{eq:ubasis}
    \mathcal{L}_4^\prime=&L_1\langle u_\mu u^\mu\rangle\langle u_\nu u^\nu\rangle+L_2\langle u_\mu u_\nu\rangle\langle u^\mu u^\nu\rangle+L_3\langle\chi_+\rangle\langle u^\mu u_\mu\rangle+iL_4\langle f_+^{\mu\nu}u_\mu u_\nu\rangle+L_5\langle f_+^{\mu\nu}f_{+\mu\nu}\rangle\notag\\
    &+ L_6\langle f_-^{\mu\nu}f_{-\mu\nu}\rangle+L_7\langle\chi_+\chi_+\rangle+L_8\langle\chi_-\chi_-\rangle+L_9\langle\chi_+\rangle\langle\chi_+\rangle+L_{10}\langle\chi_-\rangle\langle\chi_-\rangle\;,
\end{align}
where $\langle \cdots \rangle$ denotes the trace operation over the matrices, same as the notation ``Tr'', and $L_i$ $(i=1,\cdots,10)$ denote the LECs.
Besides, for the external tensor source, the chiral Lagrangian has also been constructed in the $u$ basis~\cite{Cata:2007ns}, which is given by
\begin{align}
\label{eq:tensor-ubasis}
\Delta \mathcal{L}_{4}^\prime =\Lambda_1\left\langle t_{+}^{\mu \nu} f_{+\mu \nu}\right\rangle-i \Lambda_2\left\langle t_{+}^{\mu \nu} u_\mu u_\nu\right\rangle+\Lambda_3\left\langle t_{+}^{\mu \nu} t_{\mu \nu}^{+}\right\rangle+\Lambda_4\left\langle t_{+}^{\mu \nu}\right\rangle^2\;,
\end{align}
where $\Lambda_i$ $(i=1,2,3,4)$ are the LECs of the chiral operators. Thus, there are 14 independent chiral operators for the interactions defined in Eq.~\eqref{eq:ext-interactions} at $p^4$ order.  
While one can obtain the same results in the LR and $u$ bases, it is shown that deriving the mesonic chiral Lagrangians of $\mathcal{O}(p^6)$~\cite{Bijnens:1999sh,Bijnens:2001bb} and $\mathcal{O}(p^8)$~\cite{Bijnens:2018lez,Bijnens:2023hyv,Li:2024ghg} for the electroweak interactions is easier by using $(u^\mu\;,\chi_\pm\;,f_\pm^{\mu\nu}\;,t_\pm^{\mu\nu})$.

\section{Conventional spurion method}
\label{sec:spurion-conventional}

\subsection{Setup}

We analyze quark-current interactions by categorizing them into distinct types based on their Lorentz and chiral transformation properties. In the chiral basis, the vector and axial-vector interactions are expressed as
\begin{align}
\mathcal{L}_{V,A}^q &= \left(\bar{q}_R \gamma_\mu \lambda_1^\mu q_R + \bar{q}_L \gamma_\mu \lambda_2^\mu q_L\right) + \text{h.c.}\nn \\
&= \bar{q}_R \gamma_\mu \lambda_R^\mu q_R + \bar{q}_L \gamma_\mu \lambda_L^\mu q_L\;,
\end{align}
where the Hermitian spurions are defined as $\lambda_R^\mu \equiv \lambda_1^\mu + \lambda_1^{\mu \dagger}$ and $\lambda_L^\mu \equiv \lambda_2^\mu + \lambda_2^{\mu\dagger}$. Here, we use $\lambda_1^\mu$ and $\lambda_2^\mu$ to denote non-Hermitian spurions.

For scalar and pseudo-scalar interactions, we have
\begin{align}
\mathcal{L}_{S,P}^q &= \left(\bar{q}_L \lambda_1 q_R + \bar{q}_R \lambda_2 q_L\right) + \text{h.c.} \nn\\
&= \bar{q}_L \lambda^\dagger q_R + \bar{q}_R \lambda q_L\;,
\end{align}
where the spurion $\lambda \equiv \lambda_2 + \lambda_1^\dagger$, and its Hermitian conjugate $\lambda^\dagger \equiv \lambda_1 + \lambda_2^\dagger$.

The tensor current interactions take the form as follows
\begin{align}
\mathcal{L}_T^q &= \bar{q}_L \sigma^{\mu\nu} \lambda_{1\mu\nu} q_R + \bar{q}_R \sigma^{\mu\nu} \lambda_{2,\mu\nu} q_L + \text{h.c.}\nn \\
&= \bar{q}_L \sigma^{\mu\nu} \lambda_{\mu\nu}^\dagger q_R + \bar{q}_R \sigma^{\mu\nu} \lambda_{\mu\nu} q_L\;,
\end{align}
where the spurion $\lambda_{\mu\nu} \equiv \lambda_{2,\mu\nu} + \lambda_{1,\mu\nu}^\dagger$, and its Hermitian conjugate $\lambda_{\mu\nu}^\dagger \equiv \lambda_{1,\mu\nu} + \lambda_{2,\mu\nu}^\dagger$.

Given the transformation properties of the chiral quark fields, the spurions transform under chiral symmetry as
\begin{align}
    \lambda_R^\mu &\to R \lambda_R^\mu R^\dagger\;, & \lambda_L^\mu &\to L \lambda_L^\mu L^\dagger\;,\nn\\
    \lambda &\to R \lambda L^\dagger\;, & \lambda^\dagger &\to L \lambda^\dagger R^\dagger\;,\nn\\
    \lambda_{\mu\nu} &\to R \lambda_{\mu\nu} L^\dagger\;, & \lambda_{\mu\nu}^\dagger &\to L \lambda_{\mu\nu}^\dagger R^\dagger\;.
\end{align}

By comparing Eq.~\eqref{eq:q-transform} with Eq.~\eqref{eq:u-matrix}, we establish the LO matching from quark to meson fields through the following substitution~\cite{Liao:2019gex}:
\begin{align}
\label{eq:quark-LO}
q_L \to u^\dagger\;, \quad q_R \to u\;, \quad \bar{q}_L \to u\;, \quad \bar{q}_R \to u^\dagger.
\end{align}

We introduce the covariant derivative~\cite{Manohar:1983md},
\begin{align}
\label{eq:derivative-u}
D_\mu = \partial_\mu - i \mathcal{V}_\mu\;, \quad \mathcal{V}_\mu = \frac{i}{2} \left(u^\dagger \partial_\mu u + u \partial_\mu u^\dagger\right)\;.
\end{align}
The chiral transformation of the connection field $\mathcal{V}_\mu$ is
\begin{align}
    \mathcal{V}_\mu \to K \mathcal{V}_\mu K^\dagger + i K \partial_\mu K^\dagger\;,
\end{align}
such that
\begin{align}
    D_\mu u &\to \partial_\mu (K u L^\dagger) - i (K \mathcal{V}_\mu K^\dagger + i K \partial_\mu K^\dagger) K u L^\dagger = K D_\mu u L^\dagger, \\
    D_\mu u^\dagger &\to \partial_\mu (K u^\dagger R^\dagger) - i (K \mathcal{V}_\mu K^\dagger + i K \partial_\mu K^\dagger) K u^\dagger R^\dagger = K D_\mu u^\dagger R^\dagger.
\end{align}
The second-order covariant derivatives similarly transform as
\begin{align}
   \mathcal D^2 u &\to K \mathcal D^2 u L^\dagger\;, &
   \mathcal D^2 u^\dagger &\to K \mathcal D^2 u^\dagger R^\dagger\;,
\end{align}
where $\mathcal D^2 \equiv D^\mu D_\mu$ or $D^\mu D_\nu$.

Thus, the matching from quark to meson at higher orders is
\begin{itemize}
    \item next-to-leading order (NLO)~\cite{Liao:2019gex}:
        \begin{align}
        \label{eq:quark-NLO}
        q_L \to (D_\mu u)^\dagger\;, \quad q_R \to (D_\mu u^\dagger)^\dagger\;, \quad \bar{q}_L \to D_\mu u\;, \quad \bar{q}_R \to D_\mu u^\dagger\;,
        \end{align}
    \item  next-to-next-to-leading order (NNLO):
        \begin{align}
        \label{eq:quark-NNLO}
        q_L \to (\mathcal D^2 u)^\dagger, \quad q_R \to (\mathcal D^2 u^\dagger)^\dagger, \quad \bar{q}_L \to\mathcal D^2 u, \quad \bar{q}_R \to\mathcal D^2 u^\dagger.
        \end{align}
\end{itemize}

It should be emphasized that the mapping from single quark to meson fields in Eqs.~\eqref{eq:quark-LO}~\eqref{eq:quark-NLO}\eqref{eq:quark-NNLO} provides a feasible way of implementing the matching in the conventional spurion method. More comprehensive procedure can be found in the wider literature, see e.g. Refs.~\cite{Grinstein:1985ut,Savage:1998yh,Prezeau:2003xn,Graesser:2016bpz,Cirigliano:2017ymo,Cirigliano:2018yza,Akdag:2022sbn}.

\subsection{Matching to the chiral Lagrangian}

The construction of chiral Lagrangians from fundamental quark operators requires careful treatment of both the symmetry properties and the derivative expansion. In this section, we present the matching for vector, axial-vector, scalar, pseudo-scalar, and tensor currents up to $\mathcal{O}(p^4)$ in the chiral power counting. Using the conventional spurion method, we demonstrate how quark-level operators map to their hadronic counterparts. Particular attention is given to comparing results obtained in the conventional spurion method with those derived via the external source method.

\subsubsection{Vector and axial-vector currents}

The matching for quark-level vector/axial-vector interactions is obtained as
\begin{align}
\label{eq:spurion-matching-va}
\mathcal{L}_{V,A}^q = \bar{q}_R \gamma_\mu \lambda_R^\mu q_R + \bar{q}_L \gamma_\mu \lambda_L^\mu q_L\to \mathcal{L}_{V,A}^{(d)}  \;,
\end{align}
where $d$ represents the number of derivatives.

At LO in the chiral expansion, i.e. $p^2$ order, we have  $d=1$ and 
\begin{align}
\mathcal{L}_{V,A}^{(1)} &= \Tr\left[u^\dagger \lambda_R^\mu (D_\mu u^\dagger)^\dagger + u\lambda_L^\mu (D_\mu u)^\dagger\right] g_{V,1}^{(1)} \notag \\
&\quad + \Tr\left[D_\mu u^\dagger \lambda_R^\mu u + D_\mu u \lambda_L^\mu u^\dagger\right] g_{V,2}^{(1)}\;.
\end{align}
Using the identity in Eq.~\eqref{eq:identity1}, it simplifies to
\begin{align}
\label{eq:conventional-VA-LO}
\mathcal{L}_{V,A}^{(1)} &= \Tr\left[\lambda_R^\mu u D_\mu u^\dagger + \lambda_L^\mu u^\dagger D_\mu u\right] \tilde{g}_V^{(1)}\;,
\end{align}
where $\tilde{g}_V^{(1)} \equiv g_{V,2}^{(1)} - g_{V,1}^{(1)}$.

Expressed in terms of $U$ fields, one has
\begin{align}
\mathcal{L}_{V,A}^{(1)} &= \frac{1}{2} \Tr\left[\lambda_R^\mu U \partial_\mu U^\dagger + \lambda_L^\mu U^\dagger \partial_\mu U\right] \tilde{g}_V^{(1)}\;.
\end{align}
Decomposing $\lambda_{R/L}^\mu = v^\mu \pm a^\mu$ gives
\begin{align}
\mathcal{L}_V^{(1)} &= \frac{1}{2} \Tr\left(v^\mu [U, \partial_\mu U^\dagger]\right) \tilde{g}_V^{(1)}\;, \\
\mathcal{L}_A^{(1)} &= \frac{1}{2} \Tr\left(a^\mu \{U, \partial_\mu U^\dagger\}\right) \tilde{g}_V^{(1)}\;,
\end{align}
which agree with Eq.~\eqref{eq:source_va} when $D_\mu \to \partial_\mu$.

At NNLO in the chiral expansion, i.e. $p^4$ order, the matching proceeds through the correspondence 
\begin{align}
    \bar{q}_L \gamma_\mu \lambda_L^\mu q_L \to \Tr\left(D^\nu D_\nu u \lambda_L^\mu (D_\mu u)^\dagger\right) \supset \frac{1}{4} \Tr\left(u \partial^\nu U^\dagger \partial_\nu U \lambda_L^\mu (D_\mu u)^\dagger\right)\;.
\end{align}
By using the identity in Eq.~\eqref{eq:identity2}, one has 
\begin{align}
     \Tr\left(u \partial^\nu U^\dagger \partial_\nu U \lambda_L^\mu (D_\mu u)^\dagger\right) =  \Tr\left(\partial^\nu U^\dagger \partial_\nu U \lambda_L^\mu (\partial_\mu U)^\dagger U\right)\;.
\end{align}
which aligns with the $l_1$ term in Eq.~\eqref{eq:source_p4} after considering the Cayley-Hamilton relation and relate the single trace and double trace terms (see Appendix~\ref{sec:Cayley-Hamilton}). The $l_2$ term emerges similarly through the replacement $\bar{q}_L \to D_\mu D_\nu u$ and $q_L \to D^\nu u$.

\subsubsection{Scalar and pseudo-scalar currents}

The matching for quark-level scalar and pseudo-scalar interactions  is obtained as
\begin{align}  
\label{eq:spurion-matching-sp}
\mathcal{L}_{S,P}^q = \bar{q}_L \lambda^\dagger q_R + \bar{q}_R \lambda q_L  \to  \mathcal{L}_{S,P}^{(d)} \;.
\end{align}  

At $\mathcal{O}(p^2)$, the chiral Lagrangian is  
\begin{align}  
\label{eq:conventional-SP-LO}
\mathcal{L}_{S,P}^{(0)} &= \Tr\left[u \lambda^\dagger u + u^\dagger \lambda u^\dagger\right] g_S^{(0)} \;.  
\end{align}  

At $\mathcal{O}(p^4)$, the derivative expansion yields  
\begin{align}  
\mathcal{L}_{S,P}^{(2)} &= \Tr\left[D_\mu u \lambda^\dagger (D^\mu u^\dagger)^\dagger + D_\mu u^\dagger \lambda (D^\mu u)^\dagger\right] g_S^{(2)} \;.  
\end{align}  
Matching at $p^4$ order also generates terms through the replacement $\bar{q}_L \to D^\mu D_\mu u$, giving  
\begin{align}  
\bar{q}_L \lambda^\dagger q_R &\to \Tr\left(D^\nu D_\nu u \lambda^\dagger u\right) \supset \frac{1}{4} \Tr\left(u \partial^\nu U^\dagger \partial_\nu U \lambda^\dagger u\right) \;,  
\end{align}  
which corresponds to the $l_4$ term in Eq.~\eqref{eq:source_p4}.

Using the identities in Appendix~\ref{app:identities}, one can express the chiral Lagrangians
in terms of the $U$ field,  
\begin{align}  
\mathcal{L}_{S,P}^{(0)} &= \Tr\left[\lambda^\dagger U + \lambda U^\dagger\right] g_S^{(0)} \;, \\  
\mathcal{L}_{S,P}^{(2)} &= -\frac{1}{4} \Tr\left[\partial^\mu U \partial_\mu U^\dagger (U \lambda^\dagger + \lambda U^\dagger)\right] g_S^{(2)} \;.
\end{align}  
 For $\lambda = -(s + ip)$ and $\lambda^\dagger = -(s - ip)$, they agree with the scalar term in Eq.~\eqref{eq:source_vector} at LO,  
and the $l_4$ term in Eq.~\eqref{eq:source_p4} at NLO with $D_\mu \to \partial_\mu$, respectively. At the LO, we can easily obtain that the LECs satisfy $g_S^{(0)}=-F_0^2 B/2$.

From Eq.~\eqref{eq:vielbein}, when $r_\mu =0$ and $\ell_\mu = 0$, the vielbein reduces to
\begin{align}
\label{eq:uhat}
    \hat{u}_\mu = i\left(u^{\dagger} \partial_\mu u-u \partial_\mu u^{\dagger}\right)\;.
\end{align}
Using this building block, 
    \begin{align}
        \hat u^\mu \hat u_\mu = - u^\dagger \partial^\mu U u^\dagger u^\dagger \partial_\mu U u^\dagger = u^\dagger \partial^\mu U \partial_\mu U^\dagger  u\;,
    \end{align}
we can obtain
    \begin{align}
    \Tr \left[ \partial^\mu U 
     \partial_\mu U^\dagger \left( U \lambda^\dagger + \lambda U^\dagger \right) \right]
     &= \Tr \left[ \left(u \lambda^\dagger u + u^\dagger \lambda u^\dagger \right) \hat u^\mu \hat u_\mu \right]\;.
    \end{align}

\subsubsection{Tensor currents}
\label{sec:tensor-current}

The matching for quark-level tensor interaction is obtained as
\begin{align}
\label{eq:tensormatching}
 \mathcal{L}_{T}^q =   \bar q_L \sigma^{\mu\nu} \lambda_{\mu\nu}^\dagger q_R + \bar q_R \sigma^{\mu\nu} \lambda_{\mu\nu} q_L \to \mathcal{L}_{T}^{(d)} \;.
\end{align}
Under the $C$ transformation, we have
\begin{align}  
\label{eq:C-tensor-quark}
\mathcal{L}_{T}^q &\stackrel{C}{\longrightarrow} - \left( \bar{q}_L \sigma^{\mu\nu} \lambda_{\mu\nu}^{\dagger c} q_R + \bar{q}_R \sigma^{\mu\nu} \lambda_{\mu\nu}^c q_L \right)\;.
\end{align}  
Here, we have introduced
\begin{align}
\label{eq:lambda-c}
    \lambda_{\mu\nu}^c = C \lambda_{\mu\nu}^T C^{-1}\;,\quad \lambda_{\mu\nu}^{\dagger c} = C \lambda_{\mu\nu}^{\dagger T} C^{-1}\;,
\end{align}
where $C = i \gamma^2 \gamma^0$ is the usual charge conjugation matrix. Further details on the $C$ transformation of the spurions are provided in Sec.~\ref{sec:C-transformation}.

We obtain that the chiral Lagrangian at $\mathcal{O}(p^4)$ includes
\begin{align}  
\mathcal{L}_{T}^{(2)} &\supset \Tr\left[ D^\nu u \lambda_{\mu\nu}^\dagger (D^\mu u^\dagger)^\dagger + D^\nu u^\dagger \lambda_{\mu\nu} (D^\mu u)^\dagger\right] g_T^{(2)}.  
\end{align}  
By using Eq.~\eqref{eq:u-mu-relations}, the above expression becomes
\begin{align}  
\mathcal{L}_{T}^{(2)} 
&\supset -\frac{1}{4} \Tr\left[ (u \lambda_{\mu\nu}^\dagger u + u^\dagger \lambda_{\mu\nu} u^\dagger) \hat{u}^\mu \hat{u}^\nu \right] g_T^{(2)}\;. 
\end{align}  
The chiral Lagrangian $\mathcal{L}_T^{(2)}$ is required to have the same $C$ property as the quark-level Lagrangian $\mathcal{L}_T^q$, so other terms in chiral Lagrangian must be included. 

We observe that the interactions in Eq.~\eqref{eq:tensormatching} can be equivalently expressed as
\begin{align}  
\mathcal{L}_{T}^q =  - \left(\bar{q}_L \sigma^{\nu\mu} \lambda_{\mu\nu}^\dagger q_R + \bar{q}_R \sigma^{\nu\mu} \lambda_{\mu\nu} q_L\right) \;,
\end{align}  
which is matched to the chiral operators
\begin{align}  
\label{eq:conventional-T-NLO-2}
\mathcal{L}_{T}^{(2)} & \supset - \Tr\left[ 
D^\mu u \lambda_{\mu\nu}^\dagger (D^\nu u^\dagger)^\dagger + D^\mu u^\dagger \lambda_{\mu\nu} (D^\nu u)^\dagger\right] \tilde g_T^{(2)}\;.  
\end{align}

To preserve the $C$ property in the matching, we have $\tilde g_T^{(2)} = g_T^{(2)}$,  and the chiral Lagrangian at $\mathcal{O}(p^4)$ is expressed as
\begin{align}  
\mathcal{L}_{T}^{(2)} &\supset \Tr\left[ D^\nu u \lambda_{\mu\nu}^\dagger (D^\mu u^\dagger)^\dagger + D^\nu u^\dagger \lambda_{\mu\nu} (D^\mu u)^\dagger\right.\nn\\ 
&\quad \left. -
D^\mu u \lambda_{\mu\nu}^\dagger (D^\nu u^\dagger)^\dagger - D^\mu u^\dagger \lambda_{\mu\nu} (D^\nu u)^\dagger\right] g_T^{(2)}.  
\end{align}  
By using Eq.~\eqref{eq:u-mu-relations}, the above expression becomes
\begin{align}  
\label{eq:conventional-T-NLO}
\mathcal{L}_{T}^{(2)} 
&\supset -\frac{1}{4} \Tr\left[ (u \lambda_{\mu\nu}^\dagger u + u^\dagger \lambda_{\mu\nu} u^\dagger) [\hat{u}^\mu, \hat{u}^\nu] \right] g_T^{(2)}\;. 
\end{align}

Under the $C$ transformation,
\begin{align}  
(u \lambda_{\mu\nu}^\dagger u + u^\dagger \lambda_{\mu\nu} u^\dagger) &\stackrel{C}{\longrightarrow} (u^T \lambda_{\mu\nu}^{\dagger c} u^T + u^{\dagger T}  \lambda_{\mu\nu}^c u^{\dagger T})\;, \\  
[\hat{u}^\mu, \hat{u}^\nu] &\stackrel{C}{\longrightarrow} [\hat{u}^{\mu T}, \hat{u}^{\nu T}]\;,  
\end{align}  
so that
\begin{align}  
\Tr\left[ (u \lambda_{\mu\nu}^\dagger u + u^\dagger \lambda_{\mu\nu} u^\dagger) [\hat{u}^\mu, \hat{u}^\nu] \right] &\stackrel{C}{\longrightarrow} -\Tr\left[ (u \lambda_{\mu\nu}^{\dagger c} u + u^\dagger \lambda_{\mu\nu}^c u^\dagger) [\hat{u}^\mu, \hat{u}^\nu] \right]\;,
\label{eq:C-tensor}  
\end{align} 
which 
transforms in the same way under $C$ transformation as $\mathcal{L}_T^q$.

Moreover, the quark-level tensor interaction can also be written in a different Lorentz structure as follows 
\begin{align}
\label{eq:translation}
    \mathcal{L}_T^q
    =&\, \frac{i}{2}\varepsilon^{\mu\nu\alpha\beta}(\bar q_L \sigma_{\alpha\beta} \lambda_{\mu\nu}^\dagger q_R - \bar q_R \sigma_{\alpha\beta}  \lambda_{\mu\nu} q_L)\;,
\end{align}
by using the relation
\begin{align}
\label{eq:epsilon}
    \sigma^{\mu \nu} \gamma^5=\frac{i}{2} \varepsilon^{\mu \nu \alpha\beta} \sigma_{\alpha \beta}\;,
\end{align}
where $\varepsilon^{\mu \nu \lambda \rho}$ is the Levi-Civita tensor with $\varepsilon^{0123} = +1$.
Thus, we can obtain additional terms of 
the chiral Lagrangian at $\mathcal{O}(p^4)$:
\begin{align}  
\mathcal{L}_{T}^{(2)} &\supset \dfrac{i}{4}\varepsilon^{\mu\nu\alpha\beta} \Tr\left[ D_\beta u \lambda_{\mu\nu}^\dagger (D_\alpha u^\dagger)^\dagger - D_\beta u^\dagger \lambda_{\mu\nu} (D_\alpha u)^\dagger\right.\nn\\ 
&\quad \left. -
D_\alpha u \lambda_{\mu\nu}^\dagger (D_\beta u^\dagger)^\dagger + D_\alpha u^\dagger \lambda_{\mu\nu} (D_\beta u)^\dagger\right] g_T^{(2)\prime}\;,
\end{align}  
which in the $u$ basis is expressed as
\begin{align}  
\label{eq:conventional-T2-NLO}
\mathcal{L}_{T}^{(2)} 
&\supset -\frac{i}{16}\varepsilon^{\mu\nu\alpha\beta} \Tr\left[ (u \lambda_{\mu\nu}^\dagger u - u^\dagger \lambda_{\mu\nu} u^\dagger) [\hat{u}_\alpha, \hat{u}_\beta] \right] g_T^{(2)\prime}\;.  
\end{align}
Summing over the $g_T^{(2)}$ and $g_T^{(2)\prime}$ terms, we obtain the chiral Lagrangian at $\mathcal{O}(p^4)$ for tensor interaction
\begin{align}
\label{eq:tensor-matching}
    \mathcal{L}_{T}^{(2)}&=-\frac{1}{8} \Tr\left[ (u \lambda_{\mu\nu}^\dagger u + u^\dagger \lambda_{\mu\nu} u^\dagger) [\hat{u}^\mu, \hat{u}^\nu] \right] g_T^{(2)}\nn\\
    &\quad -\frac{i}{16}\varepsilon^{\mu\nu\alpha\beta} \Tr\left[ (u \lambda_{\mu\nu}^\dagger u - u^\dagger \lambda_{\mu\nu} u^\dagger) [\hat{u}_\alpha, \hat{u}_\beta] \right] g_T^{(2)\prime}\;.
\end{align}

As shown in Appendix~\ref{app:tensor}, the LECs $g_T^{(2)}$ and $g_T^{(2)\prime}$ are equal, Consequently, the chiral Lagrangian simplifies to 
\begin{align}
\label{eq:tensor-X}
    \mathcal{L}_{T}^{(2)}=-\frac{1}{4} \Tr\left[ (u X_{\mu\nu}^\dagger u + u^\dagger X_{\mu\nu} u^\dagger) [\hat{u}^\mu, \hat{u}^\nu] \right] g_T^{(2)}\;,
\end{align}
where
\begin{align}
\label{eq:Xmunu}
    X_{\mu\nu}^\dagger&\equiv \dfrac{1}{2} \lambda^\dagger_{\mu\nu}+\frac{i}{4}\varepsilon_{\mu\nu\alpha\beta} \lambda^{\alpha\beta\dagger} \;,\nn\\
    X_{\mu\nu}&\equiv \dfrac{1}{2}\lambda_{\mu\nu}-\frac{i}{4}\varepsilon_{\mu\nu\alpha\beta }\lambda^{\alpha\beta}\;.
\end{align}
Note that $X_{\mu\nu}$ is not necessarily anti-symmetric under the interchange of $\mu $ and $ \nu$.
We obtain that $\mathcal{L}_{T}^{(2)}$ is consistent with the $\Lambda_2$ term in Eq.~\eqref{eq:tensor-ubasis} by identifying $X_{\mu\nu} = t_{\mu\nu} $ and $ X_{\mu\nu}^\dagger = t_{\mu\nu}^\dagger $. See more discussions in Appendix~\ref{app:tensor}.

\subsubsection{\tf{$C$}{C} transformation}
\label{sec:C-transformation}

We have used the $C$ transformation properties in constructing the chiral Lagrangians for the scalar, pseudo-scalar, vector, axial-vector and tensor interactions. It is crucial that the chiral operators have the same properties as the quark-level interactions. In the below, we provide discussions of the spurion transformation under $C$.

We consider the spurion $\lambda_{\Gamma}  \equiv \mathcal{C}_\Gamma \tau \otimes \bar l_1 \Gamma l_2$, where $l_1$ and $l_2$ are lepton fields, $\mathcal{C}_\Gamma$ is the Wilson coefficient, and $\tau$ denotes either $\tau^{p/n} \equiv (1+\tau^3)/2$ or $\tau^\pm \equiv (\tau^1 \pm i\tau^2)/2$. The explicit form of $\lambda_\Gamma$ under $C$ transformation depends on both the flavor structure and the lepton fields.

The lepton bilinear transforms as
\begin{align}
    \bar l_1 \Gamma l_2 \stackrel{C}{\longrightarrow} \left(\bar l_2 \Gamma_c l_1\right)^T\;,
\end{align}
where $ \Gamma_c \equiv C \Gamma^T C^{-1}$. One has $\Gamma_c = 1, -\gamma^\mu, \gamma^\mu\gamma^5, -\sigma^{\mu\nu}$ for $\Gamma = 1, \gamma^\mu, \gamma^\mu\gamma^5, \sigma^{\mu\nu}$.
For the neutral current with $l_1 = l_2$ and $\tau = \tau^{p/n}$, we obtain 
\begin{align}
    \lambda_\Gamma = \mathcal{C}_\Gamma \tau^{p/n} \otimes \bar l \Gamma l \stackrel{C}{\longrightarrow} \mathcal{C}_\Gamma \tau^{p/n} \otimes \left(\bar l \Gamma_c l\right)^T \equiv \lambda_\Gamma^c \;.
\end{align}
Thus, for the low-energy QCD Lagrangian, $\lambda_\Gamma^c$ simplifies to $\lambda_T^T$~\cite{Scherer:2012xha,Fettes:2000gb,Song:2024fae}. It agrees with the transformation of the external sources under $C$ transformation:
\begin{align}
    v_\mu \to - v_\mu^T, \quad a_\mu \to a_\mu^T, \quad s \to s^T, \quad p \to p^T,\quad t_{\mu\nu} \to - t_{\mu\nu}^T\;.
\end{align}
It is more involved for the charged current interactions, which include $\tau^\pm$ for the quark current. Under $C$ transformation $\tau^{\pm} \to \tau^{\mp}$, and the charged lepton current transforms as $(\bar l_1 \Gamma l_2)\rightarrow (\bar l_2 \Gamma_c l_1)$. However, the matching does not depend on the explicit form of $\lambda_\Gamma^c$; it only requires that $\lambda_\Gamma^c$ remains unchanged from the quark level to the hadronic level.

\subsubsection{Matching in the \tf{$q$}{q} basis}
\label{sec:matching_non-chiral}

In parallel to the matching in the chiral basis, one can also employ the matching of the quark-level interactions in the $q$ basis, using $\{ 1,\gamma^5\}$ rather than $\{P_L, P_R\}$. It should be noted that the chiral and $q$ bases, which depend on the chiralities of the quark fields, correspond to the LR and $u$ bases, when the chiral operators are expressed using the $U$-parameterization and $u$-parameterization, respectively. 

For the vector and axial-vector currents, the Lagrangian is expressed as
\begin{align}
\label{eq:lambva}
\mathcal{L}_{V,A}^q &= \bar{q}\gamma_\mu \lambda_V^\mu q + \bar{q}\gamma_\mu \gamma^5 \lambda_A^\mu q\;,
\end{align}
where that $\lambda_V^\mu$ and $\lambda_A^\mu$ are Hermitian. Thus we can rewrite the above expression as follows
\begin{align}  
\bar{q}\gamma_\mu \lambda_V^\mu q &= \bar{q}_R \gamma_\mu \lambda_V^\mu q_R + \bar{q}_L \gamma_\mu \lambda_V^{\mu\dagger} q_L \;, \nn\\  
\bar{q}\gamma_\mu \gamma^5 \lambda_A^\mu q &= \bar{q}_R \gamma_\mu \lambda_A^\mu q_R - \bar{q}_L \gamma_\mu \lambda_A^{\mu\dagger} q_L \;,  
\end{align}
which are formally invariant under chiral symmetry by promoting $\lambda_{V/A}^\mu$ and $\lambda_{V/A}^{\mu\dagger}$  on the right-handed side to be independent spurions. Under the chiral transformation, they behave as $\lambda_{V/A}^\mu \to R \lambda_{V/A}^\mu R^\dagger$ and $\lambda_{V/A}^{\mu\dagger} \to L \lambda_{V/A}^{\mu\dagger} L^\dagger$.

By matching quark bilinears to their mesonic counterparts, one has
\begin{align}
\label{eq:mathcing-va_non-chiral0}
\bar{q}\gamma_\mu \lambda_V^\mu q &\to \Tr\left[\lambda_V^\mu u D_\mu u^\dagger + \lambda_V^{\mu\dagger} u^\dagger D_\mu u\right]\;, \nn\\
\bar{q}\gamma_\mu \gamma^5 \lambda_A^\mu q &\to \Tr\left[\lambda_A^\mu u D_\mu u^\dagger - \lambda_A^{\mu\dagger} u^\dagger D_\mu u\right]\;.
\end{align}
Using the building block $\hat{u}_\mu$ and the relations in Eq.~\eqref{eq:u-mu-relations}, we obtain
\begin{align}
\label{eq:mathcing-va_non-chiral}
\bar{q}\gamma_\mu \lambda_V^\mu q &\to \Tr\left[(u^\dagger \lambda_V^\mu u - u\lambda_V^{\mu\dagger} u^\dagger) \hat{u}_\mu\right]\;, \nn\\
\bar{q}\gamma^\mu \gamma^5 \lambda_A^\mu q &\to \Tr\left[(u^\dagger \lambda_A^\mu u + u\lambda_A^{\mu\dagger} u^\dagger) \hat{u}_\mu\right]\;.
\end{align}
In the final step, we identify
\begin{align}
    \lambda_{V/A}^\mu = \lambda_{V/A}^{\mu\dagger} = (\lambda_R^\mu \pm \lambda_L^\mu)/2\;,
\end{align}
and obtain that the results in Eqs.~\eqref{eq:mathcing-va_non-chiral0}~\eqref{eq:mathcing-va_non-chiral} are consistent with those in Eq.~\eqref{eq:conventional-VA-LO}.

In the $q$ basis, the Lagrangian for scalar and pseudo-scalar currents is expressed as  
\begin{align}  
\label{eq:lambsp}
\mathcal{L}_{S,P}^q &= \bar{q} \lambda_S q + i \bar{q} \gamma^5 \lambda_P q \;,  
\end{align}  
where $\lambda_S$ and $\lambda_P$ are Hermitian.
Thus we can rewrite the above expression as follows
\begin{align}  
\bar{q} \lambda_S q &= \bar{q}_L \lambda_S q_R + \bar{q}_R \lambda_S^\dagger q_L \;, \nn\\  
i\bar{q} \gamma^5 \lambda_P q &= \bar{q}_L  i\lambda_P q_R - \bar{q}_R i\lambda_P^\dagger q_L \;.  
\end{align}  
Similarly, they are formally invariant under chiral symmetry given the transformation of the spurions $\lambda_{S/P} \to L \lambda_{S/P} R^\dagger$ while $\lambda_{S/P}^\dagger \to R \lambda_{S/P}^\dagger L^\dagger$.

By matching quark bilinears to their mesonic counterparts, one obtains
\begin{align}  
\label{eq:mathcing-sp_non-chiral}
\bar{q} \lambda_S q &\to \Tr\left[u \lambda_S u + u^\dagger \lambda_S^\dagger u^\dagger\right] \;,\nn \\  
\bar{q} \gamma^5 i \lambda_P q &\to \Tr\left[u i \lambda_P u + u^\dagger (i\lambda_P)^\dagger u^\dagger\right] \;.  
\end{align}  
We identify 
\begin{align}
    \lambda_{S} = \lambda_S^\dagger &= (\lambda^\dagger + \lambda)/2\;,\nn\\
   i \lambda_{P} = -(i\lambda_P)^\dagger &= (\lambda^\dagger - \lambda)/2\;,
\end{align}
and obtain that the results in Eq.~\eqref{eq:mathcing-sp_non-chiral} coincide with those in Eq.~\eqref{eq:conventional-SP-LO}.

In the $q$ basis, the Lagrangian for the tensor interactions is  
\begin{align}  
\label{eq:lambt}
\mathcal{L}_{T}^q &= \bar{q} \sigma_{\mu\nu} \lambda_T^{\mu\nu} q+ i \bar{q} \sigma_{\mu\nu}\gamma^5 \lambda_\varepsilon^{\mu\nu} q\;,
\end{align}  
where $\lambda_T^{\mu\nu}$ and $\lambda_\varepsilon^{\mu\nu}$ are Hermitian. 
The two terms can be written as
\begin{align}
    \bar{q} \sigma_{\mu\nu} \lambda_T^{\mu\nu} q &= \dfrac{1}{2} \left(\bar{q}_L \sigma_{\mu\nu} \lambda_T^{\mu\nu} q_R  + \bar{q}_R \sigma_{\mu\nu} \lambda_T^{\mu\nu \dagger} q_L \right) + \dfrac{1}{2} \left(\bar{q}_L \sigma_{\mu\nu} \lambda_T^{\mu\nu} \gamma^5 q_R  - \bar{q}_R \sigma_{\mu\nu} \lambda_T^{\mu\nu \dagger} \gamma^5 q_L \right)\;,\nn\\
    i\bar{q} \sigma_{\mu\nu} \gamma^5 \lambda_\varepsilon^{\mu\nu} q &=\frac{1}{2}(\bar{q}_L \sigma_{\mu\nu} i\lambda_\varepsilon^{\mu\nu} q_R -  \bar{q}_R \sigma_{\mu\nu} i\lambda_\varepsilon^{\mu\nu \dagger} q_L) + \frac{1}{2}(\bar{q}_L \sigma_{\mu\nu} i\lambda_\varepsilon^{\mu\nu} \gamma^5 q_R + \bar{q}_R \sigma_{\mu\nu} i\lambda_\varepsilon^{\mu\nu\dagger} \gamma^5 q_L)\;.
\end{align}

Using the relation in Eq.~\eqref{eq:epsilon}, 
we obtain 
\begin{align}  
\bar{q} \sigma_{\mu\nu} \lambda_T^{\mu\nu} q &= \frac{1}{2}(\bar{q}_L \sigma_{\mu\nu} \lambda_T^{\mu\nu} q_R + \bar{q}_R \sigma_{\mu\nu} \lambda_T^{\mu\nu\dagger} q_L)+\frac{i}{4}\varepsilon_{\mu\nu\alpha\beta}(\bar q_L \sigma^{\alpha\beta} \lambda^{\mu\nu}_T q_R - \bar q_R \sigma^{\alpha\beta} \lambda^{\mu\nu\dagger}_T q_L)\;,\notag\\
i\bar{q} \sigma_{\mu\nu} \gamma^5\lambda_\varepsilon^{\mu\nu} q&=\frac{1}{2}(\bar{q}_L \sigma_{\mu\nu} i\lambda_\varepsilon^{\mu\nu} q_R - \bar{q}_R \sigma_{\mu\nu} i\lambda_\varepsilon^{\mu\nu\dagger} q_L) + \frac{i}{4}\varepsilon_{\mu\nu\alpha\beta}(\bar{q}_L \sigma^{\alpha\beta} i\lambda_\varepsilon^{\mu\nu} q_R + \bar{q}_R \sigma^{\alpha\beta} i\lambda_\varepsilon^{\mu\nu\dagger} q_L)\;.
\end{align} 
They are formally invariant under chiral symmetry with the chiral transformation of the spurions $\lambda_T^{\mu\nu} \to L \lambda_T^{\mu\nu} R^\dagger$ while $\lambda_{T/\varepsilon}^{\mu\nu \dagger} \to R \lambda_{T/\varepsilon}^{\mu\nu \dagger} L^\dagger $.
By matching quark bilinears to their mesonic counterparts, one obtains
\begin{align}  
\label{eq:mathcing-t_non-chiral}
\bar{q} \sigma_{\mu\nu} \lambda_T^{\mu\nu} q \to& -\dfrac{1}{8}\Tr\left[ (u \lambda_T^{\mu\nu} u + u^\dagger \lambda_T^{\mu\nu\dagger} u^\dagger) [\hat{u}_\mu, \hat{u}_\nu] \right] g_{T}^{(2)} \nn\\
&\quad -\frac{i}{16}\varepsilon_{\mu\nu\alpha\beta}\Tr\left[ (u \lambda_T^{\alpha\beta} u - u^\dagger \lambda_T^{\alpha\beta\dagger} u^\dagger) [\hat{u}^\mu, \hat{u}^\nu] \right] g_{T}^{(2)\prime} \;,\notag\\
\bar{q} \sigma_{\mu\nu} \gamma^5 i\lambda_\varepsilon^{
\mu\nu
} q \to& -\dfrac{1}{8}\Tr\left[ (u i \lambda_\varepsilon^{\mu\nu} u + u^\dagger (i\lambda_\varepsilon^{\mu\nu})^{\dagger} u^\dagger) [\hat{u}_\mu, \hat{u}_\nu] \right] g_{T}^{(2)}  \nn\\
&\quad -\frac{i}{16}\varepsilon_{\mu\nu\alpha\beta}\Tr\left[ (u i\lambda_\varepsilon^{\alpha\beta} u - u^\dagger (i\lambda_\varepsilon^{\alpha\beta})^{\dagger} u^\dagger) [\hat{u}^\mu, \hat{u}^\nu] \right] g_{T}^{(2)\prime} \;.
\end{align}

Similarly, we can identify $\lambda_T^{\mu\nu} = \lambda_T^{\mu\nu\dagger} = (\lambda^{\mu\nu\dagger} + \lambda^{\mu\nu})/2$ and $i \lambda_\varepsilon^{\mu\nu} = -(i\lambda_\varepsilon^{\mu\nu})^{\dagger} = (\lambda^{\mu\nu\dagger} - \lambda^{\mu\nu})/2$,
and obtain that Eq.~\eqref{eq:mathcing-t_non-chiral} agrees with  Eq.~\eqref{eq:tensor-matching}. Besides, the LECs in the above satisfy $g_{T}^{(2)\prime} = g_{T}^{(2)}$. 
The patterns in Eqs.~\eqref{eq:mathcing-va_non-chiral}~\eqref{eq:mathcing-sp_non-chiral}~\eqref{eq:mathcing-t_non-chiral} can serve as a valuable guide for constructing the chiral Lagrangian in the systematic spurion method in Sec.~\ref{sec:Young tensor}.

\section{Systematic spurion method}
\label{sec:Young tensor}

In this section, we match the LEFT operators to the chiral Lagrangian using the systematic spurion method~\cite{Song:2025snz}.
This method is general, applicable to all types of interactions, free from redundancies, and can be extended to higher chiral powers and higher-dimensional operators without introducing other spurions. Below, we discuss the matching for the five types of interactions (scalar, pseudo-scalar, vector, axial-vector and tensor), all of which are described by the dimension-6 LEFT operators. The matching beyond the dimension-6 level will be discussed in Sec.~\ref{sec:dim7-9}.

\subsection{Setup}
\label{sec:systematic1}

In the matching from LEFT operators to chiral Lagrangian, only the two light quarks $u$ and $d$ are relevant, while the lepton part remains identical to that in LEFT. 
For the quark sector, the $u$ and $d$ quarks are combined into the doublet $q = (u,d)^T$. To construct the chiral Lagrangian systematically, we begin by associating appropriate spurion fields with the quark bilinears. Excluding baryon number-violating operators, the quarks form four types of bilinears:
\begin{equation}
\label{eq:quark-bilinears}
(\bar q_L\Gamma \Sigma_L q_L),\quad (\bar q_R\Gamma \Sigma_R q_R),\quad (\bar q_R\Gamma \Sigma q_L),\quad (\bar q_L\Gamma \Sigma^\dagger q_R)\;,
\end{equation}
where $\Gamma$ represents the possible Dirac matrix given the charilities of the quark fields, and four independent spurion fields $\Sigma_L$, $\Sigma_R$, $\Sigma$, and $\Sigma^\dagger$ are introduced to maintain $SU(2)_L\times SU(2)_R$ chiral symmetry. The leptons, being singlets under this chiral symmetry, transform trivially.

While it is possible to incorporate the quark mass into the spurions $\Sigma$ and $\Sigma^\dagger$, we find it advantageous to define the spurions
\begin{align}
   \hat \chi=&2Bm_q \;,
\end{align}
and its Hermitian conjugate $\hat \chi^\dagger$, which enables a convenient identification of the power countings~\cite{Sun:2025zuk}.

We therefore establish the transformation properties under $SU(2)_L \times SU(2)_R$ chiral symmetry for all relevant fields and spurions appearing in the LEFT operators as follows:
\begin{equation}
    \left(\begin{array}{c}
            q_L\\
            q_R\\
            \bar q_L\\
            \bar q_R\\
            \hat \chi\\
            \hat \chi^\dagger\\
            \Sigma\\
            \Sigma^\dagger\\
            \Sigma_L\\
            \Sigma_R\\
            e_L\\
            e_R\\
            \nu_L\\
            \Bar{e}_L\\
            \Bar{e}_R\\
            \Bar{\nu}_L
    \end{array}\right)\rightarrow\left(\begin{array}{c}
            Lq_L\\
            Rq_R\\
            \bar q_L L^\dagger\\
            \bar q_R R^\dagger\\
            R\hat \chi L^\dagger\\
            L\hat \chi^\dagger R^\dagger\\
            R\Sigma L^\dagger\\
            L\Sigma^\dagger R^\dagger\\          
            L\Sigma_LL^\dagger\\
            R\Sigma_RR^\dagger\\
            e_L\\
            e_R\\
            \nu_L\\
            \Bar{e}_L\\
            \Bar{e}_R\\
            \Bar{\nu}_L
    \end{array}\right)\;,\quad L\in SU(2)_L\;,\ R\in SU(2)_R\;.
\end{equation}

For instance, the dimension-6 operators involving the scalar lepton current $j_\ell = \bar{e} e$, $\bar{e} i\gamma^5 e$ or $\bar{\nu}_L e_R $ 
take the form
\begin{align}
\label{eq:dim-6_exam1}
\mathcal{L}_{\text{eff}}^{(6)} \supset \mathcal{C}_1 j_\ell (\bar{q}_R\Sigma_1 q_L) + \mathcal{C}_2 j_\ell^\dagger (\bar{q}_R\Sigma_2 q_L) + \text{h.c.}\;,
\end{align}
where we have used the subscripts 1 and 2 to distinguish two quark bilinears, and $\mathcal{C}_1$ and $\mathcal{C}_2$ represent the generally complex Wilson coefficients.

The LEFT operators in the chiral basis are then translated to those in the $q$ basis, and classified according to the $CP$ transformation properties of the quark bilinears:
\begin{align}
    C+P+:\quad& \mathcal{O}_S=j_\ell [\bar{q}(\Sigma_2^\dagger P_R + \Sigma_1 P_L) q]\;,\quad\mathcal{O}_S^\dagger=j_\ell^\dagger [\bar{q}(\Sigma_2 P_L + \Sigma_1^\dagger P_R) q]\;,\notag\\
    C+P-:\quad& \mathcal{O}_P=j_\ell [\bar{q}(\Sigma_2^\dagger P_R-\Sigma_1 P_L) q]\;,\quad\mathcal{O}_P^\dagger=j_\ell^\dagger [\bar{q}(\Sigma_2 P_L-\Sigma_1^\dagger P_R) q]\;,
    \label{eq:SP}
\end{align}
where the second terms are the Hermitian conjugates of the respective first terms. 
It is noted that the spurions do not change under the $P$ transformation, but they transform under the $C$ transformation as
\begin{align}
    A \to A_c \equiv C A^T C^{-1}\;,
\end{align}
which has been explicitly shown for the tensor interaction in Eq.~\eqref{eq:C-tensor-quark}. However, the transformation of the spurions remains unchanged from the quark level to the hadronic level. so that we can define the  $CP$ eigenstate of the LEFT operators by assuming that the spurions are invariant.

In the $CP$ eigenstate, the LEFT operators are expressed as
\begin{align}
\label{eq:dim-6_exam2}
 \mathcal{L}_{\text{eff}}^{(6)} \supset   \mathcal{C}_S\mathcal{O}_S+\mathcal{C}_P\mathcal{O}_P+{\rm h.c.}\;,
\end{align}
where the $\mathcal{C}_S=\left(\mathcal{C}_2^\dagger+\mathcal{C}_1\right)/2$ and $\mathcal{C}_P=\left(\mathcal{C}_2^\dagger-\mathcal{C}_1\right)/2$ are the corresponding Wilson coefficients in the new basis of $CP$ eigenstate.
It is noted that for neutral lepton current $j_\ell=j_\ell^\dagger$, the Hermitian conjugate parts of the operators in Eqs.~\eqref{eq:dim-6_exam1}~\eqref{eq:dim-6_exam2} are dropped.

The dimension-6 LEFT operators with vector lepton bilinear $j_\ell^\mu =  \bar e\gamma^\mu e$, $ \bar e\gamma^\mu \gamma^5 e$, $\bar \nu_L \gamma^\mu \nu_L$ or $\bar \nu_L \gamma^\mu e_L$ take the form
\begin{equation}
\mathcal{L}_{\text{eff}}^{(6)} \supset \mathcal{C}_3 j_\ell^\mu (\bar q_L \gamma_\mu\Sigma_{1L} q_L)+\mathcal{C}_4 j_\ell^{\mu } (\bar q_R\gamma_\mu\Sigma_{2R} q_R)+ {\rm h.c.} \;.  
\end{equation}
Again, we have also used the subscripts 1 and 2 to distinguish two quark bilinears, while $\mathcal C_3$ and $\mathcal{C}_4$ denote the Wilson coefficients. 
The operators can be classified according to the $CP$ transformation properties of their quark bilinears:
\begin{align}
    C-P+:\quad& \mathcal{O}_V=j_\ell^\mu [\bar q\gamma_\mu(\Sigma_{1L}P_L+\Sigma_{2R}P_R) q]\;,\quad\mathcal{O}_V^\dagger=j_\ell^{\mu \dagger} [\bar q\gamma_\mu(\Sigma_{1L}^\dagger P_L+\Sigma_{2R}^\dagger P_R) q]\;,\notag\\
    C+P-:\quad& \mathcal{O}_A=j_\ell^\mu [\bar q\gamma_\mu  (\Sigma_{2R}P_R-\Sigma_{1L}P_L)q]\;,\quad\mathcal{O}_A^\dagger= j_\ell^{\mu \dagger} [\bar q\gamma_\mu (\Sigma_{2R}^\dagger P_L-\Sigma_{1L}^\dagger P_L) q]\;.
\end{align}
The Lagrangian then becomes
  \begin{align}
 \mathcal{L}_{\text{eff}}^{(6)} \supset   \mathcal{C}_V\mathcal{O}_V+\mathcal{C}_A\mathcal{O}_A+{\rm h.c.}\;,
\end{align}  
where  $\mathcal{C}_V=\left(\mathcal{C}_4+\mathcal{C}_3\right)/2$ and $\mathcal{C}_P=\left(\mathcal{C}_4-\mathcal{C}_3\right)/2$. Different from the scalar lepton bilinear operators, the Hermitian conjugates of the spurion $\Sigma_L$ and $\Sigma_R$ are also the adjoint representation for $SU(2)_L$ and $SU(2)_R$. 
For the neutral lepton current, we also drop the hermitian conjugate parts of the operators.

The $\mathcal{C}_1$ and $\mathcal{C}_2$ terms in Eq.~\eqref{eq:dim-6_exam1} differ in the chiral properties of the quark bilinears, yet the spurions  $\Sigma_1$ and $\Sigma_2^\dagger$ can be the same matrix.
For instance, when we consider the lepton current $j_\ell=\bar\nu_L e_R$, the spurions are 
\begin{align}
\Sigma_1=\Sigma_2^\dagger=\tau^-\;,\quad\Sigma_1^\dagger=\Sigma_2=\tau^+\;,
\end{align}
where $\tau^\pm \equiv \left(\tau^1\pm i\tau^2\right)/2$ with $\tau^i$ being the Pauli matrices.

Notice that it is unnecessary to introduce spurion separately for each LEFT operator.
The effective Lagrangian at the dimension-6 level can be expressed as follows
\begin{align}
    \mathcal{L}_{\text{eff}}^{(6)} \supset \sum_{i=1}^{12} \mathcal{C}_i^{(6)}\mathcal{O}_i^{(6)} + \left( \sum_{i=13}^{17} \mathcal{C}_i^{(6)}\mathcal{O}_i^{(6)} + {\rm h.c.} \right)\;,
\end{align}
where the  dimension-6 LEFT operators~\cite{Jenkins:2017jig} are classified in the $CP$ eigenstate,
\begin{align}
\label{eq:operators-dim6}
\mathcal{O}_1^{(6)}&=(\bar{e}e)[\bar{q}(\Sigma^\dagger P_R+\Sigma P_L) q]\;,
&
\mathcal{O}_2^{(6)}&=(\bar{e}i\gamma^5e)[\bar{q}(\Sigma^\dagger P_R+\Sigma P_L) q]\;,
\notag\\
\mathcal{O}_3^{(6)}&=(\bar{e}e)[\bar{q}(\Sigma^\dagger P_R-\Sigma P_L) q]\;,
&
\mathcal{O}_4^{(6)}&=(\bar{e}i\gamma^5e)[\bar{q}(\Sigma^\dagger P_R-\Sigma P_L) q]\;,
\notag\\
\mathcal{O}_5^{(6)}&=(\bar{e}\gamma^\mu e)[\bar{q}\gamma_\mu(\Sigma_R P_R+\Sigma_L P_L) q]\;,\quad
&
\mathcal{O}_6^{(6)}&=(\bar{e}\gamma^\mu\gamma^5 e)[\bar{q}\gamma_\mu(\Sigma_R P_R+\Sigma_L P_L) q]\;,
\notag\\
\mathcal{O}_7^{(6)}&=(\bar{e}\gamma^\mu e)[\bar{q} \gamma_\mu (\Sigma_R P_R-\Sigma_L P_L) q]\;,
&
\mathcal{O}_8^{(6)}&=(\bar{e}\gamma^\mu\gamma^5 e)[\bar{q}\gamma_\mu(\Sigma_R P_R-\Sigma_L P_L) q]\;,
\notag\\
\mathcal{O}_9^{(6)}&=(\bar{\nu}_L\gamma^\mu \nu_L)[\bar{q}\gamma_\mu(\Sigma_R P_R+\Sigma_L P_L) q]\;,
&\mathcal{O}_{10}^{(6)}&=(\bar{\nu}_L\gamma^\mu \nu_L)[\bar{q}\gamma_\mu(\Sigma_R P_R-\Sigma_L P_L)  q]\;,
\notag\\
\mathcal{O}_{11}^{(6)}&=(\bar{e}\sigma^{\mu\nu}e)[\bar{q}\sigma_{\mu\nu}(\Sigma^\dagger P_R+\Sigma P_L)q]\;,
&
\mathcal{O}_{12}^{(6)}&=(\bar{e}i\gamma^5\sigma^{\mu\nu}e)[\bar{q}\sigma_{\mu\nu}(\Sigma^\dagger P_R+\Sigma P_L)q]\;,
\notag\\
\mathcal{O}_{13}^{(6)}&=(\bar{\nu}_Le_R)[\bar{q}(\Sigma^\dagger P_R+\Sigma P_L) q]\;,
&
\mathcal{O}_{14}^{(6)}&=(\bar{\nu}_Le_R)[\bar{q}(\Sigma^\dagger P_R-\Sigma P_L) q]\;,
\notag\\
\mathcal{O}_{15}^{(6)}&=(\bar{\nu}_L\gamma^\mu e_L)[\bar{q}\gamma_\mu(\Sigma_R P_R+\Sigma_L P_L) q]\;,
&
\mathcal{O}_{16}^{(6)}&=(\bar{\nu}_L\gamma^\mu e_L)[\bar{q}\gamma_\mu(\Sigma_R P_R-\Sigma_L P_L) q]\;,
\notag\\
\mathcal{O}_{17}^{(6)}&=(\bar{\nu}_L\sigma^{\mu\nu}e_R)[\bar{q}\sigma_{\mu\nu}(\Sigma^\dagger P_R+\Sigma P_L) q]\;.
\end{align}
In the above, we have only introduced four spurions: $\Sigma$ and $\Sigma^\dagger$ for the scalar/pseudo-scalar and tensor quark currents, while $\Sigma_L$ and $\Sigma_R$ for the vector/axial-vector quark currents. Note that $\Sigma^\dagger$ is not the Hermitian conjugate of $\Sigma$ by definition.
In comparison with Ref.~\cite{Jenkins:2017jig}, we have
\begin{align}
    \Sigma^\dagger = \Sigma = 
    \Sigma_R = \Sigma_L \;,
\end{align}
which is equal to $\tau^-$ for the charged lepton current, and $\tau^{p/n} \equiv (1\pm \tau^3)/2$ for the neutral lepton current. 

\subsection{Matching to the chiral Lagrangian}
\label{sec:systematic2}

Instead of treating lepton parts as external sources, we systematically derive the chiral Lagrangian for LEFT operators by incorporating leptons as additional degrees of freedom.
According to Eq.~\eqref{eq:u}, the spurions $\hat \chi$, $\hat \chi^\dagger$, $\Sigma$, $\Sigma^\dagger$, $\Sigma_L$ and $\Sigma_R$ can be dressed with $u$ and $u^\dagger$ to form 
\begin{align}
\label{eq:building}
\hat \chi_\pm&=u \,\hat \chi^\dagger \, u \pm u^\dagger \, \hat \chi \, u^\dagger  \;,\notag \\
 \Sigma_{\pm} &= u \,\Sigma^\dagger \, u \pm u^\dagger \, \Sigma \, u^\dagger  \;,\notag\\
    Q_{\pm} &= u^\dagger \,\Sigma_R   \, u \pm u \, \Sigma_L \, u^\dagger \;.
\end{align}
The fields $X=\hat \chi_\pm$, $\Sigma_\pm$ and $Q_\pm$ transform under the chiral symmetry as $X \to K X K^\dagger$, making them suitable building blocks of the chiral Lagrangian in the $u$-parameterization.

Additionally, the set of building blocks involves the vielbein
\begin{eqnarray}
\label{eq:uhat2}
\hat{u}_\mu = i\left(u^{\dagger} \mathcal D_\mu u - u \mathcal D_\mu u^{\dagger}\right)\;,
\end{eqnarray}
which also transforms as $\hat{u}_\mu \to K \hat{u}_\mu K^\dagger$,
and the lepton currents. Here, the derivative that includes the photon field is defined as
(notice the difference from $D_\mu$ in Eq.~\eqref{eq:derivative-u})
\begin{align}
\label{eq:covariant-em}
    \mathcal D_\mu^{\rm } =  \partial_\mu + i eA_\mu Q\;,
\end{align}
where $A_\mu$ denotes the photon field, $e$ is the elementary electric charge, and $Q$ represents the electromagnetic charge operator.
In contrast to the vielbein defined in the external source method (see Eq.~\eqref{eq:vielbein}), the lepton current is separated from the vielbein in the systematic spurion method\,\footnote{For simplicity, we retain the same symbol $\hat u_\mu$ for the vielbein with $r_\mu = \ell_\mu = 0$ (see Eq.~\eqref{eq:uhat}), as used for mesonic chiral operators in the conventional spurion method. }. Moreover, our formulation preserves the $U(1)_{\rm em}$ gauge invariance and explicitly includes the photon field.
Consequently, the LECs of the operators involving the photon field $A_\mu$ are related to that of the kinetic term, which is consistent with the results in the external source method as given in Eq.~\eqref{eq:source_vector} or Eq.~\eqref{eq:ubasis2}.

The lepton currents are also separated from the covariant derivative acting on the above building blocks,
\begin{align}
    \nabla_\mu X = \left[\partial_\mu+\frac{1}{2}\left(u^{\dagger} \mathcal D_\mu u+u \mathcal D_\mu u^{\dagger}\right) \right] X\;,
\end{align}
in comparison with the definition in Eq.~\eqref{eq:covariant derivative}.

\begin{table}[H]
\tabcolsep=4pt
\renewcommand\arraystretch{1.5}
\caption{The $CP$ properties of the chiral building blocks in the systematic spurion method. 
} 
\begin{center}
    \begin{tabular}{c|c|c|c|c|c|c|c}
    \hline
    \hline
           & $\Sigma_+$ & $\Sigma_-$&$Q_+$& $Q_-$&$\hat \chi_+$& $\hat \chi_-$ & $\hat u_\mu$ \\
    \hline
    $P$ & $+$ & $-$ &$+$&$-$&$+$&$-$& $-$\\
    \hline
$C$  & $+$ & $+$ &$+$&$-$ &$+$&$+$ & $+$\\
    \hline
    \hline
    \end{tabular}
    \end{center}
    \label{tab:building_block}
\end{table}

Given the chiral and $CP$ transformation properties, which are shown in Tab.~\ref{tab:building_block}, we can systematically construct the chiral Lagrangian through the Young tensor technique~\cite{Li:2020gnx,Li:2020xlh,Li:2022tec}
(see Appendix~\ref{sec:Yound}). The mapping from quark to hadronic degrees of freedom
satisfies the following rules:
\begin{itemize}
    \item The spurions in the LEFT operators remain unchanged in the matching.
    \item The $CP$ transformation properties of the LEFT and chiral operators are the same;
    \item The leptons parts
    are identical in the LEFT operators and chiral operators.
\end{itemize}

Applying these rules leads to the matching from the LEFT to $\chi$PT
as given in Tab.~\ref{tab:va} and Tab.~\ref{tab:sp},  
where we consider the one of building blocks $Q_\pm$ and $\Sigma_\pm$ for each chiral operator.

\begin{table}[H]
    \centering
        \captionsetup{justification=centering}
    \caption{The matching for the vector/axial-vector operators at $p^2$ and  $p^4$ orders.}
    \begin{tabular}{|c|c|c|}
    \hline
    \multirow{2}{*}{LEFT operator} & \multicolumn{2}{c|}{$\chi$PT operators} \\
    \cline{2-3}  
    & $\mathcal{O}(p^2)$ & $\mathcal{O}(p^4)$ \\
    \hline
        \multirow{4}{*}{$\mathcal{O}_5^{(6)} = (\bar e \gamma^\mu e)[\bar{q}\gamma_\mu(\Sigma_R P_R+\Sigma_L P_L) q]$}&\multirow{4}{*}{$(\bar e \gamma^\mu e)\langle Q_-\hat u_\mu\rangle$}&$(\bar e \gamma^\mu e)\langle Q_-\hat u_\mu\rangle\langle \hat u^\nu \hat u_\nu \rangle$ \\
         & &$(\bar e \gamma^\mu e)\langle Q_-\hat u_\nu\rangle\langle \hat u^\mu \hat u^\nu \rangle$\\
         &&$(\bar e \gamma^\mu e)\langle Q_- \hat u_\mu\rangle\langle \hat\chi_+ \rangle$\\
         &&$(\bar e\gamma^\mu e)\langle Q_+[\hat u_\mu,\hat\chi_-]\rangle$\\
         \hline
        \multirow{4}{*}{$\mathcal{O}_7^{(6)} =(\bar e \gamma^\mu e)[\bar{q}\gamma_\mu(\Sigma_R P_R-\Sigma_L P_L) q]$} &\multirow{4}{*}{$(\bar e \gamma^\mu e)\langle Q_+\hat u_\mu\rangle$}&$(\bar e \gamma^\mu e)\langle Q_+\hat u_\mu\rangle\langle \hat u^\nu \hat u_\nu \rangle$\\
         &&$(\bar e \gamma^\mu e)\langle Q_+\hat u_\nu\rangle\langle \hat u^\mu \hat u^\nu \rangle$\\
         &&$(\bar e \gamma^\mu e)\langle Q_+\hat u_\mu \rangle\langle\hat\chi_+ \rangle$\\
         &&$(\bar e\gamma^\mu e)\langle Q_-[\hat u_\mu,\hat\chi_-]\rangle$\\
         \hline
        \multirow{4}{*}{$\mathcal{O}_6^{(6)} =(\bar e \gamma^\mu\gamma^5 e)[\bar{q}\gamma_\mu(\Sigma_R P_R+\Sigma_L P_L) q]$ }&\multirow{4}{*}{$(\bar e \gamma^\mu\gamma^5 e)\langle Q_-\hat u_\mu\rangle$}&$(\bar e \gamma^\mu\gamma^5 e)\langle Q_-\hat u_\mu\rangle\langle \hat u^\nu \hat u_\nu \rangle$\\
         &&$(\bar e \gamma^\mu\gamma^5 e)\langle Q_-\hat u_\nu\rangle\langle \hat u^\mu \hat u^\nu \rangle$\\
         &&$(\bar e \gamma^\mu\gamma^5 e)\langle Q_- \hat u_\mu\rangle\langle \hat\chi_+ \rangle$\\
         &&$(\bar e\gamma^\mu e)\langle Q_+[\hat u_\mu,\hat\chi_-]\rangle$\\
         \hline
        \multirow{4}{*}{$\mathcal{O}_8^{(6)} =(\bar e \gamma^\mu\gamma^5 e)[\bar{q}\gamma_\mu(\Sigma_R P_R-\Sigma_L P_L) q]$} &\multirow{4}{*}{$(\bar e \gamma^\mu\gamma^5 e)\langle Q_+\hat u_\mu\rangle$}&$(\bar e \gamma^\mu\gamma^5 e)\langle Q_+\hat u_\mu\rangle\langle \hat u^\nu \hat u_\nu \rangle$\\
         &&$(\bar e \gamma^\mu\gamma e)\langle Q_+\hat u_\nu\rangle\langle \hat u^\mu \hat u^\nu \rangle$\\
         &&$(\bar e \gamma^\mu\gamma^5 e)\langle Q_+\hat u_\mu\rangle\langle \hat\chi_+ \rangle$\\
         &&$(\bar e\gamma^\mu e)\langle Q_-[\hat u_\mu,\hat\chi_-]\rangle$\\
         \hline
         \multirow{4}{*}{$\mathcal{O}_9^{(6)} =(\bar \nu_L \gamma^\mu \nu_L)[\bar{q}\gamma_\mu(\Sigma_R P_R+\Sigma_L P_L) q]$}&\multirow{4}{*}{$(\bar \nu_L \gamma^\mu \nu_L)\langle Q_-\hat u_\mu\rangle$}&$(\bar \nu_L \gamma^\mu \nu_L)\langle Q_-\hat u_\mu\rangle\langle \hat u^\nu \hat u_\nu \rangle$ \\
         & &$(\bar \nu_L \gamma^\mu \nu_L)\langle Q_-\hat u_\nu\rangle\langle \hat u^\mu \hat u^\nu \rangle$\\
         &&$(\bar \nu_L \gamma^\mu \nu_L)\langle Q_- \hat u_\mu\rangle\langle \hat\chi_+ \rangle$\\
         &&$(\bar \nu_L\gamma^\mu \nu_L)\langle Q_+[\hat u_\mu,\hat\chi_-]\rangle$\\
         \hline
        \multirow{4}{*}{$\mathcal{O}_{10}^{(6)} =(\bar \nu_L \gamma^\mu \nu_L)[\bar{q}\gamma_\mu(\Sigma_R P_R-\Sigma_L P_L) q]$} &\multirow{4}{*}{$(\bar \nu_L \gamma^\mu \nu_L)\langle Q_+\hat u_\mu\rangle$}&$(\bar \nu_L \gamma^\mu \nu_L)\langle Q_+\hat u_\mu\rangle\langle \hat u^\nu \hat u_\nu \rangle$\\
         &&$(\bar \nu_L \gamma^\mu \nu_L)\langle Q_+\hat u_\nu\rangle\langle \hat u^\mu \hat u^\nu \rangle$\\
         &&$(\bar \nu_L \gamma^\mu \nu_L)\langle Q_+\hat u_\mu\rangle\langle \hat\chi_+\rangle$\\
         &&$(\bar \nu_L\gamma^\mu \nu_L)\langle Q_-[\hat u_\mu,\hat\chi_-]\rangle$\\
         \hline
         \multirow{4}{*}{$\mathcal{O}_{15}^{(6)} =(\bar \nu_L \gamma^\mu e_L)[\bar{q}\gamma_\mu(\Sigma_R P_R+\Sigma_L P_L) q]$}&\multirow{4}{*}{$(\bar \nu_L \gamma^\mu e_L)\langle Q_-\hat u_\mu\rangle$}&$(\bar \nu_L \gamma^\mu e_L)\langle Q_-\hat u_\mu\rangle\langle \hat u^\nu \hat u_\nu \rangle$ \\
         & &$(\bar \nu_L \gamma^\mu e_L)\langle Q_-\hat u_\nu\rangle\langle \hat u^\mu \hat u^\nu \rangle$\\
         &&$(\bar \nu_L \gamma^\mu e_L)\langle Q_- \hat u_\mu\rangle\langle \hat\chi_+ \rangle$\\
         &&$(\bar \nu_L\gamma^\mu e_L)\langle Q_+[\hat u_\mu,\hat\chi_-]\rangle$\\
         \hline
        \multirow{4}{*}{$\mathcal{O}_{16}^{(6)} =(\bar \nu_L \gamma^\mu e_L)[\bar{q}\gamma_\mu(\Sigma_R P_R-\Sigma_L P_L) q]$} &\multirow{4}{*}{$(\bar \nu_L \gamma^\mu e_L)\langle Q_+\hat u_\mu\rangle$}&$(\bar \nu_L \gamma^\mu e_L)\langle Q_+\hat u_\mu\rangle\langle \hat u^\nu \hat u_\nu \rangle$\\
         &&$(\bar \nu_L \gamma^\mu e_L)\langle Q_+\hat u_\nu\rangle\langle \hat u^\mu \hat u^\nu \rangle$\\
         &&$(\bar \nu_L \gamma^\mu e_L)\langle Q_+\hat u_\mu\rangle\langle \hat\chi_+ \rangle$\\
         &&$(\bar \nu_L\gamma^\mu e_L)\langle Q_-[\hat u_\mu,\hat\chi_-]\rangle$\\
         \hline
    \end{tabular}
    \label{tab:va}
\end{table}

\begin{table}[H]
\tabcolsep=4pt
\renewcommand\arraystretch{1.2}
    \centering
    \caption{The matching for the scalar/pseudo-scalar and tensor operators at $p^2$ and  $p^4$ orders.}
    \begin{tabular}{|l|c|c|}
    \hline 
    \multirow{2}{*}{LEFT operator} & \multicolumn{2}{c|}{$\chi$PT operators} \\
    \cline{2-3}  
    &$\mathcal{O}(p^2)$&$\mathcal{O}(p^4)$\\
    \hline
        \multirow{3}{*}{$\mathcal{O}_1^{(6)} =(\bar e e)[\Bar{q}(\Sigma^\dagger P_R + \Sigma P_L) q]$} &\multirow{3}{*}{$(\bar e e)\langle\Sigma_+\rangle$}&$(\bar e e)\langle\Sigma_+\rangle\langle \hat u_\mu \hat u^\mu\rangle$\\
        &&$(\bar e e)\langle\Sigma_+\rangle\langle\hat \chi_+\rangle\;,~(\bar e e)\langle\Sigma_+\hat \chi_+\rangle$\\
        &&$(\bar e e)\langle\Sigma_-\rangle\langle\hat \chi_-\rangle\;,~(\bar e e)\langle\Sigma_-\hat \chi_-\rangle$\\
        \hline
    \multirow{3}{*}{$\mathcal{O}_3^{(6)} =(\bar e e)[\Bar{q}(\Sigma^\dagger P_R -\Sigma P_L) q]$}&\multirow{3}{*}{$(\bar e e)\langle\Sigma_-\rangle$}&$(\bar e e)\langle\Sigma_-\rangle\langle \hat u_\mu \hat u^\mu\rangle$\\
    &&$(\bar e e)\langle\Sigma_-\rangle\langle\hat \chi_+\rangle\;,~(\bar e e)\langle\Sigma_-\hat \chi_+\rangle$\\
    &&$(\bar e e)\langle\Sigma_+\rangle\langle\hat \chi_-\rangle\;,~(\bar e e)\langle\Sigma_+\hat \chi_-\rangle$\\
    \hline
    \multirow{3}{*}{$\mathcal{O}_2^{(6)} =(\bar e i\gamma^5 e)(\Bar{q}[\Sigma^\dagger P_R + \Sigma P_L] q)$}&\multirow{3}{*}{$(\bar ei\gamma^5 e)\langle\Sigma_+\rangle$}&$(\bar ei\gamma^5 e)\langle\Sigma_+\rangle\langle \hat u_\mu \hat u^\mu\rangle$\\
    &&$(\bar ei\gamma^5 e)\langle\Sigma_+\rangle\langle\hat \chi_+\rangle\;,~(\bar ei\gamma^5 e)\langle\Sigma_+\hat \chi_+\rangle$\\
    &&$(\bar ei\gamma^5 e)\langle\Sigma_-\rangle\langle\hat \chi_-\rangle\;,~(\bar ei\gamma^5 e)\langle\Sigma_-\hat \chi_-\rangle$\\
    \hline
    \multirow{3}{*}{$\mathcal{O}_4^{(6)} =(\bar ei\gamma^5 e)[\Bar{q}(\Sigma^\dagger P_R-\Sigma P_L) q]$}&\multirow{3}{*}{$(\bar e i\gamma^5e)\langle\Sigma_-\rangle$}&$(\bar ei\gamma^5 e)\langle\Sigma_-\rangle\langle \hat u_\mu \hat u^\mu\rangle$\\
    &&$(\bar ei\gamma^5 e)\langle\Sigma_-\rangle\langle\hat \chi_+\rangle\;,~(\bar ei\gamma^5 e)\langle\Sigma_-\hat \chi_+\rangle$\\
    &&$(\bar ei\gamma^5 e)\langle\Sigma_+\rangle\langle\hat \chi_-\rangle\;,~(\bar ei\gamma^5 e)\langle\Sigma_+\hat \chi_-\rangle$\\
    \hline
    {$\mathcal{O}_{11}^{(6)} =(\bar e \sigma^{\mu\nu} e)[\bar{q}\sigma_{\mu\nu}(\Sigma^\dagger P_R+\Sigma P_L) q]$} &{ $/$} & $(\bar e \sigma^{\mu\nu} e)\langle \Sigma_+[ \hat u_\mu, \hat u_\nu]\rangle$ \\
    \hline
    {$\mathcal{O}_{12}^{(6)} =(\bar e i \gamma^5\sigma^{\mu\nu} e)[\bar{q}\sigma_{\mu\nu}(\Sigma^\dagger P_R+\Sigma P_L) q]$} & { $/$} & $(\bar e i\gamma^5\sigma^{\mu\nu} e)\langle \Sigma_+[ \hat u_\mu, \hat u_\nu]\rangle$\\
    \hline
     \multirow{3}{*}{$\mathcal{O}_{13}^{(6)} =(\bar \nu_L e_R)[\Bar{q}(\Sigma^\dagger P_R + \Sigma P_L) q]$} &\multirow{3}{*}{$(\bar \nu_L e_R)\langle\Sigma_+\rangle$}&$(\bar \nu_L e_R)\langle\Sigma_+\rangle\langle \hat u_\mu \hat u^\mu\rangle$\\
    &&$(\bar \nu_L e_R)\langle\Sigma_+\rangle\langle\hat \chi_+\rangle\;,~(\bar \nu_L e_R)\langle\Sigma_+\hat \chi_+\rangle$\\
    &&$(\bar \nu_L e_R)\langle\Sigma_-\rangle\langle\hat \chi_-\rangle\;,~(\bar \nu_L e_R)\langle\Sigma_-\hat \chi_-\rangle$\\
     \hline
    \multirow{3}{*}{$\mathcal{O}_{14}^{(6)} =(\bar \nu_L e_R)[\Bar{q}(\Sigma^\dagger P_R -\Sigma P_L) q]$}&\multirow{3}{*}{$(\bar \nu_L e_R)\langle\Sigma_-\rangle$}&$(\bar \nu_L e_R)\langle\Sigma_-\rangle\langle \hat u_\mu \hat u^\mu\rangle$\\
    &&$(\bar \nu_L e_R)\langle\Sigma_-\rangle\langle\hat \chi_+\rangle\;,~(\bar \nu_L e_R)\langle\Sigma_-\hat \chi_+\rangle$\\
    &&$(\bar \nu_L e_R)\langle\Sigma_+\rangle\langle\hat \chi_-\rangle\;,~(\bar \nu_L e_R)\langle\Sigma_+\hat \chi_-\rangle$\\
    \hline
    {$\mathcal{O}_{17}^{(6)} =(\bar{\nu}_L\sigma^{\mu\nu}e_R)[\bar{q}\sigma_{\mu\nu}(\Sigma^\dagger P_R+\Sigma P_L) q]$} & {$/$} & $(\bar{\nu}_L\sigma^{\mu\nu}e_R)\langle \Sigma_+[ \hat u_\mu, \hat u_\nu]\rangle$  \\
    \hline
    \end{tabular}
    \label{tab:sp}
\end{table}

It is noted that for the tensor operators,
\begin{align}
    \bar{e}\gamma^{5}\sigma^{\mu\nu}e &= \dfrac{i}{2}\varepsilon^{\mu\nu\alpha\beta} \, \bar{e}\sigma_{\alpha\beta}e \;,\\
    \bar{e}i\sigma^{\mu\nu}e &=  \dfrac{i}{2}\varepsilon^{\mu\nu\alpha\beta}\,  \bar{e}i \gamma^5\sigma_{\alpha\beta}e \;,\\
    \bar{\nu}_L \sigma^{\mu\nu}e_R &=  \dfrac{i}{2}\varepsilon^{\mu\nu\alpha\beta}\,  \bar{\nu}_L \sigma_{\alpha\beta}e_R \;,
\end{align}
which are obtained from Eq.~\eqref{eq:epsilon}. Due to the last relation, the operator $\mathcal{O}_{17}^{(6)}$ can be matched to both $(\bar{\nu}_L\sigma^{\mu\nu}e_R)\langle \Sigma_+[ \hat u_\mu, \hat u_\nu]\rangle$ and $(\bar{\nu}_L\sigma^{\mu\nu}e_R)\langle \Sigma_-[ \hat u_\mu, \hat u_\nu]\rangle$, despite the fact that $\Sigma_+$ and $\Sigma_-$ possess opposite parities. 

Besides, we have the relations between the operators:
\begin{align}
    (\bar{e}i \sigma^{\mu\nu}e)[\bar{q}\sigma_{\mu\nu}(\Sigma^{\dagger}P_{R}-\Sigma P_{L})q] = (\bar{e} i \gamma^5 \sigma^{\mu\nu}e)[\bar{q}\sigma_{\mu\nu}(\Sigma^{\dagger}P_{R}+\Sigma P_{L})q] \equiv \mathcal{O}_{12}^{(6)}\;, \\
    (\bar{\nu}_{L}\sigma^{\mu\nu}e_{R})[\bar{q}\sigma_{\mu\nu}(\Sigma^{\dagger}P_{R}-\Sigma P_{L})q] = (\bar{\nu}_{L}\sigma^{\mu\nu}e_{R})[\bar{q}\sigma_{\mu\nu}(\Sigma^{\dagger}P_{R}+\Sigma P_{L})q]  \equiv \mathcal{O}_{17}^{(6)}\;.
\end{align}
Moreover, from the expressions in Eqs.~\eqref{eq:building} and \eqref{eq:uhat2}, we observe that the building blocks $\hat{\chi}_-$ and $[\hat{u}_\mu, \hat{u}_\nu]$ are anti-Hermitian. Thus,  in these cases, a factor of $i$ is included in the chiral Lagrangian, in addition to the chiral operators, as indicated in Eq.~\eqref{eq:factor-i}. 

The chiral operators up to $\mathcal{O}(p^4)$ are given in Tab.~\ref{tab:va} and Tab.~\ref{tab:sp}, which agree with those obtained using the external source method.
Below, we take the scalar operator $\mathcal{O}_1^{(6)}$ and the vector operator $\mathcal{O}_5^{(6)}$ as examples.

In the external source method, the chiral operators of $\mathcal{O}(p^2)$ 
and $\mathcal{O}(p^4)$ 
corresponding to the operator $\mathcal{O}_1^{(6)}$ can be obtained from Eqs.\eqref{eq:ubasis2}~\eqref{eq:ubasis}, respectively.
Following the notation in Ref.~\cite{Song:2025snz}, we rewrite the scalar operator
$\mathcal{O}_1^{(6)} \equiv (\bar e e)(\bar q \,{\rm T}\,
 q)$, where ${\rm T}$ is the coefficient matrix. Then the external scalar source $s = (\bar e e) \,{\rm T} $ is incorporated into the building blocks $\chi_\pm$:
 \begin{align}
     \chi_+
     =(\bar ee)(u^\dagger {\rm T}u^\dagger +u {\rm T} u) + \hat \chi_+ \;,\\
     \chi_-
     =(\bar ee)(u^\dagger {\rm T}u^\dagger -u {\rm T} u) + \hat \chi_-\;.
 \end{align}
Thus, 
the chiral Lagrangian up to $p^4$ order is given by
\begin{align}
    \mathcal{L}_{4, 1}^\prime=&\,  \frac{F^2_0}{4} (\bar{e}e) \langle u^\dagger {\rm T}u^\dagger +u {\rm T} u \rangle +L_3 (\bar{e}e) \langle u^\dagger {\rm T}u^\dagger +u {\rm T} u \rangle\langle u_\mu u^\mu\rangle \notag\\
    &+2L_7(\bar{e}e)\langle \hat \chi_+(u^\dagger {\rm T}u^\dagger +u {\rm T} u)\rangle +2L_8(\bar{e}e)\langle \hat \chi_-(u^\dagger {\rm T}u^\dagger -u {\rm T} u)\rangle \notag\\
    &+2L_9(\bar{e}e)\langle u^\dagger {\rm T}u^\dagger +u {\rm T} u \rangle\langle \hat \chi_+\rangle+2L_{10}(\bar{e}e)\langle u^\dagger {\rm T}u^\dagger -u {\rm T} u \rangle\langle \hat \chi_-\rangle \;.
\end{align}
This expression can be converted into the results in Tab.~\ref{tab:sp} by taking ${\rm T}= \mathcal{C}_1^{(6)} \Sigma^\dagger  =\mathcal{C}_1^{(6)} \Sigma $.

For the vector operator $\mathcal{O}_5^{(6)}=(\bar e\gamma_\mu e)(\bar q\gamma^\mu\,{\rm T}\,q)$, the vector external source $v_\mu=(\bar e\gamma_\mu e){\rm T}$, which is present in both $u_\mu$ and $f_\pm^{\mu\nu}$ in the external source method, see Eqs.~\eqref{eq:vielbein}~\eqref{eq:fmunu}. Thus, the relevant terms in the chiral Lagrangian up to $p^4$ order for $\mathcal{O}_5^{(6)}$ are given by
\begin{align}
    \mathcal{L}_{4, 5}^\prime=&\frac{F^2_0}{2}(\bar{e}\gamma_\mu e)\langle(u^\dagger {\rm T}u -u {\rm T} u^\dagger)\hat{u}^\mu\rangle+4L_1(\bar e\gamma_\mu e)\langle(u^\dagger {\rm T}u -u {\rm T} u^\dagger)\hat{u}^\mu\rangle\langle \hat{u}^\nu \hat{u}_\nu\rangle\notag\\
    &+4L_2(\bar e\gamma_\mu e)\langle(u^\dagger {\rm T}u -u {\rm T} u^\dagger)\hat{u}_\nu\rangle\langle \hat{u}^\mu \hat{u}^\nu\rangle+2L_3(\bar e\gamma_\mu e)\langle\chi_+\rangle\langle (u^\dagger {\rm T}u -u {\rm T} u^\dagger) \hat{u}^\mu\rangle\notag\\
    &+iL_4\langle f_+^{\mu\nu}\hat{u}_\mu \hat{u}_\nu\rangle\;.
\end{align}
This expression can be converted into the results in Tab.~\ref{tab:va} by taking ${\rm T}=\mathcal{C}_5^{(6)} \Sigma_L=\mathcal{C}_5^{(6)} \Sigma_R$. Especially, the $L_4$ term is transformed using the IBP and EOMs
\begin{align}
    i\langle f_+^{\mu\nu}\hat{u}_\mu \hat{u}_\nu\rangle&\sim i\langle [u(\partial^\mu v^\nu-\partial^\nu v^\mu) u^\dagger+u^\dagger(\partial^\mu v^\nu-\partial^\nu v^\mu) u][\hat{u}_\mu,\hat{u}_\nu]\rangle\;,\notag\\
    &\sim\langle (uv^\mu u^\dagger+u^\dagger v^\mu u)[\hat{u}_\mu,\hat \chi_-]\rangle\;,\notag\\
    &=(\bar e\gamma^\mu e)\langle Q_+[\hat{u}_\mu,\hat \chi_-]\rangle\;.
\end{align}
The treatment of other axial-vector and pseudo-scalar operators is similar to that of the operators $\mathcal{O}_1^{(6)}$ and $\mathcal{O}_5^{(6)}$. To clarify the matching for tensor operators, we have provided a detailed discussion of the matching for $\mathcal{O}_{11}^{(6)}$ using three different methods in Appendix~\ref{app:tensor}.

Moreover, in our systematic spurion method the LECs obey relations similar to those derived in the external source method~\cite{Scherer:2002tk,Scherer:2012xha}.
Specifically, the LEC for the operator $O_5^{(6)}$ is identical to that for the electromagnetic interaction in Eq.~\eqref{eq:kin} (see Appendix~\ref{sec:match-photon}).
If the quark currents related by parity, the associated LECs are identical~\cite{Cirigliano:2018yza}. Consequently, operators sharing the same quark-level chiral structure $\bar q \gamma_\mu (\Sigma_R P_R \pm \Sigma_L P_L) q$ share a common LEC. The same reasoning applies to operators with the chiral structure $\bar q (\Sigma^\dagger P_R \pm \Sigma P_L) q$. 
In other words, the LECs associated with LEFT operators containing vector and axial-vector quark currents are identical, e.g., $O_5^{(6)}$ and $O_7^{(6)}$, and the same relation holds for operators involving scalar and pseudo-scalar quark currents.
It should be emphasized that these arguments apply not only to the dimension-6 LEFT operators listed in Tab.~\ref{tab:va} and Tab.~\ref{tab:sp}, but also to higher-dimensional operators discussed in Sec.~\ref{sec:dim7-9}.

 \subsection{Conversion of results in three methods}
 \label{sec:conversion}

As detailed in Sec.~\ref{sec:spurion-conventional}, the matching results derived from the conventional spurion method and external source method have been explicitly compared. 
The correspondence between the external sources and spurions for the dimension-6 LEFT operators is
\begin{align}
    -s &= \lambda_S\;,  & ip &= i\lambda_P \;,\\
    v^\mu &= \lambda_V^\mu\;, & a^\mu &= \lambda_A^\mu\;,\\
    \label{eq:tbar-lambda}
    \bar t^{\mu\nu} &= \lambda_T^{\mu\nu}-\frac{1}{2}\varepsilon^{\mu\nu}_{~~\alpha\beta}\lambda^{\alpha\beta}_\varepsilon\;,
\end{align}
where the spurions $\lambda_{V/A}^\mu$, $\lambda_{S/P}$ and $\lambda_{T/\varepsilon}^{\mu\nu}$ are defined in Eqs.~\eqref{eq:lambva}~\eqref{eq:lambsp}~\eqref{eq:lambt}, respectively.

The conversion relations between the spurions in the conventional and our systematic methods are
\begin{align}
    \label{eq:conv-s}
    \lambda_S  & =    \left( \Sigma^\dagger P_R + \Sigma P_L\right) j_\ell+ {\rm h.c.} \;,\\
    \label{eq:conv-p}
   \lambda_P \gamma^5  & =  -i\,   \left( \Sigma^\dagger P_R - \Sigma P_L \right) j_\ell+ {\rm h.c.}\;,\\
    \label{eq:conv-v}
    \lambda_V^\mu & =    \left( \Sigma_R P_R + \Sigma_L P_L \right)j_\ell^\mu + {\rm h.c.}\;,\\
    \label{eq:conv-a}
    \lambda_A^\mu \gamma^5 & =    \left( \Sigma_R P_R - \Sigma_L P_L \right) j_\ell^\mu+ {\rm h.c.}\;,\\
    \lambda_T^{\mu\nu}  & =   \left( \Sigma^\dagger P_R + \Sigma P_L \right) j_\ell^{\mu\nu} + {\rm h.c.}\;,\\
    \label{eq:conv-eps}
    \lambda_\varepsilon^{\mu\nu}\gamma^5  & = -i\,   \left( \Sigma^\dagger P_R - \Sigma P_L \right)j^{\mu\nu}_{\ell} + {\rm h.c.}\;.
\end{align}
In the above, the lepton currents are given by
\begin{align}
    j_\ell &= \left\{ \mathcal{C}_{1[3]}^{(6)}\, \bar e e\;, \ \mathcal{C}_{2[4]}^{(6)}\,\bar e i \gamma^5 e\;,\ \mathcal{C}_{13[14]}^{(6)}\,\bar \nu_L e_R \right\}\;,\\
    j_\ell^\mu &= \left\{\mathcal{C}_{5[7]}^{(6)}\,\bar e \gamma^\mu e\;, \ \mathcal{C}_{6[8]}^{(6)}\,\bar e \gamma^\mu \gamma^5 e\;,\  \mathcal{C}_{9[10]}^{(6)}\,\bar \nu_L \gamma^\mu \nu_L\;,\ \mathcal{C}_{15[16]}^{(6)}\,\bar \nu_L \gamma^\mu e_L \right\}\;,\\
    j_\ell^{\mu\nu} &= \left\{ \mathcal{C}_{11}^{(6)}\, \bar e \sigma^{\mu\nu} e\;,\ \mathcal{C}_{12}^{(6)}\,\bar e i \gamma^5 \sigma^{\mu\nu} e\;,\ \mathcal{C}_{17}^{(6)}\,\bar \nu_L \sigma^{\mu\nu} e_R \right\}\;.
\end{align}
Note that the Hermitian conjugate terms in Eqs.~\eqref{eq:conv-s}-\eqref{eq:conv-eps} are omitted for the neutral lepton currents.
In case of the lepton current $\bar e e$, the Wilson coefficients are $\mathcal{C}_{1}^{(6)}$ and $\mathcal{C}_{3}^{(6)}$ for the scalar and pseudo-scalar sources, respectively. A similar indexing convention applies to the Wilson coefficients for other lepton currents.

A comparison of elements in the three methods for the matching from quark to hadronic operators is given in Tab.~\ref{tab:comparison}. In the table, we have compare the building blocks, bases at the quark/hadronic level, and their parameterizations across the external source method, conventional spurion method, and systematic spurion method.
Note that one can also obtain the matching in the conventional spurion method in the $q$ basis using the spurions $(\lambda_S,\lambda_P, \lambda_V^\mu, \lambda_A^\mu, \lambda_T^{\mu\nu}, \lambda_\varepsilon^{\mu\nu})$ as the building blocks.

\begin{table}[H]
\tabcolsep=4pt
\renewcommand\arraystretch{1.5}
\caption{Comparison of elements in the three methods for matching from quark to hadronic operators.
} 
\begin{center}
    \begin{tabular}{c|c|c|c}
    \hline
    \hline
    method & param. & bases & building blocks \\ 
    \hline
    external & $U$ & chiral/LR & $U\;,U^\dagger\;,\chi\;,\chi^\dagger\;, F_{L}^{\mu\nu}\;,F_{R}^{\mu\nu}$, $D_\mu$   \\
    \hline
    external & $u$ & $q$/$u$ & $u^\mu\;,\chi_\pm\;,f_\pm^{\mu\nu}\;,t_\pm^{\mu\nu}$, $\nabla_\mu$   \\
    \hline
    conventional & $u$ or $U$ & chiral/LR & $u$, $u^\dagger$, $D_\mu$, ($\lambda$, $\lambda^\dagger$, $\lambda_L$, $\lambda_R$, $\lambda^{\mu\nu}$, $\lambda^{\mu\nu \dagger}$)  \\
    \hline
    systematic & $u$ & $q$/$u$ & $\hat u^\mu$, $\nabla_\mu$, ($\hat\chi_\pm$, $\Sigma_\pm$, $Q_\pm$), $j_\ell$, $j_\ell^\mu$, $j_\ell^{\mu\nu}$, $A^\mu$   \\
    \hline
    \hline
    \end{tabular}
    \end{center}
    \label{tab:comparison}
\end{table}

\section{Matching beyond the dimension-6 level}
\label{sec:dim7-9}

While the dimension-6 LEFT operators can be systematically matched to the chiral Lagrangian using either the external source method or conventional spurion approach, going beyond the dimension-6 level introduces critical complications in two aspects:
\begin{itemize}
    \item The external source method is quite limited or inapplicable for higher-dimensional operators.
    \item The conventional spurion method, though viable in principle, requires the introduction of new spurions depending on the interactions, exacerbating redundancy issues already evident at the dimension-6 level, in four-quark interactions~\cite{Grinstein:1985ut,Savage:1998yh,Graesser:2016bpz,Cirigliano:2017ymo,Cirigliano:2018yza,Liao:2019gex,Akdag:2022sbn}, six-quark interactions~\cite{Bijnens:2017xrz,He:2021mrt}, tensor interactions~\cite{Mertens:2011ts}, and derivative interactions~\cite{Akdag:2022sbn}.
\end{itemize}
Our systematic spurion method resolves these challenges by providing a unified framework valid for
higher-dimensional operators 
while maintaining the minimal set of spurions. In this work, we will discuss the matching for dimension-7 and dimension-8 derivative operators as well as dimension-9 four-quark operators.

\subsection{Dimension-7 derivative operators}

We obtain the effective Lagrangian at the dimension-7 level as follows:
\begin{align}
    \mathcal{L}_{\text{eff}}^{(7)} \supset \sum_{i=1}^{10} \mathcal{O}_i^{(7)} + \left( \sum_{i=11}^{14} \mathcal{O}_i^{(7)} + {\rm h.c.} \right)\;,
\end{align} 
where the derivative LEFT operators~\cite{Liao:2020zyx} are classified in the $CP$ eigenstate,
\begin{align}
\mathcal{O}_{1}^{(7)}&=(\bar{e}i\overleftrightarrow{D}^\mu e)[\bar{q}\gamma_\mu(\Sigma_R P_R+\Sigma_L P_L) q]\;,
&
\mathcal{O}_{2}^{(7)}&=(\bar{e}\gamma^5\overleftrightarrow{D}^\mu e)[\bar{q}\gamma_\mu(\Sigma_R P_R+\Sigma_L P_L) q]\;,
\notag\\
\mathcal{O}_{3}^{(7)}&=(\bar{e}i\overleftrightarrow{D}^\mu e)[\bar{q}\gamma_\mu(\Sigma_R P_R-\Sigma_L P_L) q]\;,
&
\mathcal{O}_{4}^{(7)}&=(\bar{e}\gamma^5\overleftrightarrow{D}^\mu e)[\bar{q}\gamma_\mu(\Sigma_R P_R-\Sigma_L P_L) q]\;,
\notag\\
\mathcal{O}_{5}^{(7)}&=(\bar{e}\gamma^\mu e)[\bar{q}i\overleftrightarrow{D}_\mu(\Sigma^\dagger P_R+\Sigma P_L) q]\;,
&
\mathcal{O}_{6}^{(7)}&=(\bar{e}\gamma^\mu\gamma^5 e)[\bar{q}i\overleftrightarrow{D}_\mu(\Sigma^\dagger P_R+\Sigma P_L) q]\;,
\notag\\
\mathcal{O}_{7}^{(7)}&=(\bar{e}\gamma^\mu e)[\bar{q}i\overleftrightarrow{D}_\mu(\Sigma^\dagger P_R-\Sigma P_L) q]\;,
&
\mathcal{O}_{8}^{(7)}&=(\bar{e}\gamma^\mu\gamma^5 e)[\bar{q}i\overleftrightarrow{D}_\mu(\Sigma^\dagger P_R-\Sigma P_L) q]\;,\nn
\\
\mathcal{O}_{9}^{(7)}&=(\bar{\nu}_L\gamma^\mu \nu_L)[\bar{q}i\overleftrightarrow{D}_\mu(\Sigma^\dagger P_R+\Sigma P_L) q]\;,
&
\mathcal{O}_{10}^{(7)}&=(\bar{\nu}_L\gamma^\mu \nu_L)[\bar{q}i\overleftrightarrow{D}_\mu(\Sigma^\dagger P_R-\Sigma P_L) q]\;,
\notag\\
\mathcal{O}_{11}^{(7)}&=(\bar{\nu}_Li\overleftrightarrow{D}^\mu e_R)[\bar{q}\gamma_\mu(\Sigma_R P_R+\Sigma_L P_L) q]\;,
&
\mathcal{O}_{12}^{(7)}&=(\bar{\nu}_Li\overleftrightarrow{D}^\mu e_R)[\bar{q}\gamma_\mu(\Sigma_R P_R-\Sigma_L P_L) q]\;,
\notag\\
\mathcal{O}_{13}^{(7)}&=(\bar{\nu}_L\gamma^\mu e_L)[\bar{q}i\overleftrightarrow{D}_\mu(\Sigma^\dagger P_R+\Sigma P_L) q]\;,
&
\mathcal{O}_{14}^{(7)}&=(\bar{\nu}_L\gamma^\mu e_L)[\bar{q}i\overleftrightarrow{D}_\mu(\Sigma^\dagger P_R-\Sigma P_L) q]\;.
\end{align}
In the above, the Hermitian version of the covariant derivative is defined as
\begin{align}
    \bar{\psi} i \overleftrightarrow{D}_\mu \psi \equiv \bar{\psi} i \overrightarrow{D}_\mu \psi-\bar{\psi} i \overleftarrow{D}_\mu \psi\;,
\end{align}
where the quark field $\psi = u$ or $d$. More specifically, we have
\begin{align}
    \bar{\psi} \overrightarrow{D}_\mu \psi=\bar{\psi}\left(D_\mu \psi\right)\;,\quad 
    \bar{\psi} \overleftarrow{D}_\mu \psi=\left(D_\mu \bar{\psi}\right) \psi\;,\quad
    D_\mu \bar{\psi} \equiv\left(D_\mu \psi\right)^{\dagger} \gamma^0\;.
\end{align}
The explicit forms of the covariant derivatives are~\cite{Akdag:2022sbn}
\begin{align}
\label{eq:CD-A}
D_\mu \psi & =\left(\partial_\mu+i e Q A_\mu+i g_s T^A G_\mu^A\right) \psi \;,\\
D_\mu \bar{\psi} & =\bar{\psi}\left(\partial_\mu-i e Q A_\mu-i g_s T^A G_\mu^A\right)\;,
\end{align}
where $g_s$ is the strong coupling constant, $T^A$ denotes the color generators, and $G_\mu^A$ is the gluon field.
Therefore, $\bar{\psi} i \overleftrightarrow{D}_\mu \psi$ could generate the interactions between quarks and photon or gluon fields.
In case of the electron field, we have the similar Hermitian covariant derivative $ \bar{e} i \overleftrightarrow{D}_\mu e$, where the gluon field is not included.

\subsubsection{External source method}

For operators with derivatives acting on lepton fields ($\mathcal{O}_{1-4}^{(7)}$ and $\mathcal{O}_{11,12}^{(7)}$), the chiral Lagrangian can be constructed analogously to the dimension-6 operators. However, for operators with derivative acting on the quark field ($\mathcal{O}_{5-10}^{(7)}$ and $\mathcal{O}_{13,14}^{(7)}$), the external source method cannot be directly applied. By using the EOMs, these derivatives can be transferred to the lepton part, enabling matching to the chiral Lagrangian in the external source method, which has been studied in analogous contexts~\cite{Dekens:2020ttz,Haxton:2024lyc}. From Ref.~\cite{Haxton:2024lyc}, we find that
\begin{align}
\label{eq:EOMs}
    \partial^\mu\left(\bar{e} \gamma^\nu e\right)\left(\bar{\psi} \sigma_{\mu \nu} \psi\right)&=-\left(\bar{e} \gamma^\mu e\right)\left(\bar{\psi} i \overleftrightarrow{D}_\mu \psi\right)+2 m_\psi \left(\bar{e} \gamma_\mu e\right)\left(\bar{\psi} \gamma^\mu \psi\right)\;,\nn\\
    \partial^\mu\left(\bar{e} \gamma^\nu \gamma^5 e\right)\left(\bar{\psi} \sigma_{\mu \nu} \psi\right)&=-\left(\bar{e} \gamma^\mu \gamma^5 e\right)\left(\bar{\psi} i \overleftrightarrow{D}_\mu \psi\right)+2 m_\psi \left(\bar{e} \gamma_\mu \gamma^5 e\right)\left(\bar{\psi} \gamma^\mu \psi\right)\;,\nn\\
    \partial^\mu\left(\bar{e} \gamma^\nu e\right)\left(\bar{\psi} \sigma_{\mu \nu} \gamma^5 \psi\right) &= -\left(\bar{e} \gamma^\mu e\right)\left(\bar{\psi} i \gamma^5 \overleftrightarrow{D}_\mu \psi\right)\;,\nn\\
    \partial^\mu\left(\bar{e} \gamma^\nu \gamma^5 e\right)\left(\bar{\psi} \sigma_{\mu \nu} \gamma^5 \psi\right) &= -\left(\bar{e} \gamma^\mu \gamma^5 e\right)\left(\bar{\psi} i \gamma^5 \overleftrightarrow{D}_\mu \psi\right)\;.
\end{align}
Similarly, for the left-handed neutrino field, we have
\begin{align}
\label{eq:EOMs-2}
    \partial^\mu\left(\bar{\nu}_L \gamma^\nu \nu_L\right)\left(\bar{\psi}_L \sigma_{\mu \nu} \psi_R\right)&=-\left(\bar{\nu}_L \gamma^\mu \nu_L\right)\left(\bar{\psi}_L i \overleftrightarrow{D}_\mu \psi_R\right)\;,\nn\\
    \partial^\mu\left(\bar{\nu}_L \gamma^\nu e_L\right)\left(\bar{\psi}_L \sigma_{\mu \nu} \psi_R\right)&=-\left(\bar{\nu}_L \gamma^\mu e_L\right)\left(\bar{\psi}_L i \overleftrightarrow{D}_\mu \psi_R\right)\;.
\end{align}

In the external source method, the tensor source $\bar{t}^{\mu\nu}$ is formed from the lepton currents appearing in Eqs.~\eqref{eq:EOMs} and \eqref{eq:EOMs-2}, each multiplied by their corresponding Wilson coefficients. According to chiral power counting~\cite{Mateu:2007tr}, this tensor source scales as $\bar{t}^{\mu\nu} \sim \mathcal{O}(p^2)$, and leading to the contribution to chiral Lagrangian of $\mathcal{O}(p^4)$. 
In Eq.~\eqref{eq:EOMs}, the vector source $v^\mu$ consists of the terms $m_\psi (\bar{e}\gamma^\mu e)$ and $m_\psi (\bar{e}\gamma^\mu\gamma^5 e)$, both multiplied by their respective Wilson coefficients. Given that $m_\psi \sim \mathcal{O}(p^2)$, the vector source consequently scales as $v^\mu \sim \mathcal{O}(p^3)$ in the chiral expansion, and the contribution to chiral Lagrangian also appears at $\mathcal{O}(p^4)$. 

For consistency, the LO chiral Lagrangian for dimension-7 derivative operators
also appears at $p^4$ order.
Expanding the Hermitian covariant derivative,
\begin{align}
\label{eq:covariant-expansion}
\bar{\psi} i \overleftrightarrow{D}_\mu \psi = \bar{\psi} i \overrightarrow{\partial}_\mu \psi - \bar{\psi} i \overleftarrow{\partial}_\mu \psi + \text{gauge field terms}.
\end{align}
As demonstrated in Appendix~\ref{sec:dim-7_derivative} using the conventional spurion method, the chiral Lagrangian for partial derivative operators indeed first appears at $\mathcal{O}(p^4)$. 
The operators with single gluon field originating from the covariant derivative do not match to additional chiral operators below the chiral symmetry breaking scale $\Lambda_\chi$~\cite{Akdag:2022sbn}, and their effects are absorbed in the LECs. The matching for photon operators is detailed in Appendix~\ref{sec:match-photon}.

\subsubsection{Systematic spurion method}

While theoretically viable, the conventional spurion method is challenging for matching dimension-7 derivative operators due to the elimination of redundancies. In the following, we will present the matching results in our systematic spurion method.

\begin{table}[H]
    \centering
        \captionsetup{justification=centering}
   \caption{The matching for the derivative operators at $p^4$ order.}
    \begin{tabular}{|l|c|}
    \hline
    LEFT operator&$\chi$PT operators at $\mathcal{O}(p^4)$\\
    \hline
        \multirow{4}{*}{$\mathcal{O}_{5}^{(7)} = (\bar e \gamma^\mu e)[\bar{q}i\overleftrightarrow{D}_\mu(\Sigma^\dagger P_R+\Sigma P_L) q]$}&$i(\bar e \gamma^\mu e)\langle\Sigma_+[\nabla_\mu  \hat u_\nu, \hat u^\nu]\rangle$\\
        &$i(\bar e \gamma^\mu e)\langle \nabla^\nu\Sigma_+[ \hat u_\mu, \hat u_\nu]\rangle$\\
        &$\varepsilon_{\mu\nu\rho\sigma}(\bar e \gamma^\mu e)\langle \nabla^\nu\Sigma_- \hat u^\rho \hat u^\sigma\rangle$\\
        &$(\bar e \gamma^\mu e)\langle \hat u_\mu[\Sigma_+,\hat \chi_-]\rangle$\\
        &$(\bar e \gamma^\mu e)\langle \hat u_\mu[\Sigma_-,\hat \chi_+]\rangle$\\
        \hline
        \multirow{4}{*}{$\mathcal{O}_{7}^{(7)} = (\bar e \gamma^\mu e)[\bar{q}i\overleftrightarrow{D}_\mu(\Sigma^\dagger P_R-\Sigma P_L)  q]$}&$i(\bar e \gamma^\mu e)\langle\Sigma_-[\nabla_\mu  \hat u_\nu, \hat u^\nu]\rangle$\\
        &$i(\bar e \gamma^\mu e)\langle \nabla^\nu\Sigma_-[ \hat u_\mu, \hat u_\nu]\rangle$\\
        &$\varepsilon_{\mu\nu\rho\sigma}(\bar e \gamma^\mu e)\langle \nabla^\nu\Sigma_+ \hat u^\rho \hat u^\sigma\rangle$\\
        &$(\bar e \gamma^\mu e)\langle \hat u_\mu[\Sigma_-,\hat \chi_-]\rangle$\\
        &$(\bar e \gamma^\mu e)\langle \hat u_\mu[\Sigma_+,\hat \chi_+]\rangle$\\
        \hline
       \multirow{4}{*}{$\mathcal{O}_{6}^{(7)} = (\bar e \gamma^\mu\gamma^5 e)[\bar{q}i\overleftrightarrow{D}_\mu(\Sigma^\dagger P_R+\Sigma P_L) q]$} &$i(\bar e \gamma^\mu\gamma^5 e)\langle\Sigma_+[\nabla_\mu \hat  u_\nu, \hat u^\nu]\rangle$\\
       &$i(\bar e \gamma^\mu\gamma^5 e)\langle \nabla^\nu\Sigma_+[ \hat u_\mu, \hat u_\nu]\rangle$\\
       &$\varepsilon_{\mu\nu\rho\sigma}(\bar e \gamma^\mu\gamma^5 e)\langle \nabla^\nu\Sigma_- \hat u^\rho \hat u^\sigma\rangle$\\
       &$(\bar e \gamma^\mu\gamma^5 e)\langle \hat u_\mu[\Sigma_+,\hat \chi_-]\rangle$\\
        &$(\bar e \gamma^\mu\gamma^5 e)\langle \hat u_\mu[\Sigma_-,\hat \chi_+]\rangle$\\
       \hline
       \multirow{4}{*}{$\mathcal{O}_{8}^{(7)} = (\bar e \gamma^\mu\gamma^5 e)[\bar{q}i\overleftrightarrow{D}_\mu(\Sigma^\dagger P_R-\Sigma P_L) q]$ }&$i(\bar e \gamma^\mu\gamma^5 e)\langle\Sigma_-[\nabla_\mu  \hat u_\nu, \hat u^\nu]\rangle$\\
       &$i(\bar e \gamma^\mu \gamma^5e)\langle \nabla^\nu\Sigma_-[ \hat u_\mu, \hat u_\nu]\rangle$\\
       &$\varepsilon_{\mu\nu\rho\sigma}(\bar e \gamma^\mu\gamma^5 e)\langle \nabla^\nu\Sigma_+ \hat u^\rho \hat u^\sigma\rangle$\\
       &$(\bar e \gamma^\mu\gamma^5 e)\langle \hat u_\mu[\Sigma_-,\hat \chi_-]\rangle$\\
        &$(\bar e \gamma^\mu\gamma^5 e)\langle \hat u_\mu[\Sigma_+,\hat \chi_+]\rangle$\\
       \hline
       \multirow{4}{*}{$\mathcal{O}_{9}^{(7)} = (\bar{\nu}_L\gamma^\mu \nu_L)[\bar{q}i\overleftrightarrow{D}_\mu(\Sigma^\dagger P_R+\Sigma P_L) q]$} & $i(\bar{\nu}_L\gamma^\mu \nu_L)\langle\Sigma_+[\nabla_\mu  \hat u_\nu, \hat u^\nu]\rangle$\\
       &$i(\bar \nu_L \gamma^\mu \nu_L)\langle \nabla^\nu\Sigma_+[ \hat u_\mu, \hat u_\nu]\rangle$\\
       &$\varepsilon_{\mu\nu\rho\sigma}(\bar \nu_L \gamma^\mu \nu_L)\langle \nabla^\nu\Sigma_- \hat u^\rho \hat u^\sigma\rangle$\\
       &$(\bar \nu_L \gamma^\mu \nu_L)\langle \hat u_\mu[\Sigma_+,\hat \chi_-]\rangle$\\
        &$(\bar \nu_L \gamma^\mu \nu_L)\langle \hat u_\mu[\Sigma_-,\hat \chi_+]\rangle$\\
       \hline
      \multirow{4}{*}{ $\mathcal{O}_{10}^{(7)} = (\bar{\nu}_L\gamma^\mu \nu_L)[\bar{q}i\overleftrightarrow{D}_\mu(\Sigma^\dagger P_R-\Sigma P_L) q]$} & $i(\bar{\nu}_L\gamma^\mu \nu_L)\langle\Sigma_-[\nabla_\mu  \hat u_\nu, \hat u^\nu]\rangle$\\
      &$i(\bar \nu_L \gamma^\mu \nu_L)\langle \nabla^\nu\Sigma_-[ \hat u_\mu, \hat u_\nu]\rangle$\\
      &$\varepsilon_{\mu\nu\rho\sigma}(\bar \nu_L \gamma^\mu \nu_L)\langle \nabla^\nu\Sigma_+ \hat u^\rho \hat u^\sigma\rangle$\\
      &$(\bar \nu_L \gamma^\mu \nu_L)\langle \hat u_\mu[\Sigma_-,\hat \chi_-]\rangle$\\
        &$(\bar \nu_L \gamma^\mu \nu_L)\langle \hat u_\mu[\Sigma_+,\hat \chi_+]\rangle$\\
       \hline
       \multirow{4}{*}{$\mathcal{O}_{13}^{(7)} = (\bar{\nu}_L\gamma^\mu e_L)[\bar{q}i\overleftrightarrow{D}_\mu(\Sigma^\dagger P_R+\Sigma P_L) q]$} &  $i(\bar{\nu}_L\gamma^\mu e_L)\langle\Sigma_+[\nabla_\mu  \hat u_\nu, \hat u^\nu]\rangle$ \\
       &$i(\bar \nu_L \gamma^\mu e_L)\langle \nabla^\nu\Sigma_+[ \hat u_\mu, \hat u_\nu]\rangle$\\
       &$\varepsilon_{\mu\nu\rho\sigma}(\bar \nu_L \gamma^\mu e_L)\langle \nabla^\nu\Sigma_- \hat u^\rho \hat u^\sigma\rangle$\\
       &$(\bar \nu_L \gamma^\mu e_L)\langle \hat u_\mu[\Sigma_+,\hat \chi_-]\rangle$\\
        &$(\bar \nu_L \gamma^\mu e_L)\langle \hat u_\mu[\Sigma_-,\hat \chi_+]\rangle$\\
       \hline
       \multirow{4}{*}{$\mathcal{O}_{14}^{(7)} = (\bar{\nu}_L\gamma^\mu e_L)[\bar{q}i\overleftrightarrow{D}_\mu(\Sigma^\dagger P_R-\Sigma P_L) q]$} & $i(\bar{\nu}_L\gamma^\mu e_L)\langle\Sigma_-[\nabla_\mu  \hat u_\nu, \hat u^\nu]\rangle$\\
       &$i(\bar\nu_L \gamma^\mu e_L)\langle \nabla^\nu\Sigma_-[ \hat u_\mu, \hat u_\nu]\rangle$\\
       &$\varepsilon_{\mu\nu\rho\sigma}(\bar \nu_L \gamma^\mu e_L)\langle \nabla^\nu\Sigma_+ \hat u^\rho \hat u^\sigma\rangle$\\
       &$(\bar \nu_L \gamma^\mu e_L)\langle \hat u_\mu[\Sigma_-,\hat \chi_-]\rangle$\\
       &$(\bar \nu_L \gamma^\mu e_L)\langle \hat u_\mu[\Sigma_+,\hat \chi_+]\rangle$\\
        \hline
    \end{tabular}
    \label{tab:td}
\end{table}

For the operators $\mathcal{O}_{1-4}^{(7)}$ and $\mathcal{O}_{11,12}^{(7)}$, the matching is similar to the dimension-6 operators with vector and axial-vector currents. The conversion follows Eqs.~\eqref{eq:conv-v} and~\eqref{eq:conv-a}, with the lepton currents including derivative,
\begin{align}
\label{eq:dim7-external}
    j_\ell^\mu &= \alpha_{\ell \ell^\prime} \left\{\mathcal{C}_{1[3]}^{(7)}\, \bar{e} i\overleftrightarrow{D}^\mu e\;,\ \mathcal{C}_{2[4]}^{(7)}\, \bar{e} \gamma^5 \overleftrightarrow{D}^\mu e\;,\ \mathcal{C}_{11[12]}^{(7)}\, \bar{\nu}_L i\overleftrightarrow{D}^\mu e_R \right\}\;.
\end{align}

For operators with derivative acting on the quark field ($\mathcal{O}_{5-10}^{(7)}$ and $\mathcal{O}_{13,14}^{(7)}$), we obtain the chiral Lagrangian which arising at $\mathcal{O}(p^4)$, which are shown in Tab.~\ref{tab:td}. We employ the commutator $[\nabla_\mu \hat{u}_\nu, \hat{u}^\nu]$ in our construction, analogous to the tensor current case, to maintain the correct $C$-transformation properties. Details of the transformation properties under discrete symmetries and the matching using the conventional spurion method are given in Appendix~\ref{sec:dim-7_derivative}.

\subsection{Dimension-8 derivative operators}
Although the derivatives acting on the quark field can be transferred to the lepton part through the EOMs for dimension-7 operators, this is not always true for the dimension-8 derivative operators. 
The external source method is limited in its applicability due to its restriction to a specific set of five types of external sources. The conventional spurion method is cumbersome because of the inherent redundancies in its formulation. Both issues can be addressed in the systematic spurion method. 

We consider the effective Lagrangian at the dimension-8 level as follows:
\begin{align}
    \mathcal{L}_{\rm eff}^{(8)}=\sum_{i=1}^{22}\mathcal{O}_i^{(8)}+\left(\sum_{i=23}^{31}+ {\rm h.c.}\right)\;,
\end{align}
where the $\psi^4 D^2$ type operators~\cite{Li:2020tsi,Murphy:2020cly} in the $CP$ eigenstate are given by
\begin{align}
\mathcal{O}_1^{(8)}&=(\bar e\overleftrightarrow{D}^\mu\gamma^\nu e)[\bar q\overleftrightarrow{D}_\mu\gamma_\nu(\Sigma_R P_R+\Sigma_L P_L) q]\;,
&
\mathcal{O}_2^{(8)}&=(\bar e\overleftrightarrow{D}^\mu\gamma^\nu\gamma^5 e)[\bar q\overleftrightarrow{D}_\mu\gamma_\nu(\Sigma_R P_R+\Sigma_L P_L) q]\;,\notag\\
\mathcal{O}_3^{(8)}&=(\bar e\overleftrightarrow{D}^\mu\gamma^\nu e)[\bar q\overleftrightarrow{D}_\mu\gamma_\nu(\Sigma_R P_R-\Sigma_L P_L) q]\;,
&
\mathcal{O}_4^{(8)}&=(\bar e\overleftrightarrow{D}^\mu\gamma^\nu\gamma^5 e)[\bar q\overleftrightarrow{D}_\mu\gamma_\nu(\Sigma_R P_R-\Sigma_L P_L) q]\;,\notag\\
\mathcal{O}_5^{(8)}&=(\bar \nu_L\overleftrightarrow{D}^\mu\gamma^\nu \nu_L)[\bar q(\Sigma_R P_R+\Sigma_L P_L)\overleftrightarrow{D}_\mu\gamma_\nu q]\;,
&
\mathcal{O}_6^{(8)}&=(\bar \nu_L\overleftrightarrow{D}^\mu\gamma^\nu \nu_L)[\bar q(\Sigma_R P_R-\Sigma_L P_L)\overleftrightarrow{D}_\mu\gamma_\nu\gamma^5 q]\;,\notag\\
\mathcal{O}_{7}^{(8)}&=(\bar e e)D^2[\bar q(\Sigma^\dagger P_R+\Sigma P_L)q]\;,
&
\mathcal{O}_{8}^{(8)}&=(\bar e\gamma^5 e)D^2[\bar q(\Sigma^\dagger P_R+\Sigma P_L) q]\;,\notag\\
\mathcal{O}_{9}^{(8)}&=(\bar e e)D^2[\bar q(\Sigma^\dagger P_R-\Sigma P_L) q]\;,
&
\mathcal{O}_{10}^{(8)}&=(\bar e\gamma^5 e)D^2[\bar q(\Sigma^\dagger P_R-\Sigma P_L) q]\;,\notag\\
\mathcal{O}_{11}^{(8)}&=(\bar e\gamma^\mu e)D^2[\bar q\gamma_\mu(\Sigma_R P_R+\Sigma_L P_L) q]\;,
&
\mathcal{O}_{12}^{(8)}&=(\bar e\gamma^\mu\gamma^5 e)D^2[\bar q\gamma_\mu(\Sigma_R P_R+\Sigma_L P_L) q]\;,\notag\\
\mathcal{O}_{13}^{(8)}&=(\bar e\gamma^\mu e)D^2[\bar q\gamma_\mu(\Sigma_R P_R-\Sigma_L P_L) q]\;,
&
\mathcal{O}_{14}^{(8)}&=(\bar e\gamma^\mu\gamma^5 e)D^2[\bar q\gamma_\mu(\Sigma_R P_R-\Sigma_L P_L) q]\;,\notag\\
\mathcal{O}_{15}^{(8)}&=(\bar \nu_L\gamma^\mu \nu_L)D^2[\bar q\gamma_\mu(\Sigma_R P_R+\Sigma_L P_L) q]\;,
&
\mathcal{O}_{16}^{(8)}&=(\bar \nu_L\gamma^\mu \nu_L)D^2[\bar q\gamma_\mu(\Sigma_R P_R-\Sigma_L P_L) q]\;,\notag\\
\mathcal{O}_{17}^{(8)}&=(\bar e\overleftrightarrow{D}^\mu e)[\bar q\overleftrightarrow{D}_\mu(\Sigma^\dagger P_R+\Sigma P_L) q]\;,
&
\mathcal{O}_{18}^{(8)}&=(\bar e\overleftrightarrow{D}^\mu\gamma^5 e)[\bar q\overleftrightarrow{D}_\mu(\Sigma^\dagger P_R+\Sigma P_L) q]\;,\notag\\
\mathcal{O}_{19}^{(8)}&=(\bar e\overleftrightarrow{D}^\mu e)[\bar q\overleftrightarrow{D}_\mu(\Sigma^\dagger P_R-\Sigma P_L) q]\;,
&
\mathcal{O}_{20}^{(8)}&=(\bar e\overleftrightarrow{D}^\mu\gamma^5 e)[\bar q\overleftrightarrow{D}_\mu(\Sigma^\dagger P_R-\Sigma P_L) q]\;,\notag\\
\mathcal{O}_{21}^{(8)}&=(\bar e \sigma^{\mu\nu}e)D^2[\bar q\sigma_{\mu\nu}(\Sigma^\dagger P_R+\Sigma P_L) q]\;,
&
\mathcal{O}_{22}^{(8)}&=(\bar e \gamma^5\sigma^{\mu\nu}e)D^2[\bar q\sigma_{\mu\nu}(\Sigma^\dagger P_R+\Sigma P_L) q]\;,\notag\\
\mathcal{O}_{23}^{(8)}&=(\bar \nu_L\overleftrightarrow{D}^\mu\gamma^\nu e_L)[\bar q\overleftrightarrow{D}_\mu\gamma_\nu(\Sigma_R P_R+\Sigma_L P_L) q]\;,
&
\mathcal{O}_{24}^{(8)}&=(\bar \nu_L\overleftrightarrow{D}^\mu\gamma^\nu e_L)[\bar q\overleftrightarrow{D}_\mu\gamma_\nu(\Sigma_R P_R-\Sigma_L P_L) q]\;,\notag\\
\mathcal{O}_{25}^{(8)}&=(\bar \nu_L\overleftrightarrow{D}^\mu e_R)[\bar q\overleftrightarrow{D}_\mu(\Sigma^\dagger P_R+\Sigma P_L) q]\;,
&
\mathcal{O}_{26}^{(8)}&=(\bar \nu_L\overleftrightarrow{D}^\mu e_R)[\bar q\overleftrightarrow{D}_\mu(\Sigma^\dagger P_R-\Sigma P_L) q]\;,\notag\\
\mathcal{O}_{27}^{(8)}&=(\bar \nu_L\gamma^\mu e_L)D^2[\bar q\gamma_\mu(\Sigma_R P_R+\Sigma_L P_L) q]\;,
&
\mathcal{O}_{28}^{(8)}&=(\bar \nu_L\gamma^\mu e_L)D^2[\bar q\gamma_\mu(\Sigma_R P_R-\Sigma_L P_L) q]\;,\notag\\
\mathcal{O}_{29}^{(8)}&=(\bar \nu_L e_R)D^2[\bar q(\Sigma^\dagger P_R+\Sigma P_L) q]\;,
&
\mathcal{O}_{30}^{(8)}&=(\bar \nu_L e_R)D^2[\bar q(\Sigma^\dagger P_R-\Sigma P_L) q]\;,\notag\\
\mathcal{O}_{31}^{(8)}&=(\bar \nu_L\sigma^{\mu\nu}e_R)D^2[\bar q\sigma_{\mu\nu}(\Sigma^\dagger P_R+\Sigma P_L) q]\;.
\end{align}

We establish the following results:
\begin{itemize}
    \item The operators $\mathcal{O}_{7-16}^{(8)}$, $\mathcal{O}_{21-22}^{(8)}$ and $\mathcal{O}_{27-31}^{(8)}$, which involve two derivatives in the quark sector, can be matched to the chiral Lagrangian using the external source method. For example, the operator $(\bar e e)\partial^2[\bar q(\Sigma^\dagger P_R+\Sigma P_L)q]$ can match to $(\bar{e}e)\nabla^2\langle\Sigma_+\rangle$ as the scalar operators.
    \item The operators $\mathcal{O}_{17-20}^{(8)}$ and $\mathcal{O}_{25-26}^{(8)}$ can also be incorporated into the framework through the EOMs, similar to the treatment of dimension-7 operators.
    \item The operators $\mathcal{O}_{1-6}^{(8)}$ and $\mathcal{O}_{23-24}^{(8)}$ cannot be matched onto the chiral Lagrangian via the external source method.
\end{itemize}

\begin{table}[H]
    \centering
\captionsetup{justification=centering}
   \caption{The matching for the two derivatives operators at $p^4$ order.}
    \begin{tabular}{|l|c|}
    \hline
    LEFT operator &$\chi$PT operator at $\mathcal{O}(p^4)$\\
    \hline
       $\mathcal{O}_{1}^{(8)} = (\bar e\gamma^\mu\overleftrightarrow{D}^\nu e)[\bar q\gamma_\mu\overleftrightarrow{D_\nu}(\Sigma_R P_R+\Sigma_L P_L)q]$&$(\bar e\gamma^\mu\overleftrightarrow{D}^\nu e)\langle Q_+\{ \hat u_\mu, \hat u_\nu\}\rangle$\\
       $\mathcal{O}_{2}^{(8)} = (\bar e\gamma^\mu\gamma^5\overleftrightarrow{D}^\nu e)[\bar q\gamma_\mu\overleftrightarrow{D_\nu}(\Sigma_R P_R+\Sigma_L P_L)q]$&$(\bar e\gamma^\mu\gamma^5\overleftrightarrow{D}^\nu e)\langle Q_+\{ \hat u_\mu, \hat u_\nu\}\rangle$\\
       $\mathcal{O}_{3}^{(8)} = (\bar e\gamma^\mu\overleftrightarrow{D}^\nu e)[\bar q\gamma_\mu\overleftrightarrow{D_\nu}(\Sigma_R P_R-\Sigma_L P_L)q]$&$(\bar e\gamma^\mu\overleftrightarrow{D}^\nu e)\langle Q_-\{ \hat u_\mu, \hat u_\nu\}\rangle$\\
       $\mathcal{O}_{4}^{(8)} = (\bar e\gamma^\mu\gamma^5\overleftrightarrow{D}^\nu e)[\bar q\gamma_\mu\overleftrightarrow{D_\nu}(\Sigma_R P_R-\Sigma_L P_L)q]$&$(\bar e\gamma^\mu\gamma^5\overleftrightarrow{D}^\nu e)\langle Q_-\{ \hat u_\mu, \hat u_\nu\}\rangle$\\
       \hline
       $\mathcal{O}_{5}^{(8)} = (\bar \nu_L\gamma^\mu\overleftrightarrow{D}^\nu \nu_L)[\bar q\gamma_\mu\overleftrightarrow{D_\nu}(\Sigma_R P_R+\Sigma_L P_L)q]$&$(\bar \nu_L\gamma^\mu\overleftrightarrow{D}^\nu \nu_L)\langle Q_+\{ \hat u_\mu, \hat u_\nu\}\rangle$\\
       $\mathcal{O}_{23}^{(8)} = (\bar \nu_L\gamma^\mu\overleftrightarrow{D}^\nu e_L)[\bar q\gamma_\mu\overleftrightarrow{D_\nu}(\Sigma_R P_R+\Sigma_L P_L)q]$&$(\bar \nu_L\gamma^\mu\overleftrightarrow{D}^\nu e_L)\langle Q_+\{ \hat u_\mu, \hat u_\nu\}\rangle$\\
       $\mathcal{O}_{6}^{(8)} = (\bar \nu_L\gamma^\mu\overleftrightarrow{D}^\nu \nu_L)[\bar q\gamma_\mu\overleftrightarrow{D_\nu}(\Sigma_R P_R-\Sigma_L P_L)q]$&$(\bar \nu_L\gamma^\mu\overleftrightarrow{D}^\nu \nu_L)\langle Q_-\{ \hat u_\mu, \hat u_\nu\}\rangle$\\
       $\mathcal{O}_{24}^{(8)} = (\bar \nu_L\gamma^\mu\overleftrightarrow{D}^\nu e_L)[\bar q\gamma_\mu\overleftrightarrow{D_\nu}(\Sigma_R P_R-\Sigma_L P_L)q]$&$(\bar \nu_L\gamma^\mu\overleftrightarrow{D}^\nu e_L)\langle Q_-\{ \hat u_\mu, \hat u_\nu\}\rangle$\\
        \hline
    \end{tabular}
    \label{tab:td2}
\end{table}

The matching results for operators $\mathcal{O}_{1-6}^{(8)}$ involving the derivatives and $\mathcal{O}_{23-24}^{(8)}$ are presented in Table~\ref{tab:td2} which is only consistent for the $SU(3)$ case, and there is no $\chi$PT operator at $\mathcal{O}(p^4)$ for the $SU(2)$ case. 
As concrete examples, we examine the operators 
\begin{align}
    &\mathcal{O}_{1}^{(8)} = (\bar e\overleftrightarrow{D}^\mu\gamma^\nu e)[\bar q\overleftrightarrow{D}_\mu\gamma_\nu(\Sigma_R P_R+\Sigma_L P_L) q]\;,\nn\\
    &\mathcal{O}_{3}^{(8)} = (\bar e\overleftrightarrow{D}^\mu\gamma^\nu e)[\bar q\overleftrightarrow{D}_\mu\gamma_\nu(\Sigma_R P_R-\Sigma_L P_L) q]\;.
\end{align}
The chiral Lagrangian matching follows directly from their $CP$ transformation properties:
\begin{align}
    C+P+:\quad&(\bar e\overleftrightarrow{D}^\mu\gamma^\nu e)[\bar q\overleftrightarrow{D}_\mu\gamma_\nu(\Sigma_R P_R+\Sigma_L P_L) q]\rightarrow(\bar e\overleftrightarrow{D}^\mu\gamma^\nu e)\langle Q_+ \{\hat u_\mu,\hat u_\nu\}\rangle\;,\nn\\
    C-P-:\quad&(\bar e\overleftrightarrow{D}^\mu\gamma^\nu e)[\bar q\overleftrightarrow{D}_\mu\gamma_\nu (\Sigma_R P_R-\Sigma_L P_L)q]\rightarrow(\bar e\overleftrightarrow{D}^\mu\gamma^\nu e)\langle Q_- \{\hat u_\mu,\hat u_\nu\}\rangle\;.
\end{align}

The matching for dimension-8 derivative operators with the photon field is given in Appendix~\ref{sec:match-photon}.

 \subsection{Dimension-9 four-quark operators}

Our systematic spurion method is also valuable for interactions beyond operators with single quark bilinear. A key application involves operators with four-quark structures, which cannot be treated within the external source framework. To illustrate this, we consider lepton-number violating interactions at the dimension-9 level responsible for $0\nu\beta\beta$ decay,
\begin{align}
\mathcal{L}^{(9)}_{\Delta L =2} = \frac{1}{v^5}\sum_i\bigg[\left( C^{(9)}_{i\, \rm R}\, \bar e_R e^c_{R} + C^{(9)}_{i\, \rm L}\, \bar e_L e^c_{L} \right)  \, O_i^{(9)} +  C^{(9)}_i\bar e\gamma^\mu\gamma^5  e^c\, O_i^{\mu (9)}\bigg] + {\rm h.c.}\;.
\end{align}
The set of $SU(3)_c \times U(1)_{\text{em}}$-invariant four-quark two-lepton operators $O_i \bar{e}_X e_X^c$ $(X=R,L)$ and $O_i^\mu \bar{e} \gamma^\mu \gamma^5 e^c$, where $O_i$ and $O_i^\mu$ are Lorentz scalars and vectors, respectively. Following the notations Ref.~\cite{Cirigliano:2018yza} (see also Ref.~\cite{Graesser:2016bpz}), the dimension-9 four-quark operators are
\begin{align}\label{LagSca}
O_ 1^{(9)} =&  \bar{q}_L^\alpha \gamma_\mu \tau^+ q_L^\alpha \ \bar{q}_L^\beta  \gamma^\mu \tau^+ q_L^\beta\;, & O^{{(9)}\prime}_ 1  =&  \bar{q}_R^\alpha  \gamma_\mu \tau^+ q_R^\alpha \ \bar{q}_R^\beta  \gamma^\mu \tau^+ q_R^\beta      
\,\;,
\nn\\
O_ 2^{(9)}  = & \bar{q}_R^\alpha  \tau^+ q_L^\alpha \  \bar{q}_R^\beta  \tau^+ q_L^\beta\;, &     O^{{(9)}\prime}_ 2  =&  \bar{q}_L^\alpha  \tau^+ q_R^\alpha \  \bar{q}_L^\beta  \tau^+ q_R^\beta
\,\;,\nn\\
O_ 3^{(9)}  = & \bar{q}_R^\alpha  \tau^+ q_L^\beta \  \bar{q}_R^\beta  \tau^+ q_L^\alpha\;, &    O^{{(9)}\prime}_ 3  =&  \bar{q}_L^\alpha  \tau^+ q_R^\beta \  \bar{q}_L^\beta  \tau^+ q_R^\alpha
\,\;,\nn\\
O_ 4^{(9)}  = & \bar{q}_L^\alpha  \gamma_\mu \tau^+ q_L^\alpha \  \bar{q}_R^\beta  \gamma^\mu \tau^+ q_R^\beta    
\,\;,\nn\\ 
O_ 5^{(9)}  = & \bar{q}_L^\alpha  \gamma_\mu \tau^+ q_L^\beta \  \bar{q}_R^\beta  \gamma^\mu \tau^+ q_R^\alpha\;,\nn\\
O_{6}^{\mu(9)} = &\left(\bar q_L \tau^+\gamma^\mu q_L\right)\left(\bar q_L \tau^+ q_R\right)\,\;,& 
O_{6}^{\mu(9)\, \prime} =& \left(\bar q_R \tau^+\gamma^\mu q_R\right)\left(\bar q_R \tau^+ q_L\right)\,\;,
\nn\\
O_{7}^{\mu(9)} =& \left(\bar q_L T^A\tau^+\gamma^\mu q_L\right)\left(\bar q_L T^A\tau^+ q_R\right)\,\;,&  \,\,
O_{7}^{\mu(9)\, \prime} =& \left(\bar q_R T^A\tau^+\gamma^\mu q_R\right)\left(\bar q_R T^A\tau^+ q_L\right)\,\,
,\nn\\
O_{8}^{\mu(9)} = &\left(\bar q_L \tau^+\gamma^\mu q_L\right)\left(\bar q_R \tau^+ q_L\right)\,\;,& 
O_{8}^{\mu(9)\, \prime} =& \left(\bar q_R \tau^+\gamma^\mu q_R\right)\left(\bar q_L \tau^+ q_R\right)\,\
,\nn\\
O_{9}^{\mu (9)} =& \left(\bar q_L T^A\tau^+\gamma^\mu q_L\right)\left(\bar q_RT^A\tau^+ q_L\right)\,\;,& \,\,
O_{9}^{\mu (9)\, \prime} =& \left(\bar q_R T^A\tau^+\gamma^\mu q_R\right)\left(\bar q_LT^A\tau^+ q_R\right)\,\;,
\end{align}
where the operators $O_i^\prime$ and $O_i$ are related by parity. 
In the above, $\alpha$ and $\beta$ are color indices.  The operators $\mathcal{O}_{3}^{(9)}$ and $\mathcal{O}_{5}^{(9)}$ exhibit color mixing - where color indices are contracted between different quark bilinears - in contrast to their unmixed counterparts $\mathcal{O}_{3}^{(9)}$ and $\mathcal{O}_{4}^{(9)}$. Nevertheless, the mixed and unmixed operators have the same matching with independent LECs~\cite{Cirigliano:2018yza}. 
Similar to the single quark bilinear case, we can insert the spurions following Eq.~\eqref{eq:quark-bilinears}.

To construct the chiral Lagrangian systematically, we begin by reorganizing 
the two quark bilinears from the chiral basis
to $q$ basis:
\begin{align}
    \{P_L, P_R\} \otimes \{P_L, P_R\} \to 
    \{1, \gamma^5\} \otimes \{1, \gamma^5\}\;.
\end{align}
The operators can be systematically organized into minimal independent sets. For the color-unmixed operators, we establish the following correspondence:
\begin{align}
  & \{ O_1^{(9)}, O_1^{(9)\prime}, O_4^{(9)} \} \to \{ O_{VV},\, O_{AA},\, O_{VA} \}\;,\\
  & \{ O_2^{(9)}, O_2^{(9)\prime} \} \to \{ O_{SS},\, O_{PP},\, O_{SP} \}\;,\\
  &  \{ O_6^{\mu (9)}, O_6^{\mu (9)\prime}, O_8^{\mu (9)}, O_8^{\mu (9)\prime} \} \to \{ O_{VS},\, O_{AP},\, O_{VP},\, O_{AS} \}\;,
\end{align}
where the operators $O_{XY}$ in the $CP$ eigenstate are defined as
\begin{align}
\label{eq:four-quark-spurion}
    O_{XY} = \left(\bar q X \tau^+ q \right)  \left(\bar q Y \tau^+ q \right)\;,\quad X,Y=V,A,S,P\;,
\end{align}
with the color indices being omitted.  
In this notation, $V=\gamma^\mu $, $A=\gamma^\mu \gamma^5$, $S = 1$, $P = \gamma^5$.
The Lorentz indices are always contracted in these operator definitions. The conversion between operators in chiral and $q$ bases is given by
\begin{align}
    \begin{pmatrix}
        O_1^{(9)} \\
        O_1^{(9)\prime} \\
        O_4^{(9)} \\
    \end{pmatrix}
    &= \rho_1
    \begin{pmatrix}
        O_{VV} \\
        O_{AA} \\
        O_{VA} \\
    \end{pmatrix}\;, & 
    \begin{pmatrix}
        O_2^{(9)} \\
        O_2^{(9)\prime} \\
    \end{pmatrix}
    &= \rho_2
    \begin{pmatrix}
        O_{SS} \\
        O_{PP} \\
        O_{SP} \\
    \end{pmatrix}\;, &
    \begin{pmatrix}
        O_6^{\mu (9)} \\
        O_6^{\mu (9)\prime} \\
        O_8^{\mu (9)} \\
        O_8^{\mu (9)\prime} \\
    \end{pmatrix}
    &= \rho_3
    \begin{pmatrix}
        O_{VS} \\
        O_{AP} \\
        O_{VP} \\
        O_{AS} \\
    \end{pmatrix}\;,
\end{align}
where the correlation matrices are
\begin{align}
    \rho_1 = \dfrac{1}{4} 
    \begin{pmatrix}
        1 & 1 & -2\\
        1 & 1 & 2\\
        1 & -1 & 0\\
    \end{pmatrix}\;,\quad
    \rho_2 = \dfrac{1}{4}
    \begin{pmatrix}
        1 & 1 & -2\\
        1 & 1 & 2\\
    \end{pmatrix}\;,\quad
    \rho_3 = \dfrac{1}{4}
    \begin{pmatrix}
        1 & -1 & 1 & -1 \\
        1 & -1 & -1 & 1 \\
        1 & 1 & -1 & -1 \\
        1 & 1 & 1 & 1
    \end{pmatrix}\;.
\end{align}

In Ref.~\cite{Prezeau:2003xn}, the following set of operators was introduced
\begin{align}
O_{1+}^{++}& = \left(\bar{q}_L \tau^+ \gamma^\mu q_L\right)\left(\bar{q}_R \tau^+ \gamma_\mu q_R\right)\;, \\
O_{2 \pm}^{++}& = \left(\bar{q}_R \tau^+ q_L\right)\left(\bar{q}_R \tau^+ q_L\right) \pm\left(\bar{q}_L \tau^+ q_R\right) \left(\bar{q}_L \tau^+ q_R\right)\;, \\
O_{3 \pm}^{++}& = \left(\bar{q}_L \tau^+ \gamma^\mu q_L\right)\left(\bar{q}_L \tau^+ \gamma_\mu q_L\right)\pm\left(\bar{q}_R \tau^+ \gamma^\mu q_R\right)\left(\bar{q}_R \tau^+ \gamma_\mu q_R\right)\;, \\
O_{4 \pm}^{++, \mu}& = \left(\bar{q}_L \tau^+ \gamma^\mu q_L \mp \bar{q}_R \tau^+ \gamma^\mu q_R\right)\left(\bar{q}_L \tau^+ q_R-\bar{q}_R \tau^+ q_L\right)\;, \\
O_{5 \pm}^{++, \mu}& = \left(\bar{q}_L \tau^+ \gamma^\mu q_L \pm \bar{q}_R \tau^+ \gamma^\mu q_R\right)\left(\bar{q}_L \tau^+ q_R+\bar{q}_R \tau^+ q_L\right)\; .
\end{align}
They can be expressed in terms of the operators in the basis of Ref.~\cite{Cirigliano:2018yza}
\begin{align}
O_{1+}^{++}& = O_4^{(9)}\;, \\
O_{2 \pm}^{++}& = O_2^{(9)} \pm O_2^{(9)\prime}\;, \\
O_{3 \pm}^{++}& = O_1^{(9)} \pm O_1^{(9)\prime}\;, 
\end{align}
while the remaining operators are
\begin{align}
    \begin{pmatrix}
        O_{4 +}^{++, \mu} \\
        O_{4 -}^{++, \mu} \\
        O_{5 +}^{++, \mu} \\
        O_{5 -}^{++, \mu} \\
    \end{pmatrix}
    &=
    \eta_1
    \begin{pmatrix}
        O_6^{\mu (9)} \\
        O_6^{\mu (9)\prime} \\
        O_8^{\mu (9)} \\
        O_8^{\mu (9)\prime} \\
    \end{pmatrix}\;,\quad
    \eta_1 \equiv \begin{pmatrix}
        1 & 1 & -1 & -1 \\
        1 & -1 & -1 & 1 \\
        1 & 1 & 1 & 1 \\
        1 & -1 & 1 & -1
    \end{pmatrix}\;.
\end{align}
The product of the correlation matrices $\eta_1$ and $\rho_3$ gives
\begin{align}
\label{eq:O45-03}
    \begin{pmatrix}
        O_{4 +}^{++, \mu} \\
        O_{4 -}^{++, \mu} \\
        O_{5 +}^{++, \mu} \\
        O_{5 -}^{++, \mu} \\
    \end{pmatrix}
    &=
    \begin{pmatrix}
        0 & -1 & 0 & 0 \\
        0 & 0 & 1 & 0 \\
        1 & 0 & 0 & 0 \\
        0 & 0 & 0 & -1 \\
    \end{pmatrix}
    \begin{pmatrix}
        O_{VS} \\
        O_{AP} \\
        O_{VP} \\
        O_{AS} \\
    \end{pmatrix}\;,
\end{align}
which can also be obtained through direct computation.

In general, we need to distinguish the spurions for the two quark bilinears. However, for $0\nu\beta\beta$ decay, we can formally decompose
\begin{align}
    O_{VV}&=[\bar q\gamma^\mu(\Sigma_{R}P_R+\Sigma_{L}P_L) q][\bar q\gamma_\mu(\Sigma_{R}P_R+\Sigma_{L}P_L) q]\;,\\
    O_{SS}&=[\bar q (\Sigma^\dagger P_R+\Sigma P_L) q][\bar q (\Sigma^\dagger P_R+\Sigma P_L) q]\;,
\end{align}
where all of these spurions are equal to $\tau^+$, 
\begin{align}
    \Sigma_L=\Sigma_R=\Sigma=\Sigma^\dagger = \tau^+\;.
\end{align}
This is similar to the methodology in Ref.~\cite{Graesser:2016bpz}, where the four-quark operator is generally written as
\begin{align}
    T_{c d}^{a b}\left(\bar{q}^c \Gamma q_a\right)\left(\bar{q}^d \Gamma^{\prime} q_b\right)\;, \quad T_{c d}^{a b}\equiv \left(\tau^{+}\right)_c{ }^a\left(\tau^{+}\right)_d{ }^b\;,
\end{align}
with $a,b$ being the flavor indices. 
The matching of the four-quark operators is similar to that of the quark bilinear, and we do not need to introduce new building blocks.

\subsubsection{LO matching}

\begin{table}[H]
\centering
\caption{LO matching to chiral Lagrangian for the four-quark operators in Eq.~\eqref{eq:four-quark-spurion}.}
\begin{tabular}{ccl}
\hline
CP property & four-quark operator & chiral operator \\
\hline
$C+P+$ & $O_{VV}$ & $\langle Q_{-} Q_{-} \rangle$ \\
$C+P+$ & $O_{AA}$ & $\langle Q_{+} Q_{+} \rangle$ \\
$C-P-$ & $O_{VA}$ & $\langle Q_{-} Q_{+} \rangle$ \\
\hline
$C+P+$&$O_{SS}$&$\langle \Sigma_{+} \Sigma_{+} \rangle$\\
$C+P+$&$O_{PP}$&$\langle \Sigma_{-} \Sigma_{-} \rangle$\\
$C+P-$&$O_{SP}$&$\langle \Sigma_{-} \Sigma_{+} \rangle$\\
\hline
$C-P+$ & $O_{VS}$ & $\langle Q_{-}\nabla^\mu\Sigma_{-}\rangle$ \\
$C+P+$ & $O_{AP}$ & $\langle Q_{+}\nabla^\mu\Sigma_{+}\rangle$ \\
$C-P-$ & $O_{VP}$ & $\langle Q_{-}\nabla^\mu\Sigma_{+}\rangle$ \\
$C+P-$ & $O_{AS}$ & $\langle Q_{+}\nabla^\mu\Sigma_{-}\rangle$ \\
\hline
\end{tabular}
\label{tab:operator_matching}
\end{table}

The LO matching between four-quark operators in the LEFT and chiral operators is summarized in Table~\ref{tab:operator_matching}. The matching for the LEFT operators in the chiral basis is given by
\begin{align}
    \begin{pmatrix}
        O_1^{(9)} \\
        O_1^{(9)\prime} \\
        O_4^{(9)} \\
    \end{pmatrix}
    &\to 
    \begin{pmatrix}
        1 & 1 & -2\\
        1 & 1 & 2\\
        1 & -1 & 0\\
    \end{pmatrix}
    \begin{pmatrix}
        \langle Q_{-} Q_{-} \rangle \\
        \langle Q_{+} Q_{+} \rangle \\
        \langle Q_{-} Q_{+} \rangle \\
    \end{pmatrix}\;, \\[5pt]
    \begin{pmatrix}
        O_2^{(9)} \\
        O_2^{(9)\prime} \\
    \end{pmatrix}
    &\to 
    \begin{pmatrix}
        1 & 1 & -2\\
        1 & 1 & 2\\
    \end{pmatrix}
    \begin{pmatrix}
        \langle\Sigma_{+}\Sigma_{+}\rangle\\
        \langle\Sigma_{-}\Sigma_{-}\rangle\\
        \langle\Sigma_{-}\Sigma_{+}\rangle
    \end{pmatrix}\;.
\end{align}
It is noted that each four-quark operator in the chiral basis corresponds to an independent LEC after matching. 

The chiral operators can be explicitly expanded in terms of the $U$ field as follows
\begin{align}
    \langle Q_{+}Q_{+}\rangle&=
    2\langle U^\dagger \tau^+U\tau^+\rangle\;,\notag\\
    \langle Q_{-}Q_{-}\rangle&=
    -2\langle U^\dagger\tau^+U\tau^+\rangle\;,\notag\\
    \langle Q_{-}Q_{+}\rangle&=\langle U^\dagger\tau^+U\tau^+\rangle-\langle U\tau^+U^\dagger\tau^+\rangle\;,\notag\\
    \langle \Sigma_{+}\Sigma_{+}\rangle &= \langle U\tau^+U\tau^+\rangle + \langle U^\dagger\tau^+U^\dagger\tau^+\rangle \;,\notag\\
    \langle \Sigma_{-}\Sigma_{-}\rangle &= \langle U\tau^+U\tau^+\rangle + \langle U^\dagger\tau^+U^\dagger\tau^+\rangle \;,\notag\\
    \langle \Sigma_{-}\Sigma_{+}\rangle &= \langle U\tau^+U\tau^+\rangle - \langle U^\dagger\tau^+U^\dagger\tau^+\rangle \;.
\end{align}
In this case, we obtain LO matching yielding
\begin{align}
    \langle Q_{-} Q_{+} \rangle =0\;,
\end{align}
while 
\begin{align}
O_{1}^{(9)} &\to -2 \langle U^\dagger \tau^+ U \tau^+ \rangle + 2 \langle U^\dagger \tau^+ U \tau^+ \rangle =  0\;, \\
O_{4}^{(9)} &\to -2 \langle U^\dagger \tau^+ U \tau^+ \rangle - 2 \langle U^\dagger \tau^+ U \tau^+ \rangle = -4 \langle U^\dagger \tau^+ U \tau^+\rangle \;,\\
O_{2}^{(9)} &\to 2 \langle U^\dagger \tau^+ U^\dagger \tau^+ \rangle + 2 \langle U^\dagger \tau^+ U^\dagger \tau^+ \rangle = 4 \langle U\tau^+U\tau^+\rangle\;.
\end{align}
The color-mixed operators follow analogous matching relations:
\begin{align}
O_{3}^{(9)} &\to 4 \langle U\tau^+U\tau^+\rangle\;,\\
O_{5}^{(9)} &\to -4 \langle U^\dagger \tau^+ U \tau^+\rangle \;.
\end{align}
The matching for parity-conjugate operators is obtained through the substitution $U \leftrightarrow U^\dagger$, which is expressed as
\begin{align}
O_{1}^{(9)\prime} &\to 0\;, \\
O_{2}^{(9)\prime} &\to 4 \langle U^\dagger\tau^+U^\dagger\tau^+\rangle\;,\\
O_{3}^{(9)\prime} &\to 4 \langle U^\dagger\tau^+U^\dagger\tau^+\rangle\;.
\end{align}
These results are in complete agreement with those presented in Refs.~\cite{Cirigliano:2018yza,Prezeau:2003xn}.

From Tab.~\ref{tab:operator_matching} and Eq.~\eqref{eq:O45-03}, we obtain the matching for parity-even operators
\begin{align}
   O_{4 +}^{++, \mu} = - O_{AP} &\to -\langle Q_{+}\nabla^\mu\Sigma_{+}\rangle \sim  \pi^- \partial^\mu \pi^- \;,\\
   O_{5 +}^{++, \mu} = O_{VS} &\to \langle Q_{-}\nabla^\mu\Sigma_{-}\rangle \sim \pi^- \partial^\mu \pi^-\;.
\end{align}
Both contributions to the $0\nu\beta\beta$ decay amplitude are suppressed by the electron mass using
the IBPs,
rendering them negligible. This is consistent with the finding in Ref.~\cite{Prezeau:2003xn}.

\subsubsection{NLO matching}

\begin{table}[H]
\centering
\caption{NLO matching to chiral Lagrangian for the four-quark operators $\mathcal{O}_{VV}$, $\mathcal{O}_{AA}$ and $\mathcal{O}_{VA}$ in Eq.~\eqref{eq:four-quark-spurion}.}
\begin{tabular}{ccl}
\hline
CP property & four-quark operator & chiral operator \\
\hline
$C+P+$ & $O_{VV}$ & $\langle Q_{-}\hat{u}^\mu Q_{-}\hat{u}_\mu \rangle$ \\
$C+P+$ & $O_{AA}$ & $\langle Q_{+}\hat{u}^\mu Q_{+}\hat{u}_\mu \rangle$ \\
$C-P-$ & $O_{VA}$ & $\langle Q_{-}\hat{u}^\mu Q_{+}\hat{u}_\mu \rangle$ \\
\hline
\end{tabular}
\label{tab:operator_matching2}
\end{table}

The chiral operators decompose as follows:
\begin{align}    
    \langle Q_{-}\hat{u}^\mu Q_{-}\hat{u}_\mu \rangle &= A_1 - A_2 - A_3 + A_4 \;,  \\
    \langle Q_{+}\hat{u}^\mu Q_{+}\hat{u}_\mu \rangle &= A_1 + A_2 + A_3 + A_4 \;,  \\
    \langle Q_{-}\hat{u}^\mu Q_{+}\hat{u}_\mu \rangle &= A_1 + A_2 - A_3 - A_4 \;,  
\end{align}
where $A_i$ are given by
\begin{align}
    A_1 &= \langle u^\dagger \tau^+ u \hat u^\mu u^\dagger \tau^+ u \hat u_\mu \rangle\;,&
    A_2 &= \langle u^\dagger \tau^+ u \hat u^\mu u \tau^+ u^\dagger \hat u_\mu \rangle\;,\\
    A_3 &= \langle u \tau^+ u^\dagger \hat u^\mu u^\dagger \tau^+ u \hat u_\mu \rangle\;,&
    A_4 &= \langle u \tau^+ u^\dagger \hat u^\mu u \tau^+ u^\dagger \hat u_\mu \rangle\;.
\end{align}
Note that $A_2 = A_3$ due to the cyclic permutation invariance of the trace operation.

The NLO matching is 
\begin{align}
    O_{1}^{(9)} &\rightarrow \langle Q_{-}\hat{u}^\mu Q_{-}\hat{u}_\mu \rangle + \langle Q_{+}\hat{u}^\mu Q_{+}\hat{u}_\mu \rangle - 2 \langle Q_{-}\hat{u}^\mu Q_{+}\hat{u}_\mu \rangle = 4 A_4\;,\\
    O_{1}^{(9)\prime} &\rightarrow\langle Q_{-}\hat{u}^\mu Q_{-}\hat{u}_\mu \rangle + \langle Q_{+}\hat{u}^\mu Q_{+}\hat{u}_\mu \rangle + 2 \langle Q_{-}\hat{u}^\mu Q_{+}\hat{u}_\mu \rangle = 4 A_1\;.
\end{align}
Using the relations in Eq.~\eqref{eq:u-mu-relations}, we find that
\begin{align}
    A_1 &= -4 \langle \tau^+ u D^\mu u^{\dagger} \tau^+ u D_\mu u^{\dagger}\rangle = - \langle \tau^+ U \partial^\mu U^{\dagger} \tau^+ U \partial_\mu U^{\dagger}\rangle\;,\\
    A_4 &= -4 \langle \tau^+ u^{\dagger} D^\mu u \tau^+ u^{\dagger} D_\mu u\rangle =  - \langle \tau^+ U^{\dagger} \partial^\mu U \tau^+ U^{\dagger} \partial_\mu U\rangle\;,
\end{align}
where the identities in Eq.~\eqref{eq:identity2} are used to derive the second equalities. Note that the photon contributions vanish in $A_1$ and $A_4$, as verified by explicit calculation.

For the vanishing LO cases, the NLO matching gives the chiral Lagrangian:
\begin{align}
    O_{1}^{(9)} &\rightarrow -4 \langle U^\dagger \partial^\mu U \tau^+U^\dagger \partial_\mu U \tau^+\rangle\notag\;,\\
    O_{1}^{(9)\prime} &\rightarrow -4 \langle U \partial^\mu U^\dagger \tau^+U \partial_\mu U^\dagger \tau^+\rangle\;,
\end{align}
which are agreement with Refs.~\cite{Cirigliano:2018yza,Cirigliano:2017ymo}.
These terms can give rise to the interaction $\partial^\mu \pi \partial_\mu \pi ee$ that contributes to $\onbb$ decay.

\section{Discussion and conclusion}
\label{sec:conclusion}

In this work, we have extended the studies in Ref.~\cite{Song:2025snz}, and made a comprehensive comparison on various methods for the matching from the quark to hadronic operators, including the external source method, conventional spurion method, and systematic spurion method, based on the chiral symmetry $SU(2)_L \times SU(2)_R \to SU(2)_V$. While the external source method is convenient for matching dimension-6 operators in the low-energy effective field theory (LEFT),  it is quite limited or even inapplicable for higher-dimensional operators. In contrast, the conventional spurion method is straightforward and theoretically viable, but it presents challenges in eliminating redundancy in chiral Lagrangian. 

The systematic spurion method addresses these issues effectively. In this method, the LEFT operators are classified based on the $CP$ transformation properties of the quark bilinears, the chiral Lagrangian is constructed through the Young tensor technique, and the mapping from the quark to hadronic degrees of freedom satisfies three rules  as presented in Sec.~\ref{sec:systematic2}. Most importantly, we obtain the building blocks of the chiral Lagrangian with a minimal set of spurions, even for dimension-7 and dimension-8 derivative operators, and dimension-9 four-quark operators.

We have examined the matching of the dimension-6 LEFT operators that describe the scalar, pseudo-scalar, vector, axial-vector, and tensor interactions involving a single quark bilinear.
Special attention is given to the tensor interaction, as operators
in the form of $\sigma_{\mu\nu} \gamma^5$ are redundant. 
The matching results
agree very well.
The conversion relations and comparison of elements in the three methods for matching from quark to hadronic operators are given in Sec.~\ref{sec:conversion}. It is notable that matching in the conventional spurion method could serve as a valuable guide for constructing the chiral Lagrangian in the systematic spurion method.

For higher-dimensional operators, the external source
method is quite limited, if the derivative acting on the quark field is involved or more quark bilinears are included. For the dimension-7 LEFT operators with single quark bilinear, after using the equations of motion, the matching can also be performed in the external source method, similar to the dimension-6 operators. However, for dimension-8 and dimension-9 LEFT operators, this method becomes inapplicable.

As applications of the systematic spurion method, we have investigated the matching of dimension-7 and -8 derivative operators, as well as dimension-9 four-quark operators, which are closely related to new physics searches, such as neutrino-electron scatterings and neutrinoless double beta $(0\nu \beta\beta)$ decay. For the dimension-9 operators responsible for $0\nu \beta\beta$ decay, we have successfully established the matching between LEFT and the chiral Lagrangian, in agreement with existing literature. Furthermore, the use of spurion products avoids the irreducible decomposition into complicated  higher-dimensional representations of the chiral symmetry group within the systematic spurion method. This approach eliminates the need for complex decompositions of group representations, even as we consider an increasing number of quark bilinears.

Finally, we emphasize that although this work explicitly presents the matching for LEFT operators to the mesonic chiral Lagrangian in the $SU(2)$ case, the systematic spurion method for matching is quite general. It can also be applied to the nucleon sector, scenarios involving $SU(3)$ or lepton flavor violation. Besides, operators that involve quarks in other EFTs can also be matched to the chiral Lagrangian in this approach, such as the LEFTs extended with dark matter~\cite{Fitzpatrick:2012ix, Cirigliano:2012pq,Bishara:2016hek,Korber:2017ery,Bishara:2017pfq,deVries:2023sux} or sterile neutrino~\cite{Chala:2020vqp,Li:2020lba,Li:2021tsq}.

\begin{acknowledgments}
We thank the anonymous referees for their careful reading and insightful suggestions. We would like to thank Xiao-Dong Ma, Hao Sun, and De-Liang Yao for their helpful discussions.
This work is supported by the National Science Foundation of China under Grants No.~12347105, No.~12375099, No.~12505127 and No.~12047503, and the National Key Research and Development Program of China Grant No.~2020YFC2201501, No.~2021YFA0718304.
GL is also supported by the Guangdong Basic and Applied Basic Research Foundation (2024A1515012668), and SYSU startup funding. 
\end{acknowledgments}

\appendix

\section{Cayley-Hamilton relation}
\label{sec:Cayley-Hamilton}

The Cayley-Hamilton theorem is a general result applicable to any square matrix. Here, we follow the results in Ref.~\cite{Bijnens:1999sh}. For an arbitrary $ 2 \times 2 $ matrix $ \mathcal{A} $, the theorem states
\begin{equation}
\label{eq:ch-2}
  \mathcal{A}^2 = \langle \mathcal{A} \rangle \mathcal{A} + \frac{1}{2} \left( \langle \mathcal{A}^2 \rangle - \langle \mathcal{A} \rangle^2 \right) \mathbb{I}_{2 \times 2} \;,
\end{equation}
where $ \mathbb{I}_{2 \times 2} $ denotes the $ 2 \times 2 $ identity matrix. 

In the context of $ SU(2) $ invariants, where the relevant $ 2 \times 2 $ matrix fields are traceless (being in the adjoint representation of $ SU(2) $), Eq.~\eqref{eq:ch-2} simplifies to
\begin{equation}
  \mathcal{A}^2 = \frac{1}{2} \langle \mathcal{A}^2 \rangle \mathbb{I}_{2 \times 2} \;.
\end{equation}

By substituting $ \mathcal{A} \rightarrow \mathcal{A} + \mathcal{B} $ and extracting terms proportional to $ \mathcal{A}^{n_1} \mathcal{B}^{n_2} $ (with $ n_1 + n_2 = 2 $), we derive various relations. For instance, the Cayley-Hamilton relation for the $ \mathcal{A} \mathcal{B} $ term is
\begin{align}
    \mathcal{A} \mathcal{B} + \mathcal{B} \mathcal{A} = \frac{1}{2} \left[ \langle \mathcal{A} \mathcal{B} \rangle + \langle \mathcal{B} \mathcal{A} \rangle \right] = \langle \mathcal{A} \mathcal{B} \rangle \;.
\end{align}

Multiplying by an arbitrary $ 2 \times 2 $ matrix $ \mathcal{C} $ and taking the trace yields the fundamental Cayley-Hamilton relation
\begin{align}
\langle \mathcal{C} \mathcal{A} \mathcal{B} \rangle + \langle \mathcal{C} \mathcal{B} \mathcal{A} \rangle = \langle \mathcal{C} \rangle \langle \mathcal{A} \mathcal{B} \rangle \;,
\end{align}
where $ \mathcal{C} $ is a general $ 2 \times 2 $ matrix, while $ \mathcal{A} $ and $ \mathcal{B} $ are traceless $ 2 \times 2 $ matrices.

As an example, in the $SU(2)$ case, the chiral operators at $p^4$ order for the LEFT operator $\mathcal{O}_1^{(6)}$ satisfy the relation
\begin{align}
(\bar e e)\langle\Sigma_+\rangle\langle \hat u^\mu\hat u_\mu\rangle = 2(\bar e e)\langle\Sigma_+ \hat u^\mu\hat u_\mu\rangle \;,
\end{align}
such that $(\bar{e}e)\langle\Sigma_\pm \hat{u}^\mu\hat{u}_\mu\rangle$ is redundant and not included in Tab.~\ref{tab:sp}.
In contrast, for the $SU(3)$ case, single-trace operators remain independent and need to be included in the chiral Lagrangian.

\section{Some useful identities}
\label{app:identities}

The following identities facilitate comparison between results derived from the external source method and conventional spurion techniques:
\begin{align}
\label{eq:identity1}
    \left(D_\mu u\right) u^{\dagger}&=-u\left(D_\mu u\right)^{\dagger}\;,&
    u^{\dagger} \left(D_\mu u\right) &=-\left(D_\mu u\right)^{\dagger}u\;,\nn\\
    \left(u D_\mu u^\dagger \right)^{\dagger} & = - u D_\mu u^\dagger\;,&
    \left(u^\dagger D_\mu u \right)^{\dagger} &= - u^\dagger D_\mu u\;,
\end{align}
and
\begin{align}
\label{eq:identity2}
    u D_\mu u^\dagger &= U \partial_\mu U^\dagger/2 \;,&
    u^\dagger D_\mu u &= U^\dagger \partial_\mu U/2\;,\nn\\
    u^\dagger D_\mu u^\dagger &= \partial_\mu U^\dagger/2\;,&
    u D_\mu u &= \partial_\mu U/2\;.
\end{align}
Note that the relations in the first row in Eqs.~\eqref{eq:identity1}~\eqref{eq:identity2} were also given in Ref.~\cite{Liao:2019gex}.

The following identities are useful for  comparing results between the systematic and conventional spurion methods:
\begin{align}
\label{eq:u-mu-relations}
    u D^\mu u^\dagger &= \dfrac{i}{2} u \hat u^\mu u^\dagger\;,&
    u^\dagger D^\mu u &= - \dfrac{i}{2} u^\dagger \hat u^\mu u\;,\\
    u^\dagger D^\mu u^\dagger &= \dfrac{i}{2} u^\dagger \hat u^\mu u^\dagger\;,&
    u D^\mu u &= - \dfrac{i}{2} u \hat u^\mu u\;,
\end{align}
where $\hat u^\mu$ is defined in Eq.~\eqref{eq:uhat}.

\section{Discussions and examples of tensor interactions}
\label{app:tensor}

In this appendix, we will provide expanded discussions on matching for quark tensor interactions. 
Given the tensor interactions in Eq.~\eqref{eq:lambt}, we have 
\begin{align}
    \mathcal{L}_{T}^q &= \bar{q} \sigma_{\mu\nu} \lambda_T^{\mu\nu} q -\dfrac{1}{2}\bar{q} \sigma_{\mu\nu} \varepsilon^{\mu\nu}{}_{\alpha\beta} \lambda_\varepsilon^{\alpha\beta} q\;,
\end{align}
by using the relation in Eq.~\eqref{eq:epsilon}, which implies that the operator of the form $\sigma_{\mu\nu} \gamma^5$ is redundant.  The tensor interaction at the quark level in the external source method is
\begin{align}
\label{eq:tensor-source}
    \mathcal{L}_{\text{ext},\, T} = \bar{q} \sigma_{\mu \nu} \bar{t}^{\mu \nu} q\;,
\end{align}
where $\bar{t}^{\mu \nu}$ can be obtained from $\lambda_T^{\mu\nu}$ and $\lambda_\varepsilon^{\mu\nu}$ as in Eq.~\eqref{eq:tbar-lambda}.

Introducing the chiral projectors
\begin{align}
    P_R^{\mu \nu \lambda \rho} = \frac{1}{4}\left(g^{\mu \lambda} g^{\nu \rho} - g^{\nu \lambda} g^{\mu \rho} + i \varepsilon^{\mu \nu \lambda \rho}\right) = \left(P_L^{\mu \nu \lambda \rho}\right)^{\dagger}\;,
\end{align}
one can express the tensor interaction as~\cite{Cata:2007ns}
\begin{align}
\label{eq:tensor_Lag}
    \mathcal{L}_{\text{ext},\, T} =  \bar{q}_L \sigma_{\mu \nu} t^{\mu \nu \dagger} q_R + \bar{q}_R \sigma_{\mu \nu} t^{\mu \nu} q_L\;,
\end{align}
with the relations
\begin{align}
    \label{eq:tensor1}
   \bar{t}^{\mu\nu} &= P_L^{\mu\nu \lambda \rho} t_{\lambda \rho} + P_R^{\mu\nu \lambda \rho} t_{\lambda \rho}^\dagger\;,\\
   \label{eq:tensor2}
   t^{\mu \nu} &= P_L^{\mu \nu \lambda \rho} \bar{t}_{\lambda \rho}\;.
\end{align}

The projectors $P_{L/R}^{\mu \nu \lambda \rho}$ are anti-symmetric under the interchange of $\mu $ and $ \nu$. Thus the tensor sources $\bar{t}^{\mu\nu}$ and $t^{\mu \nu}$ in Eqs.~\eqref{eq:tensor1}~\eqref{eq:tensor2}  are anti-symmetric. However, when considering BSM interactions, they may not be anti-symmetric. In such cases, one could first anti-symmetrize the interactions as in Ref.~\cite{Li:2021phq}.

From Eqs.~\eqref{eq:tensor}~\eqref{eq:tensor-ubasis}, we obtain that the interaction with mesons in the chiral Lagrangian is
\begin{align}
\label{eq:tensor-p4-pi}
    \Delta \mathcal{L}_{4,\pi}^\prime = - \dfrac{i}{2} \Lambda_2 \langle ( u^{\dagger}t^{\mu\nu}u^{\dagger} + ut^{\mu\nu\dagger}u) [\hat u_\mu, \hat u_\nu] \rangle\;.
\end{align}

We find that it is possible to introduce the spurions $X_{\mu\nu}$ and $X_{\mu\nu}^\dagger$ in a different way compared to that in Sec.~\ref{sec:tensor-current}. Starting from the quark-level tensor interactions, we rewrite
\begin{align}
    \lambda_{\mu\nu}(\bar q_R\sigma^{\mu\nu}q_L)
    =\lambda_{\mu\nu}(\bar q_R\sigma^{\mu\nu}P_Lq_L)\;,\quad
    \lambda_{\mu\nu}^\dagger(\bar q_L\sigma^{\mu\nu}q_R) 
    =\lambda_{\mu\nu}^\dagger(\bar q_L\sigma^{\mu\nu}P_Rq_R)\;.
\end{align}
Using the relation in Eq.~\eqref{eq:epsilon}, we obtain
\begin{align}
    \lambda_{\mu\nu}(\bar q_R\sigma^{\mu\nu}q_L) = X_{\mu\nu}\left(\bar q_R\sigma^{\mu\nu}q_L \right)\;,\quad
    \lambda_{\mu\nu}^\dagger(\bar q_L\sigma^{\mu\nu}q_R)= X_{\mu\nu}^\dagger\left(\bar q_L\sigma^{\mu\nu}q_R \right)\;,
\end{align}
where $X_{\mu\nu}$ and $X_{\mu\nu}^\dagger$ are expressed in terms of $\lambda_{\mu\nu}$ and $\lambda_{\mu\nu}^\dagger$ as given in Eq.~\eqref{eq:Xmunu}. 
These two interactions are related by parity, so that the
at the corresponding LECs $g_T^{(2)}$ and $g_T^{(2)} \prime$ are equal, leaving only one independent LEC.

To compare the matching for tensor interactions using different methods, we consider the dimension-6 LEFT operator $\mathcal{O}_{11}^{(6)}$, which describes the neutral-current interaction between electron and $u$ quark, 
\begin{align}
    \mathcal{L}_{\rm eff, 11}^{(6)} = \mathcal{C}_{11}^{(6)}
    \left(\bar{e} \sigma^{\mu \nu} e\right) \left(\bar{u} \sigma_{\mu \nu} u\right) = \mathcal{C}_{11}^{(6)}
    \left(\bar{e} \sigma^{\mu \nu} e\right) \left(\bar{q} \sigma_{\mu \nu} \tau^p q\right)\;.
\end{align}
The tensor source is defined as follows
\begin{align}
    \bar{t}_{\mu\nu} &= \mathcal{C}_{11}^{(6)} \left(\bar{e} \sigma^{\mu \nu} e\right) \tau^p \;.
\end{align}
Using Eq.~\eqref{eq:tensor2}, we obtain
\begin{align}
\label{eq:tensor-missing}
   t^{\mu\nu} &=  P_L^{\mu \nu \lambda \rho} \bar{t}_{\lambda \rho} = \frac{1}{2} \mathcal{C}_{11}^{(6)} \bar{e} \left(\sigma^{\mu \nu} - \frac{i}{2} \varepsilon^{\mu\nu\alpha\beta} \sigma_{\alpha\beta}\right) e\,  \tau^p \;,\nn\\
   t^{\mu\nu\dagger} &= P_R^{\mu \nu \lambda \rho} \bar{t}_{\lambda \rho} = \frac{1}{2} \mathcal{C}_{11}^{(6)} \bar{e} \left(\sigma^{\mu \nu} + \frac{i}{2} \varepsilon^{\mu\nu\alpha\beta} \sigma_{\alpha\beta}\right) e\, \tau^p\;.
\end{align}

In the chiral basis, the interactions are expressed as
\begin{align}
    \mathcal{L}_{\rm eff, 11}^{(6)} = \mathcal{C}_{11}^{(6)}
    \left(\bar{e} \sigma^{\mu \nu} e\right) \left(\bar{q}_L \sigma_{\mu \nu} \tau^p q_R\right) +  \mathcal{C}_{11}^{(6)}
    \left(\bar{e} \sigma^{\mu \nu} e\right) \left(\bar{q}_R \sigma_{\mu \nu} \tau^p q_L\right)\;,
\end{align}
then spurions in Eq.~\eqref{eq:tensormatching} are given by
\begin{align}
    \lambda^{\mu\nu\dagger} = \lambda^{\mu\nu} =  \mathcal{C}_{11}^{(6)} \bar{e} \sigma^{\mu \nu} e \, \tau^p\;.
\end{align}

From Eq.~\eqref{eq:Xmunu}, we find
\begin{align}
    X^{\mu\nu\dagger}& = \frac{1}{2}\lambda_{\mu\nu}^\dagger+\frac{i}{4}\varepsilon_{\mu\nu\alpha\beta}\lambda^{\alpha\beta\dagger}=\dfrac{1}{2} \mathcal{C}_{11}^{(6)} \bar{e} \left(\sigma^{\mu \nu} + \frac{i}{2} \varepsilon^{\mu\nu\alpha\beta} \sigma_{\alpha\beta}\right)e \, \tau^p =t^{\mu\nu \dagger}\;,\nn\\
    X^{\mu\nu}& = \frac{1}{2}\lambda_{\mu\nu}-\frac{i}{4}\varepsilon_{\mu\nu\alpha\beta}\lambda^{\alpha\beta}=\dfrac{1}{2} \mathcal{C}_{11}^{(6)} \bar{e} \left(\sigma^{\mu \nu} - \frac{i}{2} \varepsilon^{\mu\nu\alpha\beta} \sigma_{\alpha\beta}\right)e \, \tau^p =t^{\mu\nu }\;.
\end{align}
The matched chiral Lagrangian is provided in Eq.~\eqref{eq:tensor-X}. 
Therefore, 
the results obtained using the external source method and the systematic spurion method are consistent with the following correspondence
\begin{align}
    g_T^{(2)} = 2i \Lambda_2\;,
\end{align}
where the factor $i$ arises because $[\hat u_\mu, \hat u_\nu]$ is anti-Hermitian.

Moreover, the combination of $\lambda^{\mu\nu}$ and $\lambda^{\mu\nu\dagger}$ yields 
\begin{align}
    \lambda_T^{\mu\nu} = \mathcal{C}_{11}^{(6)} \bar{e} \sigma^{\mu \nu} e \, \tau^p = \bar t^{\mu\nu}\;,\quad
    \lambda_\varepsilon^{\mu\nu} = 0\;.
\end{align}
Thus, the matching in the conventional spurion method can also be achieved in the $q$ basis, as indicated in Eq.~\eqref{eq:mathcing-t_non-chiral}. Since $\lambda^\dagger =\lambda = \lambda_T$, we can express 
\begin{align}
    \mathcal{L}_{\rm eff, 11}^{(6)}  \to & -\frac{1}{8}\mathrm{Tr}\left[(u\lambda^{\mu\nu\dagger}u+u^{\dagger}\lambda^{\mu\nu}u^{\dagger})[\hat{u}_{\mu},\hat{u}_{\nu}]\right]g_{T}^{(2)} \nn\\
 & -\frac{i}{16}\varepsilon_{\mu\nu\alpha\beta}\operatorname{Tr}\left[(u\lambda^{\alpha\beta\dagger}u-u^\dagger\lambda^{\alpha\beta}u^\dagger)[\hat{u}^\mu,\hat{u}^\nu]\right]g_T^{(2)}\;,
\end{align}
which agrees with the results from the external source method.

In the systematic spurion method, we have
\begin{align}
    \mathcal{L}_{\rm eff, 11}^{(6)} = \mathcal{C}_{11}^{(6)}(\bar{e}\sigma^{\mu\nu}e)[\bar{q}\sigma^{\mu\nu}(\Sigma^{\dagger}P_{R}+\Sigma P_{L})q]\;,
\end{align}
where $\Sigma^\dagger = \Sigma = \tau^p$. The matching at $p^4$ order is given by
\begin{align}
\label{eq:factor-i}
    \mathcal{L}_{\rm eff, 11}^{(6)}  \to &\, \mathcal{C}_{11}^{(6)} \left[ (\bar{e}\sigma^{\mu\nu}e)\langle\Sigma_+[\hat{u}_\mu,\hat{u}_\nu]\rangle  i \beta_1 + (\bar{e}\gamma^{5}\sigma^{\mu\nu}e)\langle\Sigma_{-}[\hat{u}_{\mu},\hat{u}_{\nu}]\rangle i \beta_2 \right]\;,
\end{align}
where we have used the relation in Eq.~\eqref{eq:epsilon}. Here, $\beta_1$ and $\beta_2$ are the LECs, and the building blocks $\Sigma_\pm$ are defined in Eq.~\eqref{eq:building}.
Note that we have included the factor $i$ has been included to maintain the Hermiticity of the chiral Lagrangian.
Similar to the discussion in Sec.~\ref{sec:systematic2}, operators characterized by the quark-level chiral structure $(\Sigma^\dagger P_R \pm \Sigma P_L)$ share a common LEC, which implies $\beta_1 = \beta_2 \equiv \beta$.

Utilizing the relation in Eq.~\eqref{eq:epsilon}, we find that the results from the conventional spurion method are consistent, and the LECs satisfy
\begin{align}
    i\beta = -\dfrac{1}{8} g_T^{(2) }\;.
\end{align}

Similarly, for the dimension-6 operator $\mathcal{O}_{12}^{(6)}$, we have
\begin{align}
    \mathcal{L}_{\rm eff, 12}^{(6)} = &\, \mathcal{C}_{12}^{(6)}(\bar{e}i \gamma^5 \sigma^{\mu\nu}e)[\bar{q}\sigma^{\mu\nu}(\Sigma^{\dagger}P_{R}+\Sigma P_{L})q]\;,\nn\\
    \to & \, \mathcal{C}_{12}^{(6)} \left[ (\bar{e}i \gamma^5\sigma^{\mu\nu}e)\langle\Sigma_+[\hat{u}_\mu,\hat{u}_\nu]\rangle  \beta + (\bar{e} i\sigma^{\mu\nu}e)\langle\Sigma_{-}[\hat{u}_{\mu},\hat{u}_{\nu}]\rangle \beta \right]\;.
\end{align}

If the lepton interactions are defined in the chiral basis, 
\begin{align}
    \mathcal{L}_{\rm eff,\, LR}^{(6)} = \left[ \mathcal{C}_{LR}(\bar{e}_L\sigma^{\mu\nu}e_R) +\mathcal{C}_{LR}^\dagger (\bar{e}_R \sigma^{\mu\nu}e_L)\right][\bar{q}\sigma^{\mu\nu}(\Sigma^{\dagger}P_{R}+\Sigma P_{L})q]\;,
\end{align}
where the Wilson coefficient $\mathcal{C}_{LR}$ is generally complex. The matching for the terms $\bar{e}_L\sigma^{\mu\nu}e_R$ and $\bar{e}_R \sigma^{\mu\nu}e_L$ results in
\begin{align}
    &(\bar{e}_L\sigma^{\mu\nu}e_R) [\bar{q}\sigma^{\mu\nu}(\Sigma^{\dagger}P_{R}+\Sigma P_{L})q] \nn\\
    &\quad\to (\bar{e}_L\sigma^{\mu\nu}e_R)\langle\Sigma_+[\hat{u}_\mu,\hat{u}_\nu]\rangle i \beta + (\bar{e}_L\sigma^{\mu\nu}e_R)\langle\Sigma_-[\hat{u}_\mu,\hat{u}_\nu]\rangle i \beta\;,\nn\\
    &(\bar{e}_R\sigma^{\mu\nu}e_L) [\bar{q}\sigma^{\mu\nu}(\Sigma^{\dagger}P_{R}+\Sigma P_{L})q] \nn\\
    &\quad\to (\bar{e}_R\sigma^{\mu\nu}e_L)\langle\Sigma_+[\hat{u}_\mu,\hat{u}_\nu]\rangle i \beta - (\bar{e}_R\sigma^{\mu\nu}e_L)\langle\Sigma_-[\hat{u}_\mu,\hat{u}_\nu]\rangle i \beta\;.
\end{align}
Thus, we obtain 
\begin{align}
    \mathcal{L}_{\rm eff,\, LR}^{(6)}  \to & \,\left[ \mathcal{C}_{LR}(\bar{e}_L\sigma^{\mu\nu}e_R) +\mathcal{C}_{LR}^\dagger (\bar{e}_R \sigma^{\mu\nu}e_L)\right] \langle\Sigma_+[\hat{u}_\mu,\hat{u}_\nu]\rangle i \beta\nn\\
    &+ \left[ \mathcal{C}_{LR}(\bar{e}_L\sigma^{\mu\nu}e_R) -\mathcal{C}_{LR}^\dagger (\bar{e}_R \sigma^{\mu\nu}e_L)\right] \langle\Sigma_-[\hat{u}_\mu,\hat{u}_\nu]\rangle i \beta\;.
\end{align}
From the above, we see that the LEC $\beta$ is real.

Assuming both operators $\mathcal{O}_{11}^{(6)}$ and $\mathcal{O}_{12}^{(6)}$ are present, 
\begin{align}
    \mathcal{C}_{LR} = \mathcal{C}_{11}^{(6)}+i\,\mathcal{C}_{12}^{(6)}\;,
\end{align}
where $\mathcal{C}_{11}^{(6)}$ and $\mathcal{C}_{12}^{(6)}$ are real. 
The matching result becomes
\begin{align}
    \mathcal{L}_{\rm eff,\, LR}^{(6)}  \to & \, \mathcal{C}_{11}^{(6)} \left[ (\bar{e}\sigma^{\mu\nu}e)\langle\Sigma_+[\hat{u}_\mu,\hat{u}_\nu]\rangle  i \beta + (\bar{e}\gamma^{5}\sigma^{\mu\nu}e)\langle\Sigma_{-}[\hat{u}_{\mu},\hat{u}_{\nu}]\rangle i \beta \right]\nn\\
   &+ \mathcal{C}_{12}^{(6)} \left[ (\bar{e}i \gamma^5\sigma^{\mu\nu}e)\langle\Sigma_+[\hat{u}_\mu,\hat{u}_\nu]\rangle  i \beta + (\bar{e} i\sigma^{\mu\nu}e)\langle\Sigma_{-}[\hat{u}_{\mu},\hat{u}_{\nu}]\rangle i \beta \right]\;.
\end{align}
Therefore, we can obtain the same results if the operators are defined with chiral lepton fields.

If the operators are defined with chiral quark fields~\cite{Jenkins:2017jig}
\begin{align}
    \mathcal{L}_{{\rm eff}, eu}^{(6)} = \mathcal{O}_{eu}^{T,RR} (\bar{e}_{L}\sigma^{\mu\nu}e_{R})(\bar{u}_{L}\sigma_{\mu\nu}u_{R}) + {\rm h.c.}\;,
\end{align}
we can use the relation in Eq.~\eqref{eq:epsilon} to show 
\begin{align}
    \mathcal{L}_{{\rm eff},\, eu}^{(6)} = \mathcal{C}_{eu}^{T,RR} (\bar{e}_{L}\sigma^{\mu\nu}e_{R})(\bar{u}\sigma_{\mu\nu}u) + {\rm h.c.}\;.
\end{align}
This result can also be derived using the Fierz identity. For instance, as shown in Ref.~\cite{Liao:2020jmn}, the following relation holds for Dirac fields $\psi_{1,2,3,4}$:
\begin{align}
    (\bar{\psi}_1\sigma^{\mu\nu}P_R\psi_2)(\bar{\psi}_3\sigma_{\mu\nu}P_L \psi_4)=0\;.
\end{align}
Then, one can perform the matching for $\mathcal{L}_{{\rm eff},\, eu}^{(6)}$ as that for $\mathcal{L}_{\rm eff,\, LR}^{(6)}$.

The above discussion can also be applied to the operator $\mathcal{O}_{17}^{(6)}$, which describes the charged-current neutrino non-standard interaction,
\begin{align}
    \mathcal{L}_{\rm eff, 17}^{(6)} 
    &\equiv \mathcal{C}_{17}^{(6)} \left(\bar{\nu}_L \sigma^{\mu \nu} e_R\right) \left(\bar{d} \sigma_{\mu \nu} u\right) + {\rm h.c.}\;.
\end{align}
In the basis of Ref.~\cite{Jenkins:2017jig}, the interaction is written as
\begin{align}
\label{eq:nuedu}
    \mathcal{L}_{{\rm eff},\, \nu edu}^{(6)} = 
    \mathcal{C}_{\nu edu}^{T,RR}  (\bar{\nu}_{L}\sigma^{\mu\nu}e_{R})(\bar{d}_{L}\sigma_{\mu\nu}u_{R}) + {\rm h.c.}\;.
\end{align}
Using the relation in Eq.~\eqref{eq:epsilon}  or the Fierz identity, we obtain
\begin{align}
     \mathcal{L}_{{\rm eff},\, \nu edu}^{(6)} = \mathcal{C}_{\nu edu}^{T,RR} 
    \left(\bar{\nu}_L \sigma^{\mu \nu} e_R\right) \left(\bar{d} \sigma_{\mu \nu} u\right) + {\rm h.c.}\;,
\end{align}
which indicates that $\mathcal{C}_{\nu edu}^{T,RR} =\mathcal{C}_{17}^{(6)}$.
In comparison with Eq.~\eqref{eq:tensor-source}, we get
\begin{align}
    \bar t^{\mu\nu} =\mathcal{C}_{17}^{(6)} (\bar{\nu}_{L}\sigma^{\mu\nu}e_{R}) \tau^- + {\rm h.c.}\;,
\end{align}
which can also be obtained with Eqs.~\eqref{eq:tbar-lambda}~\eqref{eq:nuedu}.

Using the chiral projectors, we obtain the tensor source
\begin{align}
   t^{\mu\nu} =  P_L^{\mu \nu \lambda \rho} \bar{t}_{\lambda \rho} &= \frac{1}{2} \mathcal{C}_{17}^{(6)} \bar{\nu}_L \left(\sigma^{\mu \nu} - \frac{i}{2} \varepsilon^{\mu\nu\alpha\beta} \sigma_{\alpha\beta}\right) e_R\,  \tau^- \nn\\
   &\quad +  \frac{1}{2} \mathcal{C}_{17}^{(6)\dagger} \bar{e}_R \left(\sigma^{\mu \nu} - \frac{i}{2} \varepsilon^{\mu\nu\alpha\beta} \sigma_{\alpha\beta}\right) \nu_L\,  \tau^+ \;,\nn\\
   t^{\mu\nu\dagger} = P_R^{\mu \nu \lambda \rho} \bar{t}_{\lambda \rho} &= \frac{1}{2} \mathcal{C}_{17}^{(6)} \bar{\nu}_L \left(\sigma^{\mu \nu} + \frac{i}{2} \varepsilon^{\mu\nu\alpha\beta} \sigma_{\alpha\beta}\right) e_R\, \tau^- \nn\\
   &\quad +  \frac{1}{2} \mathcal{C}_{17}^{(6)\dagger} \bar{e}_R \left(\sigma^{\mu \nu} + \frac{i}{2} \varepsilon^{\mu\nu\alpha\beta} \sigma_{\alpha\beta}\right) \nu_L\,  \tau^+ 
\end{align}
The chiral Lagrangian can thus be obtained from Eq.~\eqref{eq:tensor-p4-pi} in the external source method. 

In the conventional spurion method, the interaction is expressed in the chiral basis as follows 
\begin{align}
    \mathcal{L}_{{\rm eff}, 17}^{(6)} = \mathcal{C}_{17}^{(6)} (\bar{\nu}_{L}\sigma^{\mu\nu}e_{R})(\bar{q}_{L}\sigma_{\mu\nu} \tau^- q_{R}) + \mathcal{C}_{17}^{(6)} (\bar{\nu}_{L}\sigma^{\mu\nu}e_{R})(\bar{q}_{R}\sigma_{\mu\nu} \tau^- q_{L}) + {\rm h.c.}\;,
\end{align}
so that the spurions in Eq.~\eqref{eq:tensormatching} are
\begin{align}
    \lambda^{\mu\nu\dagger} &=  \mathcal{C}_{17}^{(6)} \bar{\nu}_L \sigma^{\mu \nu} e_R \, \tau^- + \mathcal{C}_{17}^{(6)\dagger} \bar{e}_R \sigma^{\mu \nu} \nu_L \, \tau^+ \;,\nn\\
    \lambda^{\mu\nu} &=  \mathcal{C}_{17}^{(6)\dagger} \bar{e}_R \sigma^{\mu \nu} \nu_L \, \tau^+  +   \mathcal{C}_{17}^{(6)} (\bar{\nu}_{L}\sigma^{\mu\nu}e_{R}) \tau^-\;.
\end{align}
From Eq.~\eqref{eq:Xmunu}, we have
\begin{align}
    X^{\mu\nu \dagger}& = \frac{1}{2}\lambda_{\mu\nu}^\dagger+\frac{i}{4}\varepsilon_{\mu\nu\alpha\beta}\lambda^{\alpha\beta\dagger} = t^{\mu\nu \dagger}\nn\\
    X^{\mu\nu}& = \frac{1}{2}\lambda_{\mu\nu}-\frac{i}{4}\varepsilon_{\mu\nu\alpha\beta}\lambda^{\alpha\beta}  = t^{\mu\nu} \;.
\end{align}

Besides, the combination of $\lambda^{\mu\nu}$ and $\lambda^{\mu\nu\dagger}$ gives 
\begin{align}
    \lambda_T^{\mu\nu} =\mathcal{C}_{17}^{(6)} (\bar{\nu}_{L}\sigma^{\mu\nu}e_{R}) \tau^- + {\rm h.c.} = \bar t^{\mu\nu}\;,\quad
    \lambda_\varepsilon^{\mu\nu} = 0\;.
\end{align}
Thus, the matching in the conventional spurion method can also be achieved in the $q$ basis, as indicated in Eq.~\eqref{eq:mathcing-t_non-chiral}. Since $\lambda^\dagger =\lambda = \lambda_T$, we can express 
\begin{align}
    \mathcal{L}_{\rm eff, 17}^{(6)}  \to & -\frac{1}{8}\mathrm{Tr}\left[(u\lambda^{\mu\nu\dagger}u+u^{\dagger}\lambda^{\mu\nu}u^{\dagger})[\hat{u}_{\mu},\hat{u}_{\nu}]\right]g_{T}^{(2)} \nn\\
 & -\frac{i}{16}\varepsilon_{\mu\nu\alpha\beta}\operatorname{Tr}\left[(u\lambda^{\alpha\beta\dagger}u-u^\dagger\lambda^{\alpha\beta}u^\dagger)[\hat{u}^\mu,\hat{u}^\nu]\right]g_T^{(2)}\;,
\end{align}
which agrees with the results from the external source method.

In the systematic spurion method, 
\begin{align}
\mathcal{L}_{{\rm eff}, 17} = \mathcal{C}_{17}^{(6)}(\bar{\nu}_L\sigma^{\mu\nu}e_R)[\bar{q}\sigma_{\mu\nu}(\Sigma^\dagger P_R+\Sigma P_L)q] + {\rm h.c.}\;,
\end{align}
where $\Sigma^\dagger = \Sigma = \tau^-$. The matching at $p^4$ order is 
\begin{align}
\mathcal{L}_{\mathrm{eff},17}^{(6)}\to\mathcal{C}_{17}^{(6)}\left[(\bar{\nu}_L\sigma^{\mu\nu}e_R)\langle\Sigma_{+}[\hat{u}_{\mu},\hat{u}_{\nu}]\rangle i\beta+(\bar{\nu}_L\sigma^{\mu\nu}e_R)\langle\Sigma_{-}[\hat{u}_{\mu},\hat{u}_{\nu}]\rangle i\beta\right] + {\rm h.c.}\;.
\end{align}

\section{Young tensor technique}
\label{sec:Yound}
The Young tensor technique~\cite{Li:2020gnx,Li:2020xlh,Li:2022tec} is a systematic method to construct independent group invariants explicitly. Next, we will outline the basic points of this method for the Lorentz and internal structures separately.

\subsubsection*{Lorentz Structure}

The complexification of the Lie algebra of the Lorentz group $SO(3,1)$ is the direct sum Lie algebra $SL(2,\mathbb{C})_l\oplus SL(2,\mathbb{C})_r$, thus all the irreducible representations of the Lorentz group is labeled by two half integers $(j_l,j_r)$, where $j_l$ and $j_r$ labels the irreducible representations of $SL(2,\mathbb{C})$.

If we define the $\lambda_\alpha\in (\frac{1}{2},0)$ as a left-handed spinor and $\Tilde{\lambda}_{\dot{\alpha}}\in (0,\frac{1}{2})$ as a right-handed spinor, all other fields of other irreducible representation of the Lorentz group can be expressed as their product. In the $\chi$PT, the helicities of all the degrees of freedom are at most 1, and their representations and spinor forms are
\begin{align}
\label{define}
    \text{Scalar field: }&\phi\in (0,0) \sim 1\notag \\
    \text{Left-handed spinor field: } & \psi \in (\frac{1}{2},0) \sim \lambda_\alpha\notag \\
    \text{Right-handed spinor field: } & \psi^\dagger \in (0,\frac{1}{2}) \sim \Tilde{\lambda}_{\dot{\alpha}}\notag \\
    \text{Left-handed field strength tensor: } & F_L=\frac{F-i\tilde{F}}{2} \in (1,0) \sim \lambda_\alpha\lambda_\beta \notag \\
    \text{Right-handed field strength tensor: } & F_R=\frac{F+i\tilde{F}}{2} \in (0,1) \sim \Tilde{\lambda}_{\dot{\alpha}} \Tilde{\lambda}_{\dot{\beta}}\notag \\
    \text{Derivative: } & D \in (1,1) \sim \lambda_\alpha\Tilde{\lambda}_{\dot{\alpha}} \,,
\end{align}
thus an operator can be expressed as a product of spinors with all the left- and right-handed indices contracted by the asymmetric tensors $\epsilon^{\alpha\beta}\,,\epsilon^{\dot{\alpha}\dot{\beta}}$. Defining 
\begin{equation}
    \epsilon^{\alpha\beta} \lambda^i_\beta\lambda^j_\alpha = \langle ij \rangle\,,\quad \tilde{\lambda}^i_{\dot{\alpha}} \tilde{\lambda}^j_{\dot{\beta}}\epsilon^{\dot{\beta}\dot{\alpha}} = [ij]\,,
\end{equation}
a general operator takes the form that
\begin{equation}
    \mathcal{B} = \prod^n \langle ij\rangle \prod^{\tilde{n}} [kl]\,,
\end{equation}
where $n$ and $\tilde{n}$ are the half numbers of left- and right-handed spinor indices carried by the operator respectively. 

After such an operator-amplitude correspondence, the independent Lorentz structures can be constructed via the so-called primary Young diagrams, with the redundancies such as the EOM, IBP, the Fierz identities, and so on removed automatically. For an operator of $N$ fields whose helicities are $\{h_1,h_2,\dots,h_N\}$, and $k$ derivatives, the number of the left- and right-handed indices are determined,
\begin{equation}
\label{eq:nn}
    n=\frac{k}{2}-\sum_{h_i<0}h_i\,,\quad \tilde{n}=\frac{k}{2}+\sum_{h_i>0}h_i\,,
\end{equation}
and the corresponding primary Young diagram takes the form that
\begin{equation}
    \begin{tikzpicture}
\filldraw [draw = black, fill = cyan] (10pt,10pt) rectangle (22pt,22pt);
\filldraw [draw = black, fill = cyan] (10pt,22pt) rectangle (22pt,34pt);
\draw [densely dotted] (24pt,22pt) -- (32pt,22pt);
\filldraw [draw = black, fill = cyan] (10pt,46pt) rectangle (22pt,58pt);
\filldraw [draw = black, fill = cyan] (10pt,58pt) rectangle (22pt,70pt);
\draw [densely dotted] (24pt,58pt) -- (32pt,58pt);
\draw [densely dotted] (16pt,36pt) -- (16pt,46pt);
\filldraw [draw = black, fill = cyan] (34pt,10pt) rectangle (46pt,22pt);
\filldraw [draw = black, fill = cyan] (34pt,22pt) rectangle (46pt,34pt);
\filldraw [draw = black, fill = cyan] (34pt,46pt) rectangle (46pt,58pt);
\filldraw [draw = black, fill = cyan] (34pt,58pt) rectangle (46pt,70pt);
\draw [densely dotted] (40pt,36pt) -- (40pt,46pt);
\draw [|<-] (0pt,10pt)--(0pt,34pt);
\draw [|<-] (0pt,70pt)--(0pt,46pt);
\node (n2) at (-5pt,40pt) {\small $N-2$};

\draw [|<-] (10pt,80pt)--(22pt,80pt);
\draw [|<-] (46pt,80pt)--(34pt,80pt);
\node (nt) at (28pt,80pt) {\small $\tilde{n}$};


\draw (46pt,58pt) rectangle (58pt,70pt);
\draw (46pt,46pt) rectangle (58pt,58pt);
\draw [densely dotted] (60pt,58pt) -- (70pt,58pt);
\draw (70pt,58pt) rectangle (82pt,70pt);
\draw (70pt,46pt) rectangle (82pt,58pt);
\draw [|<-] (46pt,36pt)--(58pt,36pt);
\draw [|<-] (82pt,36pt)--(70pt,36pt);
\node (n) at (64pt,36pt) {\small $n$};
    \end{tikzpicture}\,,
\end{equation}
where the cyan boxes are from right-handed indices.
According to the group theory, the semi-standard Young tableaux (SSYT) associated with the primary Young diagram form a set of bases, called y-basis. Considering a specific operator, the numbers can be filled to the Young diagrams are also determined by $h_i$ and $k$,
\begin{equation}
\label{eq:numi}
    \# i = \sum_{h_i>0}h_i -2h_i + \frac{k}{2}\,,\quad i =1\,,2\,,\dots \,, N\,,    
\end{equation}

When applying this method to the $\chi$PT, there are several remarks
\begin{itemize}
    \item The NGBs $u_\mu$ in the $\chi$PT is of the representation $(\frac{1}{2},\frac{1}{2})$ of the Lorentz group meaning its amplitude satisfies Adler's zero condition, which constrains the y-basis. Such constraints can be extracted by taking the limit $\lim_{p\rightarrow p} \mathcal{B}(p)$, where $p$ is the momentum of the NGB, and are complemented as a system of linear equations of the original y-basis. The solution space of the system of linear equations contains the Lorentz structures satisfying Adler's zero condition.
    \item The Young tensor method adopts left- and right-handed spinors $\psi/\psi^\dagger$ as fundamental building blocks, but the operators composed of them are usually not $C\,, P$ eigenstates. thus we need to combine them to form Dirac spinors so that they have specific $C\,, P$ eigenvalues. A Dirac spinor $\psi$ is a direct sum of a left-handed spinor $\psi_L$ and a right-handed spinor $\psi_R$,  
    \begin{equation}
        \psi = \left(\begin{array}{c}
\psi_L \\ \psi_R
        \end{array}\right)\,.
    \end{equation}
    Considering a simple example of operator type $u\Sigma_+ e^2$, there are two Lorentz structures (the chiral symmetry has been discarded for simplicity for now)
    \begin{equation}
        \mathcal{B}_1 = (e^\dagger_L\bar{\sigma}^\mu e_L) u_\mu\Sigma_+ \,,\quad \mathcal{B}_2 = (e_R^\dagger \sigma^\mu e_R) u_\mu \Sigma_+\,.
    \end{equation}
    Under the parity transformation, they interchange with each other, since $P$ makes left- and right-handed spinors swap, and neither of them is in the $P$ eigenstate. Recalling 
    \begin{equation}
        \gamma^\mu = \left(\begin{array}{cc}
0 & \sigma^\mu \\ \overline{\sigma}^\mu & 0
        \end{array}\right)\,,
    \end{equation}
    there is
    \begin{equation}
        \mathcal{B}_1 = (\bar e \frac{1-\gamma^5}{2}\gamma^\mu e)u_\mu\Sigma_+\,,\quad \mathcal{B}_1 = (\bar e \frac{1+\gamma^5}{2}\gamma^\mu e)u_\mu\Sigma_+\,,
    \end{equation}
    thus they can be combined to give 
    \begin{equation}
        \mathcal{B}_1 + \mathcal{B}_2 =  (\bar{e}\gamma^\mu e)u_\mu\Sigma_+ \,,\quad -\mathcal{B}_1 + \mathcal{B}_2 =  (\bar{e}\gamma^5\gamma^\mu e)u_\mu\Sigma_+ \,,
    \end{equation}
    whose $P$ eigenvalues are $-1$ and $+1$ respectively since the intrinsic party charge of $u_\mu$ is $-1$.
\end{itemize}

\subsubsection*{Internal Flavor Structure}
The flavor group here refers to the special unitary group $SU(N)$, and the case of $SU(2)$ is needed in this work.

According to the group theory, every irreducible representation of the $SU(N)$ group corresponds to a standard Young diagram. The effective operators are gauge invariants, which means they are of the trivial representation $\mathbf{1}$, which corresponds to the Young diagram of the form
\begin{equation}
    SU(2)\sim \yng(2,2)\dots\yng(1,1)\,.
\end{equation}
Thus the gauge invariants are the different Young diagrams of the shape that respect the Littlewood-Richardson rule. More details of the construction for chiral Lagrangian can be found in Ref.~\cite{Song:2024fae,Li:2024ghg}.

\subsubsection*{Example}
We take $\mathcal{O}_5^{(6)}=(\bar e \gamma^\mu e)[\bar{q}\gamma_\mu(\Sigma_RP_R+\Sigma_LP_L)q]$ in Tab.~\ref{tab:va} as an example. There would be numerous types of chiral operators for $\mathcal{O}_5^{(6)}$. We can list the type by the number of fields, and there would be two electron fields and a spurion $Q$ at least in the chiral operators. However, the type of $Qe^2$ will not correspond to the $\mathcal{O}_5^{(6)}$ due to the Lorentz structure of electron current. Then we can add building block $\nabla_\mu$ and $\hat{u}_\mu$, the corresponding types are $\nabla Qe^2$ and $\hat{u}Qe^2$. The type $\nabla Qe^2$ would be invalid as the single spurion Q cannot construct the chiral invariant. Thus, there is only one type $\hat uQe^2$ at LO. In NLO, the addition building blocks include $\nabla^3\,,\nabla^2\hat{u}\,,\nabla \hat{u}^2\,,\hat{u}^3\,,\nabla\hat\chi\,,\hat{u}\hat\chi$, which correspond to the types $\nabla^3Qe^2$\,,$\nabla^2\hat{u}Qe^2$\,,$\nabla\hat{u}^2Qe^2$\,,$\hat{u}^3Qe^2$\,,$\nabla\hat\chi Qe^2$\,,$\hat{u}\hat\chi Qe^2$\,. Moreover, there is a relation between the $\nabla Q$, $uQ$ and $\nabla \hat\chi$, $u\hat\chi$,
\begin{align}
    \label{dQ}\nabla^\mu Q_\pm&\sim [Q_\mp,\hat{u}^\mu]\,,\\
    \nabla^\mu \hat\chi_\pm&\sim \{\hat\chi_\mp,\hat{u}^\mu\}\,,
\end{align}
which is similar to Eq.~(2.6) of Ref.~\cite{Bijnens:1999sh}. Thus, the complete and independent types of chiral operators up to NLO are $\hat{u}Qe^2$\,, $\hat{u}^3Qe^2$ and $\hat{u}\hat\chi Qe^2$ which would generate the operators in Tab.~\ref{tab:va}.

In addition, we list the whole procedure of the type $\hat{u}\hat{\chi}Qe^2$ and the invalid type $\nabla\hat\chi Qe^2$. There are two classes: $\psi_R\psi_R^\dagger \phi^3 D$ and $\psi_L\psi_L^\dagger \phi^3 D$ for the type $\hat{u}\hat{\chi}Qe^2$ in the procedure where these fields can be found in Eq.~\eqref{define} . We first consider the class $\psi_R\psi_R^\dagger \phi^3 D$, there are 5 fields and 3 derivatives in the class $\psi_R\psi_R^\dagger \phi^3 D$ and their helicities are $\{-1/2,0,0,0,1/2\}$, while the last three Goldstone bosons are repeated fields. We obtain $n=1,\,\tilde n=1$ through Eq.~\eqref{eq:nn} and the corresponding primary Young diagram is
\begin{equation}
    \ydiagram{2,2,1}\,.
\end{equation}
The numbers of all indices that are used to fill the diagram can be obtained that
\begin{equation}
    \# 1=2,\,\quad\#2=1,\,\quad\#3=1,\,\quad\#4=1,\,\quad\#5=0.
\end{equation}
There are two SSYTs
\begin{align}
    \begin{ytableau}
            1&1\\
            2&3\\
            4
        \end{ytableau}\,,\quad\begin{ytableau}
            1&1\\
            2&4\\
            3
        \end{ytableau}\,,
\end{align}
and the corresponding Lorentz y-basis are
\begin{align}
    \mathcal{M}_1&=\langle13\rangle[35]\,,\notag\\
    \mathcal{M}_2&=\langle14\rangle[45]\,.
\end{align}

Furthermore, the Adler zero condition should be considered that the amplitudes should be zero whenever particle 2 soft. The Adler zero condition and translating can be summed up as a $1\times 2$ transformation matrix 
\begin{equation}
    \left(-1,1\right)\,,
\end{equation}
and then the Lorentz m-basis can be obtained
\begin{align}
    \mathcal{B}_{1R}&=\psi_{R1}\Bar{\sigma}^\mu \psi_{R5}^\dagger D_\mu \phi_2 \phi_3 \phi_4 \,,
\end{align}
Thus the Lorentz m-basis of $\psi_L\psi_L^\dagger\phi^4 D^3$ are similar to the class $\psi_R\psi_R^\dagger\phi^3 D$
\begin{align}
    \mathcal{B}_{1L}&=\psi_{L1}\Bar{\sigma}^\mu \psi_{L5}^\dagger D_\mu \phi_2 \phi_3 \phi_4\,.
\end{align}
As for the internal flavor y-basis, the $SU(2)$ tensors can be obtained by similar SSYT techniques in the $SU(2)$ case. And there is only three $SU(2)$ index in this type while the electron do not have the $SU(2)$ index.  For this type, there is only one flavor SSYT
\begin{align}
    \begin{ytableau}
        i_1&j_1&i_2\\
        j_2&i_3&j_3
    \end{ytableau}\,.
\end{align}
We can convert them to obtain the flavor m-basis
\begin{align}
    \mathcal{T}_1^m&=\epsilon^{IJK} \,,
\end{align}
and the three fields $\phi_2$, $\phi_3$ and $\phi_4$ would contain these $SU(2)$ index.

Combining the Lorentz and internal structure, the chiral operators in the type $\hat{u}\hat{\chi}Qe^2$ are
\begin{align}
    \mathcal{O}_1&=\epsilon^{IJK}(e_{L}\Bar{\sigma}^\mu e_{L}^\dagger) \hat{u}_\mu^I Q^J\hat{\chi}^K\,,\notag\\
    \mathcal{O}_2&=\epsilon^{IJK}(e_{R}\Bar{\sigma}^\mu e_{R}^\dagger) \hat{u}_\mu^I Q^J\hat{\chi}^K\,.
\end{align}
Recalling the relation
\begin{align}
        e = \left(\begin{array}{c}
e_L \\ e_R
        \end{array}\right)\,,\quad 
        \gamma^\mu = \left(\begin{array}{cc}
0 & \sigma^\mu \\ \bar{\sigma}^\mu & 0
        \end{array}\right)\,,
\end{align}
then the corresponding chiral operator of type $\hat{u}\hat{\chi}Qe^2$ for $\mathcal{O}_5^{(6)}$ with $C-P+$ properties for the hadronic case is
\begin{align}
    O_{1}&=(\bar e\gamma^\mu e) \langle  \chi_-[Q_+,\hat{u}_\mu]\rangle\,.
\end{align}
Similarly, the corresponding chiral operator of type $ \nabla\chi Qe^2$ for $\mathcal{O}_5^{(6)}$ with $C-P+$ properties for the hadronic case is
\begin{align}
    O_{1}^{\rm R}=(\bar e\gamma^\mu e)\langle\hat{\chi}_- \nabla_\mu Q_-\rangle\,.
\end{align}
Then the operator $O_{1}^{\rm R}$ can be translated to $O_1$ through the Eq.\eqref{dQ} which means $O_{1}^{\rm R}$ is redundant.

\section{Matching for photon operators}
\label{sec:match-photon}

In this appendix, we discuss the matching for the operators involving one or two photon fields. In the external source method, the photon field $ A^\mu $ is incorporated in the vector source; while in the systematic spurion method, $ A^\mu $ is separated from the lepton currents.

We first consider operators involving a photon field that appears in the covariant derivative $ D^\mu = \partial^\mu + ie Q A^\mu $.
The matching for these operators is analogous to that of dimension-6 operators.

For instance, considering the interaction from the kinetic term
\begin{align}
\label{eq:kin}
    \mathcal{L}_{\rm kin} = \bar q i\gamma^\mu D_{\mu} q \sim - e A_{\mu}  (\bar{q} \gamma^\mu Q q)\;,
\end{align}
which is matched to the chiral operator $\langle \hat u^\mu \hat u_\mu \rangle$ at order of $p^2$.
The corresponding LEC is identical to that of $\mathcal{O}_5^{(6)}$, as both chiral operators arise from the same underlying quark bilinear structure.

The photon field can also appear in the field strength tensor $F^{\mu\nu} = \partial^\mu A^\nu - \partial^\nu A^\mu$. The relevant dimension-5 and dimension-7 operators involving $F^{\mu\nu}$ are
\begin{align}
 \mathcal{O}_{1}^{(5)}  =&F^{\mu\nu}[\bar q(\Sigma^\dagger P_R+\Sigma P_L)\sigma_{\mu\nu} q] \;,&
 \mathcal{O}_{2}^{(5)}  =&\tilde{F}^{\mu\nu}[\bar q(\Sigma^\dagger P_R+\Sigma P_L)\sigma_{\mu\nu}q]\;, \\
 \mathcal{O}_{15}^{(7)}  =&F^{\mu\nu}F_{\mu\nu}[\bar q(\Sigma^\dagger P_R+\Sigma P_L)q]\;,&
 \mathcal{O}_{16}^{(7)}  =&F^{\mu\nu}F_{\mu\nu}[\bar q(\Sigma^\dagger P_R-\Sigma P_L) q]\;,\\
 \mathcal{O}_{17}^{(7)}  =&\tilde{F}^{\mu\nu}F_{\mu\nu}[\bar q(\Sigma^\dagger P_R+\Sigma P_L)q]\;,&
 \mathcal{O}_{18}^{(7)}  =&\tilde{F}^{\mu\nu}F_{\mu\nu}[\bar q(\Sigma^\dagger P_R-\Sigma P_L)q]\;.
\end{align}
Here, the dual field strength is defined as $\tilde{F}^{\mu\nu}=\frac{1}{2}\varepsilon^{\mu\nu\alpha\beta}F_{\alpha\beta}$. By separating the field strength, the matching of the quark bilinears can be treated analogously to the dimension-6 operators, which are given explicitly in Tab.~\ref{tab:A}. 
For the dimension-8 operators involving $F^{\mu\nu}$ and four quarks, the matching is similar to dimension-6 LEFT operators.

\begin{table}[H]
    \centering
\captionsetup{justification=centering}
   \caption{The matching for field strength operators at $p^2$ and  $p^4$ orders.}
    \begin{tabular}{|l|c|c|}
    \hline 
    \multirow{2}{*}{LEFT operator} & \multicolumn{2}{c|}{$\chi$PT operators} \\
    \cline{2-3}  
    &$\mathcal{O}(p^2)$&$\mathcal{O}(p^4)$\\
    \hline
    \multirow{2}{*}{$\mathcal{O}_{1}^{(5)} = F^{\mu\nu}[\bar q(\Sigma^\dagger P_R+\Sigma P_L)\sigma_{\mu\nu} q]$} &\multirow{2}{*}{ $/$} & $F^{\mu\nu}\langle \Sigma_+[ \hat u_\mu, \hat u_\nu]\rangle$  \\
    &&$\tilde{F}^{\mu\nu}\langle \Sigma_-[ \hat u_\mu, \hat u_\nu]\rangle$\\
    \hline
    \multirow{2}{*}{$\mathcal{O}_{2}^{(5)} =\tilde{F}^{\mu\nu}[\bar q(\Sigma^\dagger P_R+\Sigma P_L)\sigma_{\mu\nu} q]$} & \multirow{2}{*}{$/$} & $\tilde{F}^{\mu\nu}\langle \Sigma_+[ \hat u_\mu, \hat u_\nu]\rangle$  \\
    &&$F^{\mu\nu}\langle \Sigma_-[ \hat u_\mu, \hat u_\nu]\rangle$\\
    \hline
     \multirow{2}{*}{$\mathcal{O}_{15}^{(7)} =F^{\mu\nu}F_{\mu\nu}[\Bar{q}(\Sigma P_R+\Sigma^\dagger P_L) q]$} &\multirow{2}{*}{$F^{\mu\nu}F_{\mu\nu}\langle\Sigma_+\rangle$}&$F^{\mu\nu}F_{\mu\nu}\langle\Sigma_+\rangle\langle \hat u_\mu \hat u^\mu\rangle$\\
     &&$F^{\mu\nu}F_{\mu\nu}\langle\Sigma_+\rangle\langle \chi_+\rangle$\\
     \hline
    \multirow{2}{*}{$\mathcal{O}_{16}^{(7)} =F^{\mu\nu}F_{\mu\nu}[\Bar{q}(\Sigma^\dagger P_R-\Sigma P_L) q]$}&\multirow{2}{*}{$F^{\mu\nu}F_{\mu\nu}\langle\Sigma_-\rangle$}&$F^{\mu\nu}F_{\mu\nu}\langle\Sigma_-\rangle\langle \hat u_\mu \hat u^\mu\rangle$\\
    &&$F^{\mu\nu}F_{\mu\nu}\langle\Sigma_-\rangle\langle \chi_+\rangle$\\
    \hline
    \multirow{2}{*}{$\mathcal{O}_{17}^{(7)} =\tilde{F}^{\mu\nu}F_{\mu\nu}[\Bar{q}(\Sigma P_R +\Sigma^\dagger P_L) q]$} &\multirow{2}{*}{$\tilde{F}^{\mu\nu}F_{\mu\nu}\langle\Sigma_+\rangle$}&$\tilde{F}^{\mu\nu}F_{\mu\nu}\langle\Sigma_+\rangle\langle \hat u_\mu \hat u^\mu\rangle$\\
    &&$\tilde{F}^{\mu\nu}F_{\mu\nu}\langle\Sigma_+\rangle\langle \chi_+\rangle$\\
    \hline
    \multirow{2}{*}{$\mathcal{O}_{18}^{(7)} =\tilde{F}^{\mu\nu}F_{\mu\nu}[\Bar{q}(\Sigma^\dagger P_R-\Sigma P_L) q]$}&\multirow{2}{*}{$\tilde{F}^{\mu\nu}F_{\mu\nu}\langle\Sigma_-\rangle$}&$\tilde{F}^{\mu\nu}F_{\mu\nu}\langle\Sigma_-\rangle\langle \hat u_\mu \hat u^\mu\rangle$\\
    &&$\tilde{F}^{\mu\nu}F_{\mu\nu}\langle\Sigma_-\rangle\langle \chi_+\rangle$\\
    \hline
    \end{tabular}
    \label{tab:A}
\end{table}

\section{Matching for derivative operators}
\label{sec:dim-7_derivative}

In this appendix, we will give details of matching the dimension-7 derivative operators with single quark bilinear to the chiral Lagrangian in the conventional spurion method. 

The interaction with one partial derivative acting on the quark field is expressed as
\begin{align}
\mathcal{L}_{D,D5}^q &= \bar{q}_L \lambda_1^\mu \overleftrightarrow{\partial}_\mu q_R + \bar{q}_R \lambda_2^\mu \overleftrightarrow{\partial}_\mu q_L + \text{h.c.} \\
&= \bar{q}_L (-i\lambda^\mu)^\dagger \overleftrightarrow{\partial}_\mu q_R - \bar{q}_R (-i\lambda^\mu) \overleftrightarrow{\partial}_\mu q_L\;,
\end{align}
where the Hermitian spurion is $\lambda^\mu \equiv i(\lambda_1^{\mu\dagger} - \lambda_2^\mu)$. The quark-level Lagrangian is matched to chiral Lagrangian $\mathcal{L}_{D,D5}^{(d)}$, 
\begin{align}
    \mathcal{L}_{D,D5}^q &= i \left( \bar{q}_L \lambda^{\mu\dagger} \partial_\mu q_R - \partial_\mu \bar{q}_L \lambda^{\mu\dagger} q_R \right) + \text{h.c.} \to \mathcal{L}_{D,D5}^{(d)}\;.
\end{align}
where $d$ represents the number of derivatives.

At $\mathcal{O}(p^2)$, the chiral Lagrangian is
\begin{align}
\mathcal{L}_{D,D5}^{(1)} &= i \Tr\left[ u \lambda^{\mu\dagger} (D^\mu u^\dagger)^\dagger \right] g_{D,1}^{(1)} 
- i \Tr\left[ D^\mu u \lambda^{\mu\dagger} u \right] g_{D,2}^{(1)} + \text{h.c.}\;.
\end{align}
Expressed in terms of $U$ or $\hat{u}^\mu$, one has
\begin{align}
\mathcal{L}_{D,D5}^{(1)} &= \frac{i}{2} \Tr\left( \lambda^{\mu\dagger} \partial_\mu U - \lambda^\mu \partial_\mu U^\dagger \right) \tilde{g}_{D}^{(1)} \;, \quad \tilde{g}_{D}^{(1)} \equiv g_{D,1}^{(1)} - g_{D,2}^{(1)}\;,
\end{align}
or equivalently,
\begin{align}
\mathcal{L}_{D,D5}^{(1)} &= \frac{1}{2} \Tr\left[\left( u\lambda^{\mu\dagger} u + u^\dagger \lambda^\mu u^\dagger \right) \hat{u}_\mu \right] \tilde{g}_{D}^{(1)}\;.
\end{align}

Similar to tensor current in Sec.~\ref{sec:tensor-current}, we consider the $C$ transformation, 
\begin{align}
    \mathcal{L}_{D,D5}^q &\stackrel{C}{\longrightarrow} -i \left(\bar{q}_L \lambda_\mu^{c \dagger} 
 \overleftrightarrow{\partial}^\mu  q_R + \bar{q}_R \lambda_\mu^c\overleftrightarrow{\partial}^\mu  q_L\right)\;,  
\end{align}
where
$\lambda_\mu^c$ and $\lambda_\mu^{c\dagger}$ are defined as in Eq.~\eqref{eq:lambda-c}.
The above relation shows that the chiral Lagrangian $\mathcal{L}_D^{(1)}$ 
has wrong sign under the $C$ transformation.

Analogous to the tensor current case, we introduce the commutator, which transforms as
\begin{align}
\left[ \nabla_\mu \hat{u}_\nu, \hat{u}^\nu \right] &\stackrel{C}{\longrightarrow} \left[ (\nabla_\mu \hat{u}_\nu)^T, \left(\hat{u}^\nu\right)^T \right]\;,
\end{align}
yielding the correct transformation property as follows:
\begin{align}
    \Tr \left[ \left( u \lambda_\mu^{\dagger} u + u^\dagger \lambda_\mu u^\dagger\right) [\nabla^\mu \hat{u}_\nu, \hat{u}^\nu \right]  \stackrel{C}{\longrightarrow}&\ - \Tr \left[ \left( u \lambda_\mu^{\dagger c} u + u^\dagger \lambda^c_\mu u^\dagger\right) [\nabla^\mu \hat{u}_\nu, \hat{u}^\nu] \right]\;.
\end{align}
The consistent chiral Lagrangian emerges at $\mathcal{O}(p^4)$ with $d=3$, which is expressed as
\begin{align}
\label{eq:derivative}
\mathcal{L}_{D,D5}^{(3)} &= \Tr \left[ \left( u\lambda^{\mu\dagger} u + u^\dagger \lambda^\mu u^\dagger \right) \left[ \nabla_\mu \hat{u}_\nu, \hat{u}^\nu \right] \right] \tilde{g}_{D}^{(3)}.
\end{align}

Notice that in case of vanishing fied strength, $\nabla_\mu \hat{u}_\nu - \nabla_\nu \hat{u}_\mu = 0$, and $ \left[ \nabla_\nu \hat{u}_\mu, \hat{u}^\nu \right]$ is redundant in constructing the chiral Lagrangian. On the other hand, the commutator $\left[  \hat{u}_\mu, \nabla_\nu \hat{u}^\nu \right]$ gives rise to an independent chiral operator, which corresponds to the chiral operators with $\left\langle\hat{u}_\mu\left[\Sigma_{+}, \chi_{-}\right]\right\rangle$ or $\left\langle\hat{u}_\mu\left[\Sigma_{-}, \chi_{-}\right]\right\rangle$.

In the $q$ basis, we obtain
\begin{align}
\mathcal{L}_{D,D5}^q &= \bar{q} \lambda_D^\mu \overleftrightarrow{\partial}_\mu q + \bar{q} \gamma^5 \lambda_{D5}^\mu \overleftrightarrow{\partial}_\mu q\;,
\end{align}
where both $\lambda_D^\mu$ and $\lambda_{D5}^\mu$ are Hermitian. 

We can rewrite the above expression as follows
\begin{align}
\bar{q} \lambda_D^\mu \overleftrightarrow{\partial}_\mu q &= \bar{q}_L \lambda_D^\mu \overleftrightarrow{\partial}_\mu q_R + \bar{q}_R \lambda_D^{\mu\dagger} \overleftrightarrow{\partial}_\mu q_L\;, \\
\bar{q} \gamma^5 \lambda_{D5}^\mu \overleftrightarrow{\partial}_\mu q &= \bar{q}_L \lambda_{D5}^\mu \overleftrightarrow{\partial}_\mu q_R - \bar{q}_R \lambda_{D5}^{\mu\dagger} \overleftrightarrow{\partial}_\mu q_L\;,
\end{align}
which are formally invariant under chiral symmetry given the transformation of the 
spurions $\lambda_{D,D5}^\mu \to L \lambda_{D,D5}^\mu R^\dagger$ and  $\lambda_{D,D5}^{\mu\dagger} \to R \lambda_{D,D5}^{\mu\dagger} L^\dagger$.

The matched chiral operators become
\begin{align}
\bar{q} \lambda_D^\mu \overleftrightarrow{\partial}_\mu q &\to \Tr \left[ \left( u\lambda_D^\mu u + u^\dagger \lambda_D^{\mu\dagger} u^\dagger \right) \left[ \nabla_\mu \hat{u}_\nu, \hat{u}^\nu \right] \right], \nn \\
\bar{q} \gamma^5 \lambda_{D5}^\mu \overleftrightarrow{\partial}_\mu q &\to \Tr \left[ \left( u\lambda_{D5}^\mu u - u^\dagger \lambda_{D5}^{\mu\dagger} u^\dagger \right) \left[ \nabla_\mu \hat{u}_\nu, \hat{u}^\nu \right] \right].
\end{align}
We identify
\begin{align}
    \lambda_{D}^\mu = \lambda_D^{\mu\dagger} &= i(\lambda^{\mu\dagger} + \lambda^\mu)/2\;,\nn\\
    \lambda_{D5}^\mu = \lambda_{D5}^{\mu\dagger} &= i(\lambda^{\mu\dagger} - \lambda^\mu)/2\;,
\end{align}
and obtain that the matched chiral operators agree with Eq.~\eqref{eq:derivative}.

\bibliographystyle{JHEP}
\bibliography{reference}

@article{Scherer:2002tk,
    author = "Scherer, Stefan",
    editor = "Negele, John W. and Vogt, E. W.",
    title = "{Introduction to chiral perturbation theory}",
    eprint = "hep-ph/0210398",
    archivePrefix = "arXiv",
    reportNumber = "MKPH-T-02-09",
    journal = "Adv. Nucl. Phys.",
    volume = "27",
    pages = "277",
    year = "2003"
}

@article{Wilczek:1979hc,
    author = "Wilczek, Frank and Zee, A.",
    title = "{Operator Analysis of Nucleon Decay}",
    reportNumber = "Print-79-0709 (PRINCETON)",
    doi = "10.1103/PhysRevLett.43.1571",
    journal = "Phys. Rev. Lett.",
    volume = "43",
    pages = "1571--1573",
    year = "1979"
}

@article{Ellis:1979hy,
    author = "Ellis, John R. and Gaillard, Mary K. and Nanopoulos, Dimitri V.",
    title = "{On the Effective Lagrangian for Baryon Decay}",
    reportNumber = "CERN-TH-2749, LAPP-TH-06",
    doi = "10.1016/0370-2693(79)90477-5",
    journal = "Phys. Lett. B",
    volume = "88",
    pages = "320--324",
    year = "1979"
}

@article{Weinberg:1979sa,
    author = "Weinberg, Steven",
    title = "{Baryon and Lepton Nonconserving Processes}",
    reportNumber = "HUTP-79-A050",
    doi = "10.1103/PhysRevLett.43.1566",
    journal = "Phys. Rev. Lett.",
    volume = "43",
    pages = "1566--1570",
    year = "1979"
}

@article{Weinberg:1980bf,
    author = "Weinberg, Steven",
    title = "{Varieties of Baryon and Lepton Nonconservation}",
    reportNumber = "HUTP-80/A023",
    doi = "10.1103/PhysRevD.22.1694",
    journal = "Phys. Rev. D",
    volume = "22",
    pages = "1694",
    year = "1980"
}

@article{Abbott:1980zj,
    author = "Abbott, L. F. and Wise, Mark B.",
    title = "{The Effective Hamiltonian for Nucleon Decay}",
    reportNumber = "SLAC-PUB-2487",
    doi = "10.1103/PhysRevD.22.2208",
    journal = "Phys. Rev. D",
    volume = "22",
    pages = "2208",
    year = "1980"
}

@article{Kaymakcalan:1983uc,
    author = "Kaymakcalan, Omer and Lo, Chong-Huah and Wali, Kameshwar C.",
    title = "{Chiral Lagrangian for Proton Decay}",
    reportNumber = "SU-4217-250, COO-3533-250",
    doi = "10.1103/PhysRevD.29.1962",
    journal = "Phys. Rev. D",
    volume = "29",
    pages = "1962",
    year = "1984"
}

@article{Liao:2020jmn,
    author = "Liao, Yi and Ma, Xiao-Dong",
    title = "{An explicit construction of the dimension-9 operator basis in the standard model effective field theory}",
    eprint = "2007.08125",
    archivePrefix = "arXiv",
    primaryClass = "hep-ph",
    doi = "10.1007/JHEP11(2020)152",
    journal = "JHEP",
    volume = "11",
    pages = "152",
    year = "2020"
}

@article{Li:2021phq,
    author = "Li, Tong and Ma, Xiao-Dong and Schmidt, Michael A. and Zhang, Rui-Jia",
    title = "{Implication of J/\ensuremath{\psi}\textrightarrow{}(\ensuremath{\gamma}+)invisible for the effective field theories of neutrino and dark matter}",
    eprint = "2104.01780",
    archivePrefix = "arXiv",
    primaryClass = "hep-ph",
    reportNumber = "CPPC-2021-05",
    doi = "10.1103/PhysRevD.104.035024",
    journal = "Phys. Rev. D",
    volume = "104",
    number = "3",
    pages = "035024",
    year = "2021"
}

@article{Low:2022iim,
    author = "Low, Ian and Shu, Jing and Xiao, Ming-Lei and Zheng, Yu-Hui",
    title = "{Amplitude/operator basis in chiral perturbation theory}",
    eprint = "2209.00198",
    archivePrefix = "arXiv",
    primaryClass = "hep-ph",
    doi = "10.1007/JHEP01(2023)031",
    journal = "JHEP",
    volume = "01",
    pages = "031",
    year = "2023"
}

@article{Sun:2025zuk,
    author = "Sun, Hao and Wang, Yi-Ning and Yu, Jiang-Hao",
    title = "{Chiral Effective Field Theories for Strong and Weak Dynamics}",
    eprint = "2501.14018",
    archivePrefix = "arXiv",
    primaryClass = "hep-ph",
    month = "1",
    year = "2025"
}

@article{Carmona:2021xtq,
    author = "Carmona, Adrian and Lazopoulos, Achilleas and Olgoso, Pablo and Santiago, Jose",
    title = "{Matchmakereft: automated tree-level and one-loop matching}",
    eprint = "2112.10787",
    archivePrefix = "arXiv",
    primaryClass = "hep-ph",
    doi = "10.21468/SciPostPhys.12.6.198",
    journal = "SciPost Phys.",
    volume = "12",
    number = "6",
    pages = "198",
    year = "2022"
}

@article{Fuentes-Martin:2022jrf,
    author = {Fuentes-Mart\'\i{}n, Javier and K\"onig, Matthias and Pag\`es, Julie and Thomsen, Anders Eller and Wilsch, Felix},
    title = "{A proof of concept for matchete: an automated tool for matching effective theories}",
    eprint = "2212.04510",
    archivePrefix = "arXiv",
    primaryClass = "hep-ph",
    reportNumber = "MITP-22-105, TUM-HEP-1443/22, ZU-TH-58/22",
    doi = "10.1140/epjc/s10052-023-11726-1",
    journal = "Eur. Phys. J. C",
    volume = "83",
    number = "7",
    pages = "662",
    year = "2023"
}

@article{Scholer:2023bnn,
    author = "Scholer, Oliver and de Vries, Jordy and Gr\'af, Luk\'a\v{s}",
    title = "{\ensuremath{\nu}DoBe \textemdash{} A Python tool for neutrinoless double beta decay}",
    eprint = "2304.05415",
    archivePrefix = "arXiv",
    primaryClass = "hep-ph",
    doi = "10.1007/JHEP08(2023)043",
    journal = "JHEP",
    volume = "08",
    pages = "043",
    year = "2023"
}

@article{Graf:2025cfk,
    author = "Gr\'af, Luk\'a\v{s} and Hati, Chandan and Mart\'\i{}n-Gal\'an, Ana and Scholer, Oliver",
    title = "{Importance of Loop Effects in Probing Lepton Number Violation}",
    eprint = "2504.00081",
    archivePrefix = "arXiv",
    primaryClass = "hep-ph",
    month = "3",
    year = "2025"
}

@article{Liao:2025qwp,
    author = "Liao, Yi and Ma, Xiao-Dong and Wang, Hao-Lin and Zhao, Xiang",
    title = "{RGE solver for the complete dim-7 SMEFT interactions and its application to $0\nu\beta\beta$ decay}",
    eprint = "2505.06499",
    archivePrefix = "arXiv",
    primaryClass = "hep-ph",
    month = "5",
    year = "2025"
}

@article{Bijnens:1999sh,
    author = "Bijnens, Johan and Colangelo, Gilberto and Ecker, Gerhard",
    title = "{The Mesonic chiral Lagrangian of order p**6}",
    eprint = "hep-ph/9902437",
    archivePrefix = "arXiv",
    reportNumber = "LU-TP-99-02, UWTHPH-1999-02, ZU-TH-9-99",
    doi = "10.1088/1126-6708/1999/02/020",
    journal = "JHEP",
    volume = "02",
    pages = "020",
    year = "1999"
}

@article{Mertens:2011ts,
    author = "Mertens, Philippe and Smith, Christopher",
    title = "{The s ---\ensuremath{>} d gamma decay in and beyond the Standard Model}",
    eprint = "1103.5992",
    archivePrefix = "arXiv",
    primaryClass = "hep-ph",
    doi = "10.1007/JHEP08(2011)069",
    journal = "JHEP",
    volume = "08",
    pages = "069",
    year = "2011"
}

@article{Grinstein:1985ut,
    author = "Grinstein, Benjamin and Rey, Soo-Jong and Wise, Mark B.",
    title = "{{CP} Violation in Charged Kaon Decay}",
    reportNumber = "CALT-68-1286",
    doi = "10.1103/PhysRevD.33.1495",
    journal = "Phys. Rev. D",
    volume = "33",
    pages = "1495",
    year = "1986"
}

@article{He:2021mrt,
    author = "He, Xiao-Gang and Ma, Xiao-Dong",
    title = "{An EFT toolbox for baryon and lepton number violating dinucleon to dilepton decays}",
    eprint = "2102.02562",
    archivePrefix = "arXiv",
    primaryClass = "hep-ph",
    doi = "10.1007/JHEP06(2021)047",
    journal = "JHEP",
    volume = "06",
    pages = "047",
    year = "2021"
}

@article{Bijnens:2017xrz,
    author = "Bijnens, Johan and Kofoed, Erik",
    title = "{Chiral perturbation theory for neutron\textendash{}antineutron oscillations}",
    eprint = "1710.04383",
    archivePrefix = "arXiv",
    primaryClass = "hep-ph",
    reportNumber = "LU-TP-17-30",
    doi = "10.1140/epjc/s10052-017-5411-7",
    journal = "Eur. Phys. J. C",
    volume = "77",
    number = "12",
    pages = "867",
    year = "2017"
}

@article{Savage:1998yh,
    author = "Savage, Martin J.",
    title = "{Pionic matrix elements in neutrinoless double Beta decay}",
    eprint = "nucl-th/9811087",
    archivePrefix = "arXiv",
    reportNumber = "NT-UW-99-02",
    doi = "10.1103/PhysRevC.59.2293",
    journal = "Phys. Rev. C",
    volume = "59",
    pages = "2293--2296",
    year = "1999"
}

@article{Cirigliano:2017ymo,
    author = "Cirigliano, V. and Dekens, W. and Graesser, M. and Mereghetti, E.",
    title = "{Neutrinoless double beta decay and chiral $SU(3)$}",
    eprint = "1701.01443",
    archivePrefix = "arXiv",
    primaryClass = "hep-ph",
    reportNumber = "LA-UR-17-20043",
    doi = "10.1016/j.physletb.2017.04.020",
    journal = "Phys. Lett. B",
    volume = "769",
    pages = "460--464",
    year = "2017"
}

@article{Akdag:2022sbn,
    author = "Akdag, Hakan and Kubis, Bastian and Wirzba, Andreas",
    title = "{C and CP violation in effective field theories}",
    eprint = "2212.07794",
    archivePrefix = "arXiv",
    primaryClass = "hep-ph",
    doi = "10.1007/JHEP06(2023)154",
    journal = "JHEP",
    volume = "06",
    pages = "154",
    year = "2023"
}

@article{Haxton:2024lyc,
    author = "Haxton, Wick and McElvain, Kenneth and Menzo, Tony and Rule, Evan and Zupan, Jure",
    title = "{Effective theory tower for \ensuremath{\mu} \textrightarrow{} e conversion}",
    eprint = "2406.13818",
    archivePrefix = "arXiv",
    primaryClass = "hep-ph",
    reportNumber = "LA-UR-24-24937, N3AS-24-023",
    doi = "10.1007/JHEP11(2024)076",
    journal = "JHEP",
    volume = "11",
    pages = "076",
    year = "2024"
}

@article{Song:2025snz,
    author = "Song, Chuan-Qiang and Sun, Hao and Yu, Jiang-Hao",
    title = "{Systematic Spurion Matching between Low Energy EFT and Chiral Lagrangian}",
    eprint = "2501.09787",
    archivePrefix = "arXiv",
    primaryClass = "hep-ph",
    month = "1",
    year = "2025"
}

@article{Dekens:2020ttz,
    author = "Dekens, Wouter and de Vries, Jordy and Fuyuto, Kaori and Mereghetti, Emanuele and Zhou, Guanghui",
    title = "{Sterile neutrinos and neutrinoless double beta decay in effective field theory}",
    eprint = "2002.07182",
    archivePrefix = "arXiv",
    primaryClass = "hep-ph",
    reportNumber = "LA-UR-20-21376",
    doi = "10.1007/JHEP06(2020)097",
    journal = "JHEP",
    volume = "06",
    pages = "097",
    year = "2020"
}

@article{Gasser:1984gg,
    author = "Gasser, J. and Leutwyler, H.",
    title = "{Chiral Perturbation Theory: Expansions in the Mass of the Strange Quark}",
    reportNumber = "CERN-TH-3798",
    doi = "10.1016/0550-3213(85)90492-4",
    journal = "Nucl. Phys. B",
    volume = "250",
    pages = "465--516",
    year = "1985"
}

@article{Jenkins:2017jig,
    author = "Jenkins, Elizabeth E. and Manohar, Aneesh V. and Stoffer, Peter",
    title = "{Low-Energy Effective Field Theory below the Electroweak Scale: Operators and Matching}",
    eprint = "1709.04486",
    archivePrefix = "arXiv",
    primaryClass = "hep-ph",
    doi = "10.1007/JHEP03(2018)016",
    journal = "JHEP",
    volume = "03",
    pages = "016",
    year = "2018",
    note = "[Erratum: JHEP 12, 043 (2023)]"
}

@article{Gasser:1987rb,
    author = "Gasser, J. and Sainio, M. E. and Svarc, A.",
    title = "{Nucleons with chiral loops}",
    reportNumber = "BUTP-87-17",
    doi = "10.1016/0550-3213(88)90108-3",
    journal = "Nucl. Phys. B",
    volume = "307",
    pages = "779--853",
    year = "1988"
}

@article{Dekens:2018pbu,
    author = "Dekens, Wouter and Jenkins, Elizabeth E. and Manohar, Aneesh V. and Stoffer, Peter",
    title = "{Non-perturbative effects in $\mu \to e \gamma$}",
    eprint = "1810.05675",
    archivePrefix = "arXiv",
    primaryClass = "hep-ph",
    doi = "10.1007/JHEP01(2019)088",
    journal = "JHEP",
    volume = "01",
    pages = "088",
    year = "2019"
}

@article{Li:2024ghg,
    author = "Li, Xuan-He and Sun, Hao and Tang, Feng-Jie and Yu, Jiang-Hao",
    title = "{Complete CP eigen-bases of mesonic chiral Lagrangian up to p$^{8}$-order}",
    eprint = "2404.14152",
    archivePrefix = "arXiv",
    primaryClass = "hep-ph",
    doi = "10.1007/JHEP08(2024)189",
    journal = "JHEP",
    volume = "08",
    pages = "189",
    year = "2024"
}

@article{Song:2024fae,
    author = "Song, Chuan-Qiang and Sun, Hao and Yu, Jiang-Hao",
    title = "{Complete CP-eigen bases of meson-baryon chiral lagrangian up to p$^{5}$-order}",
    eprint = "2404.15047",
    archivePrefix = "arXiv",
    primaryClass = "hep-ph",
    doi = "10.1007/JHEP09(2024)171",
    journal = "JHEP",
    volume = "09",
    pages = "171",
    year = "2024"
}

@article{Mateu:2007tr,
    author = "Mateu, V. and Portoles, J.",
    title = "{Form-factors in radiative pion decay}",
    eprint = "0706.1039",
    archivePrefix = "arXiv",
    primaryClass = "hep-ph",
    reportNumber = "IFIC-07-29, FTUV-07-0607",
    doi = "10.1140/epjc/s10052-007-0393-5",
    journal = "Eur. Phys. J. C",
    volume = "52",
    pages = "325--338",
    year = "2007"
}

@article{Manohar:1983md,
    author = "Manohar, Aneesh and Georgi, Howard",
    title = "{Chiral Quarks and the Nonrelativistic Quark Model}",
    reportNumber = "HUTP-83/A042a",
    doi = "10.1016/0550-3213(84)90231-1",
    journal = "Nucl. Phys. B",
    volume = "234",
    pages = "189--212",
    year = "1984"
}

@book{Scherer:2012xha,
    author = "Scherer, Stefan and Schindler, Matthias R.",
    title = "{A Primer for Chiral Perturbation Theory}",
    doi = "10.1007/978-3-642-19254-8",
    isbn = "978-3-642-19253-1",
    volume = "830",
    year = "2012"
}

@article{Liao:2020zyx,
    author = "Liao, Yi and Ma, Xiao-Dong and Wang, Quan-Yu",
    title = "{Extending low energy effective field theory with a complete set of dimension-7 operators}",
    eprint = "2005.08013",
    archivePrefix = "arXiv",
    primaryClass = "hep-ph",
    doi = "10.1007/JHEP08(2020)162",
    journal = "JHEP",
    volume = "08",
    pages = "162",
    year = "2020"
}

@article{Georgi:1993mps,
    author = "Georgi, H.",
    title = "{Effective field theory}",
    doi = "10.1146/annurev.ns.43.120193.001233",
    journal = "Ann. Rev. Nucl. Part. Sci.",
    volume = "43",
    pages = "209--252",
    year = "1993"
}

@article{Fettes:2000gb,
    author = "Fettes, Nadia and Meissner, Ulf-G. and Mojzis, Martin and Steininger, Sven",
    title = "{The Chiral effective pion nucleon Lagrangian of order p**4}",
    eprint = "hep-ph/0001308",
    archivePrefix = "arXiv",
    reportNumber = "FZJ-IKP-TH-2000-04",
    doi = "10.1006/aphy.2000.6059",
    journal = "Annals Phys.",
    volume = "283",
    pages = "273--302",
    year = "2000",
    note = "[Erratum: Annals Phys. 288, 249--250 (2001)]"
}

@article{Cata:2007ns,
    author = "Cata, O. and Mateu, V.",
    title = "{Chiral perturbation theory with tensor sources}",
    eprint = "0705.2948",
    archivePrefix = "arXiv",
    primaryClass = "hep-ph",
    reportNumber = "FTUV-07-05-21, IFIC-07-23",
    doi = "10.1088/1126-6708/2007/09/078",
    journal = "JHEP",
    volume = "09",
    pages = "078",
    year = "2007"
}

@article{Li:2020tsi,
    author = "Li, Hao-Lin and Ren, Zhe and Xiao, Ming-Lei and Yu, Jiang-Hao and Zheng, Yu-Hui",
    title = "{Low energy effective field theory operator basis at d \ensuremath{\leq} 9}",
    eprint = "2012.09188",
    archivePrefix = "arXiv",
    primaryClass = "hep-ph",
    doi = "10.1007/JHEP06(2021)138",
    journal = "JHEP",
    volume = "06",
    pages = "138",
    year = "2021"
}

@article{Murphy:2020cly,
    author = "Murphy, Christopher W.",
    title = "{Low-Energy Effective Field Theory below the Electroweak Scale: Dimension-8 Operators}",
    eprint = "2012.13291",
    archivePrefix = "arXiv",
    primaryClass = "hep-ph",
    doi = "10.1007/JHEP04(2021)101",
    journal = "JHEP",
    volume = "04",
    pages = "101",
    year = "2021"
}

@article{Cirigliano:2018yza,
    author = "Cirigliano, V. and Dekens, W. and de Vries, J. and Graesser, M. L. and Mereghetti, E.",
    title = "{A neutrinoless double beta decay master formula from effective field theory}",
    eprint = "1806.02780",
    archivePrefix = "arXiv",
    primaryClass = "hep-ph",
    reportNumber = "LA-UR-18-24895, Nikhef 2018-023, NIKHEF-2018-023, DESY-18-072",
    doi = "10.1007/JHEP12(2018)097",
    journal = "JHEP",
    volume = "12",
    pages = "097",
    year = "2018"
}

@article{Graesser:2016bpz,
    author = "Graesser, Michael L.",
    title = "{An electroweak basis for neutrinoless double $\beta$ decay}",
    eprint = "1606.04549",
    archivePrefix = "arXiv",
    primaryClass = "hep-ph",
    reportNumber = "LA-UR-16-23550",
    doi = "10.1007/JHEP08(2017)099",
    journal = "JHEP",
    volume = "08",
    pages = "099",
    year = "2017"
}

@article{Prezeau:2003xn,
    author = "Prezeau, Gary and Ramsey-Musolf, M. and Vogel, Petr",
    title = "{Neutrinoless double beta decay and effective field theory}",
    eprint = "hep-ph/0303205",
    archivePrefix = "arXiv",
    doi = "10.1103/PhysRevD.68.034016",
    journal = "Phys. Rev. D",
    volume = "68",
    pages = "034016",
    year = "2003"
}

@article{Li:2020gnx,
    author = "Li, Hao-Lin and Ren, Zhe and Shu, Jing and Xiao, Ming-Lei and Yu, Jiang-Hao and Zheng, Yu-Hui",
    title = "{Complete set of dimension-eight operators in the standard model effective field theory}",
    eprint = "2005.00008",
    archivePrefix = "arXiv",
    primaryClass = "hep-ph",
    doi = "10.1103/PhysRevD.104.015026",
    journal = "Phys. Rev. D",
    volume = "104",
    number = "1",
    pages = "015026",
    year = "2021"
}

@article{Li:2020xlh,
    author = "Li, Hao-Lin and Ren, Zhe and Xiao, Ming-Lei and Yu, Jiang-Hao and Zheng, Yu-Hui",
    title = "{Complete set of dimension-nine operators in the standard model effective field theory}",
    eprint = "2007.07899",
    archivePrefix = "arXiv",
    primaryClass = "hep-ph",
    doi = "10.1103/PhysRevD.104.015025",
    journal = "Phys. Rev. D",
    volume = "104",
    number = "1",
    pages = "015025",
    year = "2021"
}

@article{Li:2022tec,
    author = "Li, Hao-Lin and Ren, Zhe and Xiao, Ming-Lei and Yu, Jiang-Hao and Zheng, Yu-Hui",
    title = "{Operators for generic effective field theory at any dimension: on-shell amplitude basis construction}",
    eprint = "2201.04639",
    archivePrefix = "arXiv",
    primaryClass = "hep-ph",
    doi = "10.1007/JHEP04(2022)140",
    journal = "JHEP",
    volume = "04",
    pages = "140",
    year = "2022"
}

@article{Weinberg:1968de,
    author = "Weinberg, Steven",
    title = "{Nonlinear realizations of chiral symmetry}",
    doi = "10.1103/PhysRev.166.1568",
    journal = "Phys. Rev.",
    volume = "166",
    pages = "1568--1577",
    year = "1968"
}

@article{Weinberg:1978kz,
    author = "Weinberg, Steven",
    editor = "Deser, S.",
    title = "{Phenomenological Lagrangians}",
    reportNumber = "HUTP-78-A051A",
    doi = "10.1016/0378-4371(79)90223-1",
    journal = "Physica A",
    volume = "96",
    number = "1-2",
    pages = "327--340",
    year = "1979"
}

@article{Gasser:1983yg,
    author = "Gasser, J. and Leutwyler, H.",
    title = "{Chiral Perturbation Theory to One Loop}",
    reportNumber = "CERN-TH-3689",
    doi = "10.1016/0003-4916(84)90242-2",
    journal = "Annals Phys.",
    volume = "158",
    pages = "142",
    year = "1984"
}

@article{Lindner:2016wff,
    author = "Lindner, Manfred and Rodejohann, Werner and Xu, Xun-Jie",
    title = "{Coherent Neutrino-Nucleus Scattering and new Neutrino Interactions}",
    eprint = "1612.04150",
    archivePrefix = "arXiv",
    primaryClass = "hep-ph",
    doi = "10.1007/JHEP03(2017)097",
    journal = "JHEP",
    volume = "03",
    pages = "097",
    year = "2017"
}

@article{Li:2024iij,
    author = "Li, Gang and Song, Chuan-Qiang and Tang, Feng-Jie and Yu, Jiang-Hao",
    title = "{Constraints on neutrino nonstandard interactions from COHERENT, PandaX-4T and XENONnT}",
    eprint = "2409.04703",
    archivePrefix = "arXiv",
    primaryClass = "hep-ph",
    doi = "10.1103/PhysRevD.111.035002",
    journal = "Phys. Rev. D",
    volume = "111",
    number = "3",
    pages = "035002",
    year = "2025"
}

@article{deVries:2023sux,
    author = {de Vries, J. and K\"orber, C. and Nogga, A. and Shain, S.},
    title = "{Dark matter scattering off $ ^{4}$He in chiral effective field theory}",
    eprint = "2310.11343",
    archivePrefix = "arXiv",
    primaryClass = "hep-ph",
    doi = "10.1140/epjc/s10052-024-13477-z",
    journal = "Eur. Phys. J. C",
    volume = "84",
    number = "11",
    pages = "1138",
    year = "2024"
}

@article{Buchoff:2015qwa,
    author = "Buchoff, Michael I. and Wagman, Michael",
    title = "{Perturbative Renormalization of Neutron-Antineutron Operators}",
    eprint = "1506.00647",
    archivePrefix = "arXiv",
    primaryClass = "hep-ph",
    doi = "10.1103/PhysRevD.93.016005",
    journal = "Phys. Rev. D",
    volume = "93",
    number = "1",
    pages = "016005",
    year = "2016",
    note = "[Erratum: Phys.Rev.D 98, 079901 (2018)]"
}

@article{Hoferichter:2020osn,
    author = "Hoferichter, Martin and Men\'endez, Javier and Schwenk, Achim",
    title = "{Coherent elastic neutrino-nucleus scattering: EFT analysis and nuclear responses}",
    eprint = "2007.08529",
    archivePrefix = "arXiv",
    primaryClass = "hep-ph",
    reportNumber = "INT-PUB-20-026",
    doi = "10.1103/PhysRevD.102.074018",
    journal = "Phys. Rev. D",
    volume = "102",
    number = "7",
    pages = "074018",
    year = "2020"
}

@article{Du:2020dwr,
    author = "Du, Yong and Li, Hao-Lin and Tang, Jian and Vihonen, Sampsa and Yu, Jiang-Hao",
    title = "{Non-standard interactions in SMEFT confronted with terrestrial neutrino experiments}",
    eprint = "2011.14292",
    archivePrefix = "arXiv",
    primaryClass = "hep-ph",
    reportNumber = "ACFI-T20-15",
    doi = "10.1007/JHEP03(2021)019",
    journal = "JHEP",
    volume = "03",
    pages = "019",
    year = "2021"
}

@article{Bischer:2019ttk,
    author = "Bischer, Ingolf and Rodejohann, Werner",
    title = "{General neutrino interactions from an effective field theory perspective}",
    eprint = "1905.08699",
    archivePrefix = "arXiv",
    primaryClass = "hep-ph",
    doi = "10.1016/j.nuclphysb.2019.114746",
    journal = "Nucl. Phys. B",
    volume = "947",
    pages = "114746",
    year = "2019"
}

@article{Altmannshofer:2018xyo,
    author = "Altmannshofer, Wolfgang and Tammaro, Michele and Zupan, Jure",
    title = "{Non-standard neutrino interactions and low energy experiments}",
    eprint = "1812.02778",
    archivePrefix = "arXiv",
    primaryClass = "hep-ph",
    doi = "10.1007/JHEP11(2021)113",
    journal = "JHEP",
    volume = "09",
    pages = "083",
    year = "2019",
    note = "[Erratum: JHEP 11, 113 (2021)]"
}

@article{AristizabalSierra:2018eqm,
    author = "Aristizabal Sierra, D. and De Romeri, Valentina and Rojas, N.",
    title = "{COHERENT analysis of neutrino generalized interactions}",
    eprint = "1806.07424",
    archivePrefix = "arXiv",
    primaryClass = "hep-ph",
    doi = "10.1103/PhysRevD.98.075018",
    journal = "Phys. Rev. D",
    volume = "98",
    pages = "075018",
    year = "2018"
}

@article{Farzan:2018gtr,
    author = "Farzan, Yasaman and Lindner, Manfred and Rodejohann, Werner and Xu, Xun-Jie",
    title = "{Probing neutrino coupling to a light scalar with coherent neutrino scattering}",
    eprint = "1802.05171",
    archivePrefix = "arXiv",
    primaryClass = "hep-ph",
    doi = "10.1007/JHEP05(2018)066",
    journal = "JHEP",
    volume = "05",
    pages = "066",
    year = "2018"
}

@article{Haxton:2022piv,
    author = "Haxton, W. C. and Rule, Evan and McElvain, Ken and Ramsey-Musolf, Michael J.",
    title = "{Nuclear-level effective theory of \ensuremath{\mu}\textrightarrow{}e conversion: Formalism and applications}",
    eprint = "2208.07945",
    archivePrefix = "arXiv",
    primaryClass = "nucl-th",
    doi = "10.1103/PhysRevC.107.035504",
    journal = "Phys. Rev. C",
    volume = "107",
    number = "3",
    pages = "035504",
    year = "2023"
}

@article{Cirigliano:2022ekw,
    author = "Cirigliano, Vincenzo and Fuyuto, Kaori and Ramsey-Musolf, Michael J. and Rule, Evan",
    title = "{Next-to-leading order scalar contributions to \ensuremath{\mu}\textrightarrow{}e conversion}",
    eprint = "2203.09547",
    archivePrefix = "arXiv",
    primaryClass = "hep-ph",
    reportNumber = "ACFI-T22-04, INT-PUB-22-009, LA-UR-21-324-32420",
    doi = "10.1103/PhysRevC.105.055504",
    journal = "Phys. Rev. C",
    volume = "105",
    number = "5",
    pages = "055504",
    year = "2022"
}

@article{Rule:2021oxe,
    author = "Rule, Evan and Haxton, W. C. and McElvain, Ken and McElvain, Kenneth",
    title = "{Nuclear-Level Effective Theory of \ensuremath{\mu}\textrightarrow{}e Conversion}",
    eprint = "2109.13503",
    archivePrefix = "arXiv",
    primaryClass = "hep-ph",
    doi = "10.1103/PhysRevLett.130.131901",
    journal = "Phys. Rev. Lett.",
    volume = "130",
    number = "13",
    pages = "131901",
    year = "2023"
}

@article{Bartolotta:2017mff,
    author = "Bartolotta, Anthony and Ramsey-Musolf, Michael J.",
    title = "{Coherent $\mu-e$ conversion at next-to-leading order}",
    eprint = "1710.02129",
    archivePrefix = "arXiv",
    primaryClass = "hep-ph",
    reportNumber = "CALT-TH-2017-005, ACFI-T17-21",
    doi = "10.1103/PhysRevC.98.015208",
    journal = "Phys. Rev. C",
    volume = "98",
    number = "1",
    pages = "015208",
    year = "2018"
}

@article{Cirigliano:2017azj,
    author = "Cirigliano, Vincenzo and Davidson, Sacha and Kuno, Yoshitaka",
    title = "{Spin-dependent $\mu \to e$ conversion}",
    eprint = "1703.02057",
    archivePrefix = "arXiv",
    primaryClass = "hep-ph",
    reportNumber = "LA-UR-17-21718, OUHEP-17-1",
    doi = "10.1016/j.physletb.2017.05.053",
    journal = "Phys. Lett. B",
    volume = "771",
    pages = "242--246",
    year = "2017"
}

@article{Crivellin:2017rmk,
    author = "Crivellin, Andreas and Davidson, Sacha and Pruna, Giovanni Marco and Signer, Adrian",
    title = "{Renormalisation-group improved analysis of $\mu\to e$ processes in a systematic effective-field-theory approach}",
    eprint = "1702.03020",
    archivePrefix = "arXiv",
    primaryClass = "hep-ph",
    reportNumber = "PSI-PR-17-01, ZU-TH-01-17",
    doi = "10.1007/JHEP05(2017)117",
    journal = "JHEP",
    volume = "05",
    pages = "117",
    year = "2017"
}

@article{Liao:2019gex,
    author = "Liao, Yi and Ma, Xiao-Dong and Wang, Hao-Lin",
    title = "{Effective field theory approach to lepton number violating decays $K^\pm\rightarrow \pi^\mp l^{\pm}l^{\pm}$: short-distance contribution}",
    eprint = "1909.06272",
    archivePrefix = "arXiv",
    primaryClass = "hep-ph",
    doi = "10.1007/JHEP01(2020)127",
    journal = "JHEP",
    volume = "01",
    pages = "127",
    year = "2020"
}

@article{Bishara:2017pfq,
    author = "Bishara, Fady and Brod, Joachim and Grinstein, Benjamin and Zupan, Jure",
    title = "{From quarks to nucleons in dark matter direct detection}",
    eprint = "1707.06998",
    archivePrefix = "arXiv",
    primaryClass = "hep-ph",
    reportNumber = "DO-TH-17-10, OUTP-17-07P, CERN-TH-2017-157",
    doi = "10.1007/JHEP11(2017)059",
    journal = "JHEP",
    volume = "11",
    pages = "059",
    year = "2017"
}

@article{Korber:2017ery,
    author = {K\"orber, C. and Nogga, A. and de Vries, J.},
    title = "{First-principle calculations of Dark Matter scattering off light nuclei}",
    eprint = "1704.01150",
    archivePrefix = "arXiv",
    primaryClass = "hep-ph",
    doi = "10.1103/PhysRevC.96.035805",
    journal = "Phys. Rev. C",
    volume = "96",
    number = "3",
    pages = "035805",
    year = "2017"
}

@article{Bishara:2016hek,
    author = "Bishara, Fady and Brod, Joachim and Grinstein, Benjamin and Zupan, Jure",
    title = "{Chiral Effective Theory of Dark Matter Direct Detection}",
    eprint = "1611.00368",
    archivePrefix = "arXiv",
    primaryClass = "hep-ph",
    reportNumber = "DO-TH-16-28, OUTP-16-24P, CERN-TH-2016-259",
    doi = "10.1088/1475-7516/2017/02/009",
    journal = "JCAP",
    volume = "02",
    pages = "009",
    year = "2017"
}

@article{Cirigliano:2012pq,
    author = "Cirigliano, Vincenzo and Graesser, Michael L. and Ovanesyan, Grigory",
    title = "{WIMP-nucleus scattering in chiral effective theory}",
    eprint = "1205.2695",
    archivePrefix = "arXiv",
    primaryClass = "hep-ph",
    doi = "10.1007/JHEP10(2012)025",
    journal = "JHEP",
    volume = "10",
    pages = "025",
    year = "2012"
}

@article{Fitzpatrick:2012ix,
    author = "Fitzpatrick, A. Liam and Haxton, Wick and Katz, Emanuel and Lubbers, Nicholas and Xu, Yiming",
    title = "{The Effective Field Theory of Dark Matter Direct Detection}",
    eprint = "1203.3542",
    archivePrefix = "arXiv",
    primaryClass = "hep-ph",
    doi = "10.1088/1475-7516/2013/02/004",
    journal = "JCAP",
    volume = "02",
    pages = "004",
    year = "2013"
}

@article{Li:2020lba,
    author = "Li, Tong and Ma, Xiao-Dong and Schmidt, Michael A.",
    title = "{General neutrino interactions with sterile neutrinos in light of coherent neutrino-nucleus scattering and meson invisible decays}",
    eprint = "2005.01543",
    archivePrefix = "arXiv",
    primaryClass = "hep-ph",
    doi = "10.1007/JHEP07(2020)152",
    journal = "JHEP",
    volume = "07",
    pages = "152",
    year = "2020"
}

@article{Grzadkowski:2010es,
    author = "Grzadkowski, B. and Iskrzynski, M. and Misiak, M. and Rosiek, J.",
    title = "{Dimension-Six Terms in the Standard Model Lagrangian}",
    eprint = "1008.4884",
    archivePrefix = "arXiv",
    primaryClass = "hep-ph",
    reportNumber = "IFT-9-2010, TTP10-35",
    doi = "10.1007/JHEP10(2010)085",
    journal = "JHEP",
    volume = "10",
    pages = "085",
    year = "2010"
}

@article{Li:2021tsq,
    author = "Li, Hao-Lin and Ren, Zhe and Xiao, Ming-Lei and Yu, Jiang-Hao and Zheng, Yu-Hui",
    title = "{Operator bases in effective field theories with sterile neutrinos: d \ensuremath{\leq} 9}",
    eprint = "2105.09329",
    archivePrefix = "arXiv",
    primaryClass = "hep-ph",
    doi = "10.1007/JHEP11(2021)003",
    journal = "JHEP",
    volume = "11",
    pages = "003",
    year = "2021"
}

@article{Chala:2020vqp,
    author = "Chala, Mikael and Titov, Arsenii",
    title = "{One-loop matching in the SMEFT extended with a sterile neutrino}",
    eprint = "2001.07732",
    archivePrefix = "arXiv",
    primaryClass = "hep-ph",
    doi = "10.1007/JHEP05(2020)139",
    journal = "JHEP",
    volume = "05",
    pages = "139",
    year = "2020"
}

@article{Fearing:1994ga,
    author = "Fearing, H. W. and Scherer, S.",
    title = "{Extension of the chiral perturbation theory meson Lagrangian to order p(6)}",
    eprint = "hep-ph/9408346",
    archivePrefix = "arXiv",
    reportNumber = "TRI-PP-94-68",
    doi = "10.1103/PhysRevD.53.315",
    journal = "Phys. Rev. D",
    volume = "53",
    pages = "315--348",
    year = "1996"
}

@article{Bijnens:2001bb,
    author = "Bijnens, J. and Girlanda, L. and Talavera, P.",
    title = "{The Anomalous chiral Lagrangian of order p**6}",
    eprint = "hep-ph/0110400",
    archivePrefix = "arXiv",
    reportNumber = "LU-TP-01-34, DFPD-01-TH-25, CPT-2001-P4256",
    doi = "10.1007/s100520100887",
    journal = "Eur. Phys. J. C",
    volume = "23",
    pages = "539--544",
    year = "2002"
}

@article{Ebertshauser:2001nj,
    author = "Ebertshauser, T. and Fearing, H. W. and Scherer, S.",
    title = "{The Anomalous chiral perturbation theory meson Lagrangian to order p**6 revisited}",
    eprint = "hep-ph/0110261",
    archivePrefix = "arXiv",
    reportNumber = "MKPH-T-01-22, TRI-PP-01-34",
    doi = "10.1103/PhysRevD.65.054033",
    journal = "Phys. Rev. D",
    volume = "65",
    pages = "054033",
    year = "2002"
}

@article{Bijnens:2018lez,
    author = "Bijnens, Johan and Hermansson-Truedsson, Nils and Wang, Si",
    title = "{The order p$^{8}$ mesonic chiral Lagrangian}",
    eprint = "1810.06834",
    archivePrefix = "arXiv",
    primaryClass = "hep-ph",
    reportNumber = "LU TP 18-34",
    doi = "10.1007/JHEP01(2019)102",
    journal = "JHEP",
    volume = "01",
    pages = "102",
    year = "2019"
}

@article{Bijnens:2023hyv,
    author = "Bijnens, Johan and Hermansson-Truedsson, Nils and Ruiz-Vidal, Joan",
    title = "{The anomalous chiral Lagrangian at order p$^{8}$}",
    eprint = "2310.20547",
    archivePrefix = "arXiv",
    primaryClass = "hep-ph",
    doi = "10.1007/JHEP01(2024)009",
    journal = "JHEP",
    volume = "01",
    pages = "009",
    year = "2024"
}

\end{document}